\documentclass[12pt, letterpaper]{extarticle}

\usepackage{subcaption}
\usepackage{fullpage}
\usepackage[switch]{lineno}
\usepackage{amsmath}
\usepackage{amssymb}
\usepackage{rotating}
\usepackage{array}
\usepackage{mathtools}
\usepackage[ruled]{algorithm2e}
\usepackage{algorithmic}
\usepackage{bm}
\usepackage{breqn}
\usepackage{comment}
\usepackage{enumitem}
\usepackage{graphics}
\usepackage{graphicx}
\usepackage{latexsym}
\usepackage{mathrsfs}
\usepackage{morefloats}
\usepackage{nicefrac}
\usepackage{authblk}
\usepackage{pifont}
\usepackage{times}
\usepackage{xcolor}
\usepackage[para,online,flushleft]{threeparttable}
\usepackage{booktabs}

\usepackage[hyphens]{url}
\usepackage{hyperref}
\hypersetup{colorlinks=false,breaklinks=true}

\usepackage{etoolbox}
\AtBeginEnvironment{quote}{\par\singlespacing\small}

\title{Causal evidence of racial and institutional biases in accessing paywalled articles and scientific data}

\author[1+]{Hazem Ibrahim}
\author[1+]{Fengyuan Liu}
\author[1]{Khalid Mengal}
\author[2]{Wisam Alshaibi}
\author[2]{\\Aaron R. Kaufman}
\author[1*]{Yasir Zaki}
\author[1*]{Talal Rahwan}

\affil[1]{\normalsize Computer Science, Science Division, New York University Abu Dhabi, UAE.}
\affil[2]{\normalsize Social Science Division, New York University Abu Dhabi, UAE}
\affil[+]{\footnotesize Joint first author}
\affil[*]{\footnotesize Corresponding authors. E-mail: \{yasir.zaki,talal.rahwan\}@nyu.edu}
\date{}

\renewcommand{\bf}{}

\usepackage{setspace}
\usepackage{colortbl}
\usepackage{placeins}
\usepackage{multirow}
\usepackage{longtable}
\usepackage{float}
\usepackage{multicol}
\usepackage{supertabular}
\usepackage[T1]{fontenc}
\usepackage[graphicx]{realboxes}
\usepackage{makecell}
\usepackage{tabularx}
\usepackage{arydshln}
\newcolumntype{C}[1]{>{\centering\arraybackslash}p{#1}}

\begin{document}

\maketitle 

\baselineskip22pt

\begin{abstract}
\noindent

Scientific progress fundamentally depends on researchers' ability to access and build upon the work of others. Yet, a majority of published work remains behind expensive paywalls, limiting access to universities that can afford subscriptions. Furthermore, even when articles are accessible, the underlying datasets could be restricted, available only through a “reasonable request” to the authors. One way researchers could overcome these barriers is by relying on informal channels, such as emailing authors directly, to obtain paywalled articles or restricted datasets. However, whether these informal channels are hindered by racial and/or institutional biases remains unknown. Here, we combine survey data, semi-structured interviews, large-scale observational analysis, and two randomized audit experiments to examine racial and institutional disparities in access to scientific knowledge. Surveying a sample of researchers worldwide, those in the Global South report markedly lower institutional access to the published literature and depend more heavily on informal channels---such as emailing authors directly---to obtain papers and data; interviews with researchers from across the Global South elaborate the workarounds they rely on and the geopolitical and racialized frictions they encounter. Our analysis of 250 million articles reveals that researchers in the Global South cite paywalled papers at significantly lower rates than their Global North counterparts---a gap associated with reduced knowledge breadth and scholarly impact. Using citation-context classification, we further find that papers whose data is available only upon request are less likely to be cited for the purpose of reusing their data, and that this penalty falls disproportionately on researchers in the Global South. To interrogate the mechanisms underlying these phenomena, we conduct two randomized email audit studies in which fictional PhD students differing in racial background and institutional affiliation request access to paywalled articles (N = 18,000) and datasets (N = 16,000). 
We find that racial identity influences response rate to both paywalled article and dataset requests, whereas institutional affiliation influences access to datasets. These findings reveal how informal gatekeeping can perpetuate structural inequities in science, highlighting the need for stronger data-sharing mandates and more equitable open access policies.
\end{abstract}

\clearpage
\section*{Introduction}

Scientific progress is fundamentally cumulative, with later advances relying on access to the knowledge that came before. As Newton famously remarked, ``If I have seen further, it is by standing on the shoulders of giants.''
This steady accumulation of knowledge relies on the ability of scientists to access, scrutinize, and build upon the work of others. Without such access to publications, data, code, and methodologies, the scientific enterprise becomes siloed or even stalls. In many ways, knowledge is a public good: its use by one researcher does not diminish its value to others, and broad access can accelerate collective progress. Championing this principle, the open science movement has emerged to ensure that knowledge is widely and equitably accessible: by removing barriers to published articles, encouraging the sharing of datasets where ethically and legally feasible, and promoting transparent research practices, open science promises not only faster discovery but also more reproducible and inclusive research.~\cite{smith2017knowledge, shu2018such, mckiernan2016open, national2018open, allen2019open}. 

In practice, however, many obstacles to this ideal remain. A substantial portion of the scientific literature continues to be published behind paywalls~\cite{boai2002budapest, piwowar2018state, brainard2024open}, limiting access to institutions with expensive subscriptions. Even when articles are accessible, the underlying data may not be. Despite journals increasingly mandating the inclusion of Data Availability Statements (DAS)~\cite{federer2018data}, as well as the proliferation of research data repositories (e.g., figshare~\cite{figshareFigshareCredit}, dataverse~\cite{harvard_dataverse}, huggingface~\cite{huggingfaceHuggingFace}), compliance with data sharing policies is inconsistent. Many authors offer data ``upon request'', yet studies have shown that such requests often go unanswered~\cite{milham2018assessment, tenopir2011data, tenopir2020data, germaine2024lack, tedersoo2021data, gabelica2022many, jiao2024data}. Even when data requests receive responses, access is not always unconditional: authors may offer to exchange data for co-authorship, or even demand financial compensation. As scientific {research is becoming more computationally intensive}, datasets are ever more crucial to the advancement of science~\cite{hey2009fourth,kitchin2014data,leonelli2020scientific}. The accessibility of data is not merely a convenience but a critical component for the replication, validation, and extension of scientific findings~\cite{milham2018assessment}. As such, informal discretionary practices create uncertainty and may introduce new barriers that formal open science policies were designed to eliminate. 

Whether these barriers affect all researchers equally remains an open question. Much of the infrastructure of global science is concentrated in the Global North~\cite{wagner2024developing}, among institutions with greater funding~\cite{rakotonarivo2023global}, access to subscription databases~\cite{mueller2020does}, and established research networks~\cite{mannocci2019geographical,pillai2018next}. In contrast, scholars at institutions in the Global South often operate with fewer resources and limited formal access~\cite{gaillard2010measuring, bezuidenhout2017beyond, albornoz2018framing}. Even as open science policies expand, their benefits may be unequally distributed, potentially reinforcing existing disparities in who can fully participate in scientific discovery. This possibility carries particular importance given the growing relevance of Global South researchers to fields that address global challenges. In disciplines such as global health, development economics, and climate adaptation, local context and sociocultural expertise are critical. When researchers closest to the problem are unable to access the literature or data needed to contribute, it not only limits individual opportunity, it impoverishes science itself~\cite{liu2023non,wapman2022quantifying}.

The broader literature on racial bias and academic gatekeeping shows why disparities in access may arise\footnote{We acknowledge, with no small irony, that our conceptualization of race itself is limited by the centralization of academic studies of race in the Global North. The structure of racial hierarchy and its importance are deeply contextual, varying widely across time and space, and even subnationally. We necessarily coarsen this nuance for the sake of our experiments, and define race as per the American Psychological Association: ``The social construction and categorization of people based on perceived shared physical traits that result in the maintenance of a sociopolitical hierarchy'' \cite{vandenbos2007apa}.}. Audit studies in labor economics consistently find that White applicants receive more callbacks than equally qualified Black or minority candidates~\cite{altonji1999race,pager2005walking,pager2009discrimination,stauffer2005existence,lahey2018discrimination,shore2021influence,banerjee2018large,gaddis2015discrimination}. In response to these patterns, some individuals engage in ``resume whitening'' to conceal racial cues~\cite{kang2016whitened}. Similar dynamics appear within academia. Non-White scientists are less likely to serve on editorial boards, tend to spend more time under review, receive fewer citations compared to textually similar papers, and are less frequently mentioned in the media~\cite{liu2023non,peng2024author}. Prestige-based inequalities amplify these racial disparities. High-prestige scientists receive more favorable peer reviews~\cite{smirnova2023bias}, and experimental evidence shows that prestige cues affect acceptance outcomes. For example, one study found that a manuscript attributed to a Nobel laureate received higher acceptance rates than an identical manuscript attributed to an early-career researcher~\cite{huber2022nobel}. At the institutional level, elite universities attract greater funding~\cite{szell2015research,ma2015anatomy}, supply a disproportionate share of the academic workforce~\cite{wapman2022quantifying}, and produce research that spreads more rapidly~\cite{morgan2018prestige}. Single-anonymous peer review can intensify these advantages by systematically benefiting prominent researchers and institutions~\cite{smirnova2023bias,tomkins2017reviewer}, which contributes to long-standing hierarchies in scientific evaluation~\cite{clauset2015systematic}. Together, this literature suggests that both racial identity and institutional prestige may shape who gains access to the materials, networks, and opportunities required to participate fully in scientific research.

\subsection*{Our contributions}
 
We start by studying whether and how key hurdles targeted by open science---paywalled papers and the lack of accessible data---disproportionately burden researchers in the Global South.\footnote{For the purposes of this paper, we define ``Global South'' in accordance with UN Trade and Development (see Methods for more details). However, we acknowledge that this binary definition contrasts with other, more nuanced conceptualizations of global inequality \cite{sud2022southern}, such as the World Bank's four-category scale from high-income to low-income, or theoretically-motivated constructions like Marxism's imperial core versus imperial periphery \cite{tickner2013core}.}. In Study~1, we survey a sample of researchers worldwide and conduct qualitative semi-structured interviews with researchers from universities across the Global South, providing direct insight into the lived experience of researchers navigating constraints in access to publications and data. Surveyed researchers in the Global South report markedly lower institutional access to the published literature, encounter paywalls more frequently, and rely more heavily on informal channels to obtain papers and data. Participants in our interviews describe the criticality of replication data and code to their research and teaching, the hoops they jump through to access paywalled papers, and the deteriorating funding environments that have damaged their research capacity in recent months and years.
 
Motivated by these researchers' stories, Studies~2 and 4 assess whether these individual experiences are reflected in aggregate patterns of scholarly citation. Study~2 draws on OpenAlex~\cite{priem2022openalex}, a large bibliometric database that has indexed over 250 million articles to date, to examine whether researchers from the Global South cite paywalled articles at different rates than those in the Global North. Study~4 turns to data access: using the Semantic Scholar database, we retrieve Data Availability Statements for 200,000 papers and classify 4 million citation contexts with a custom deep learning classifier, allowing us to determine whether a paper is cited for the purpose of reusing its data. Citation behavior, while not a direct measure of access, serves as a useful proxy: if researchers are systematically less likely to cite materials that require special access, this may reflect broader disparities in who can engage with those materials in the first place~\cite{evans2009open}. These analyses reveal stark divides: Global South researchers cite paywalled articles at lower rates than their Global North counterparts (Study~2), and are disproportionately less likely to cite a paper for the purpose of reusing its data when that data is available only upon request (Study~4).
 
Studies~3 and 5 investigate potential mechanisms behind such disparities by conducting two large-scale randomized audit experiments. Based on the extensive literature examining discrimination in non-academic contexts~ \cite{altonji1999race,pager2005walking,pager2009discrimination,stauffer2005existence,lahey2018discrimination,shore2021influence,banerjee2018large}, we hypothesize that whether one can successfully gain access to knowledge through informal channels may not only depend on what is being requested, but also on who is making the request: discrimination may play a role in Global South researchers' access to paywalled articles and restricted data. 
 
Building on recent work that distinguishes between statistical and taste-based discrimination~\cite{bohren2025systemic,celaya2023choosing}, we sent 34,000 email requests for articles and data to the corresponding authors of published papers. We randomly varied the race, country of origin, and institutional affiliation of the requesters to test whether these social signals influence recipients' willingness to share paywalled papers (Study~3) or replication data (Study~5). The randomized design allows us to isolate the causal effect of identity-related cues on researchers’ willingness to share, and to distinguish between several possible motives for discrimination. We find that racial identity significantly affects access to paywalled articles and datasets, with institutional affiliation also playing a role in accessing datasets.
 
We conclude by discussing how these interpersonal disparities may contribute to broader patterns of inequality in science. If researchers from less prestigious institutions or historically marginalized groups are less likely to receive the materials they need, they may be less able to publish, cite, and build on prior work. Over time, such disparities can compound, contributing to persistent inequalities in who participates in and benefits from the global production of knowledge. Our findings suggest that despite the promise of open science, informal access remains a key site of gatekeeping, one shaped by race and geography. To fully realize the ideals of openness and equity, the scientific community must address the challenge posed by informal access channels that continue to shape who gets to do science and who gets left behind.

\section*{Results}
\subsection*{Study~1: Global disparities in access to paywalled papers and data}
\subsubsection*{Global survey on access restrictions}

To document the practical challenges researchers face when accessing paywalled papers and restricted datasets, we surveyed 376 active researchers affiliated with institutions across 76 countries (216 from the Global North, 160 from the Global South; see Methods for sampling and survey design details). The sample spans all major academic disciplines and career stages from PhD students to full professors, and includes researchers from 36 Global North and 40 Global South countries (Supplementary Table~1). Below, we report survey findings organized around four themes: institutional infrastructure, paywall encounters, coping strategies, and the consequences of access barriers.

\begin{figure}[t!]
\centering
\includegraphics[width=\textwidth]{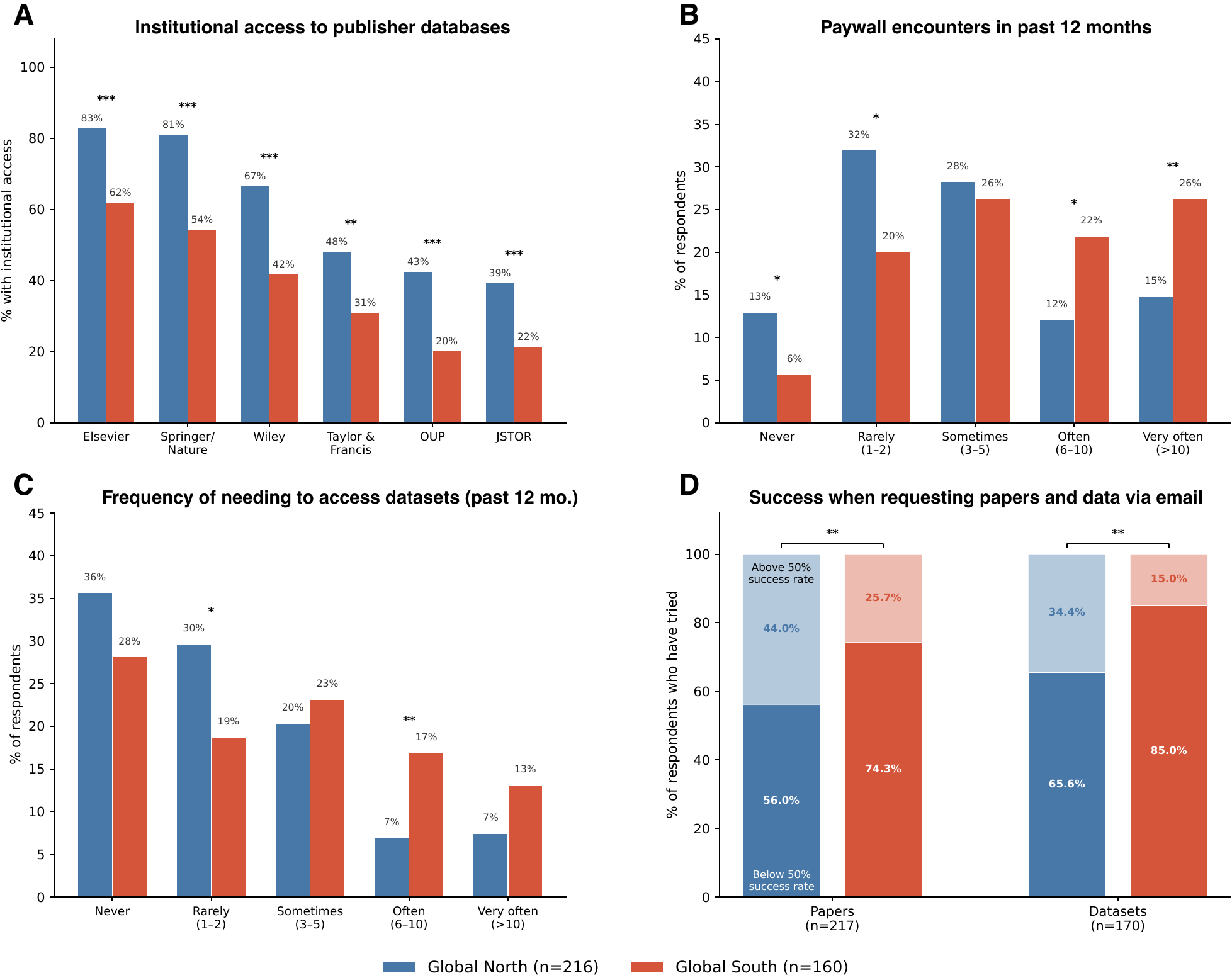}
\caption{\textbf{Survey results on access disparities between Global North and Global South researchers} ($N = 376$; Global North: $n = 216$, Global South: $n = 160$, across 76 countries). \textbf{A,} Percentage of respondents reporting institutional access to each publisher database ($^{***}p<0.001$, $^{**}p<0.01$, $^{*}p<0.05$; $\chi^2$ tests). \textbf{B,} Distribution of self-reported paywall encounter frequency in the past 12 months. Global South researchers are significantly more likely to report frequent paywall encounters ($\chi^2 = 17.16$, $p < 0.001$). \textbf{C,} Frequency of needing to access datasets associated with published papers in the past 12 months. \textbf{D,} Self-reported success rates when emailing authors, among respondents who have attempted such requests. For paywalled papers (left), success is defined as receiving a response; for datasets described as ``available upon request'' (right), as actually receiving the requested data, mirroring the wording of the survey items. Darker shading indicates the proportion reporting success less than 50\% of the time; lighter shading indicates more than 50\%. Global South researchers report significantly lower success rates for both papers ($p = 0.008$) and datasets ($p = 0.006$). Source data: Supplementary Table~26.}
\label{fig:survey}
\end{figure}
 
Global South researchers report substantially lower rates of institutional access to academic publisher databases across all major publishers (Figure~\ref{fig:survey}A). For instance, 83\% of Global North respondents report institutional access to Elsevier compared to 62\% in the Global South ($\chi^2 = 19.57$, $p < 0.001$). The gap is even more pronounced for Springer/Nature (81\% vs.\ 54\%, $\chi^2 = 29.35$, $p < 0.001$), Wiley (67\% vs.\ 42\%, $\chi^2 = 21.97$, $p < 0.001$), and Oxford University Press (43\% vs.\ 20\%, $\chi^2 = 19.55$, $p < 0.001$). These gaps are statistically significant for all six publishers shown in Figure~\ref{fig:survey}A ($p < 0.01$ in all cases; see Supplementary Table~2 for all twelve publishers surveyed).

The open-ended responses reinforce these quantitative patterns. In an open-ended question administered before any of the structured access items, respondents were asked to name the biggest challenges they face in accessing research materials, with published papers, datasets, code, and equipment offered as examples but no mention of paywalls, subscriptions, or cost. Even so, 34\% of Global South respondents spontaneously identified paywall barriers as a major obstacle, compared to 21\% in the Global North. Financial constraints were raised at comparable rates across regions (22\% vs.\ 20\%).

The tone of these descriptions differed markedly by region. Global North researchers tended to describe access barriers as occasional inconveniences within a well-functioning system: a researcher in the Netherlands wrote, ``our institute/university is well-funded so there is enough money to buy access, if needed.'' By contrast, Global South researchers described barriers as structural and persistent. A full professor in India noted that ``the huge subscription cost of journals'' and ``the exorbitant cost of publishing in open access'' jointly constrain both research inputs and outputs. A Nigerian researcher explained that the dollar-to-naira exchange rate renders Q1 and Q2 journal subscriptions effectively unaffordable. Several Argentine respondents described a recent systemic disruption in which the government withdrew national journal subscriptions entirely, with one writing: ``Scientists thus have two possibilities: restricting their bibliographic research to arXiv or open articles; or using Sci-Hub.'' Notably, Sci-Hub was mentioned exclusively by Global South respondents, suggesting that it could play a role as a critical lifeline for researchers without institutional access.
 
These infrastructure disparities translate directly into day-to-day research friction. Nearly half (48\%) of Global South researchers report encountering paywalled papers they cannot access ``often'' or ``very often'' (6 or more times in the past 12 months), compared to 27\% of Global North researchers ($\chi^2 = 17.16$, $p < 0.001$; Figure~\ref{fig:survey}B). Conversely, 13\% of Global North respondents report never encountering a paywall barrier, versus only 6\% in the Global South. A similar pattern emerges for data access: 30\% of Global South researchers report frequently needing to access datasets associated with published papers (6 or more times in the past 12 months), compared to just 14\% in the Global North (Figure~\ref{fig:survey}C).
 
When unable to access a paper through institutional channels, the strategies researchers use differ by region in revealing ways (Supplementary Table~3). The most common strategy among Global North researchers is searching for preprints (52\%), whereas Global South researchers rely most heavily on ResearchGate (48\%) and Sci-Hub (48\%). Global South researchers are also more likely to ask colleagues at other institutions for help (47\% vs.\ 35\%) and less likely to use interlibrary loan (14\% vs.\ 26\%), suggesting that their institutions may lack functioning interlibrary infrastructure. Notably, 46\% of Global North researchers report giving up and using a different paper when they cannot access the one they need, compared to 36\% in the Global South---potentially reflecting that for Global North researchers, a paywall is an inconvenience with substitutes, while for Global South researchers, the inaccessible paper may be uniquely relevant.
 
Among researchers who have emailed authors to request paywalled papers, Global South researchers report significantly lower success rates. Only 26\% of Global South researchers who have tried this approach report receiving responses more than half the time, compared to 44\% of Global North researchers ($\chi^2 = 7.06$, $p = 0.008$; Figure~\ref{fig:survey}D). The gap is even larger for dataset requests: just 15\% of Global South researchers report above-50\% success rates when requesting data described as ``available upon request,'' compared to 34\% in the Global North ($\chi^2 = 7.48$, $p = 0.006$; Figure~\ref{fig:survey}D). As we show in Studies 3 and 5, these self-reported disparities are directly corroborated by causal evidence from our randomized audit experiments.
 
Access barriers produce tangible downstream effects on research. Among Global South respondents, 71\% report at least one concrete consequence of paywall barriers, compared to 64\% of Global North respondents (Supplementary Table~4). While the most commonly reported consequences---narrowing the scope of a literature review (37--38\%) and submitting papers without citing known relevant work (32--34\%)---are comparably prevalent, the more severe consequences are disproportionately borne by Global South researchers. Specifically, 9\% of Global South respondents report changing the topic of a student's thesis or dissertation due to paywall barriers, versus 2\% in the Global North, and 12\% report dropping a line of inquiry entirely, versus 9\%. For data access barriers, 17\% of Global South respondents report changing a student's thesis topic due to the inability to access data, compared to 8\% in the Global North.

At the end of the survey, respondents were asked whether they were willing to participate in follow-up interviews. To assess whether the follow-up interview pool was systematically different from the broader survey sample, we repeated the primary analyses separately for respondents who opted in ($n = 211$) and those who did not ($n = 165$). Effect directions were consistent across both subgroups for all variables, though most comparisons in the opt-out subgroup did not reach statistical significance, likely due to the smaller number of Global South respondents in that group ($n = 52$ vs.\ $n = 108$; Supplementary Table~5).

\subsubsection*{Semi-structured interviews}
 
To complement the survey with a process-level understanding of how researchers navigate access barriers, we conducted 13 semi-structured interviews with researchers drawn from the pool of survey respondents who opted in (see Methods; the demographic characteristics of these interviewees are summarized in Supplementary Table~6). These interviews revealed stark inequalities. Paywalls and restricted access to data are universal features of academic life. What makes such barriers a problem is whether institutions can absorb the cost, a problem that is particularly acute for scholars in the Global South, and one that was reflected across all the interviews we conducted. Every respondent described institutional barriers to obtaining paywalled papers and the data, findings, or sources contained therein, and all deployed various strategies---informal networks, piracy platforms, paying out of pocket, or simply changing their research topic---as practices of circumvention. We did, however, observe variation in the degree to which restricted access was experienced as problematic.
 
On one pole, half of the respondents reported the inability to access papers and data through formal institutional means as an active source of frustration and constraint. In a tone of resignation, an architecture PhD student at a Nepalese university explained that ``in our country, we don't have any subscriptions from our university\ldots I don't get any answers from the dean\ldots there is no solution\ldots what can I do?'' On the other pole, the remaining respondents treated such restrictions almost as a banality. An Indian doctoral student noted that students at his institute simply understood that they could not access many paywalled journals, and that ``you would just find some work around.'' This does not mean such impediments are inconsequential; on the contrary, it suggests that some scholars have become so habituated to the scarcity of scientific resources that the costs are no longer fully legible even to those bearing them.
 
Regardless of whether barriers were experienced as an active problem or a taken-for-granted routine, the interviews bring into focus an informal economy of circumvention that operates outside formal institutional channels. When formal access fails or is simply absent, researchers resort to a consistent repertoire of workarounds: reaching out to friends at better-resourced universities (typically in Europe or the US) to download papers for them; requesting papers directly from authors through ResearchGate or personal emails; or turning to piracy platforms such as Libgen. Drawing on personal networks relies on having developed ties to the right people in the right places, which is itself unevenly distributed and privileges those with more extensive international networks. An Indian biorobotics PhD student noted, ``if we don't have access, we ask our friends from some other big universities.'' For some, no such networked workaround exists, which in some cases led to abandoning or altering their research topic. A Nigerian PhD student explained that because his university lacks journal subscriptions and nobody responded to his emails, ``so what I did, I changed the topic to the one I'm working on now.''
 
Consistent with our experimental findings, reaching out directly to authors worked sometimes but was generally unreliable. Scholars often received no reply, or were told that the authors did not have permission to distribute the work. From patterns in the interview data, we identified geopolitical and racialized processes at work, which are substantiated in Studies~2--5. An Iranian scholar now at a leading UK university reported informal pressure discouraging him from collaborating with scholars in Iran, owing to fears of unwanted scrutiny from administrators and funders. A Nigerian scholar described a related dynamic: researchers from better-resourced institutions, he felt, were reluctant to share data or methods with him, noting that ``they don't trust Nigerians.'' The stigmatization of Nigerians---``the world judging us as one,'' in his words---generated a kind of intellectual exile in which he gave up trying to build collaborations with scholars in Europe and the US. While there may be alternative explanations for such reluctance, the accumulated perception of belonging to a stigmatized group and an ``inferior'' institution nevertheless colors one's sense of being a valued member of a global research community. A South African scholar put it directly: ``I have collaboration networks, but they're in the African countries\ldots it's very difficult finding European universities or American universities that want to partner with South African colleagues\ldots I think they view us as not being good enough.'' In short, the informal workaround economy is stratified not only by who has access to the right networks, but also by who is trusted or valued enough to be included in them---perceptions that tend to be folded into stigmatized hierarchies of nationality, institutional prestige, and race.
 
A related burden emerged around the cost of publishing. The growth of open-access publishing---widely promoted as the remedy for precisely the access barriers documented above, and the channel whose citation patterns we examine in Study~2---has in many cases simply inverted the problem, shifting the financial barrier from reader to author. Respondents reported being priced out of open-access publication by article processing charges that are insurmountable in their funding environments; others described truncating the length of their articles not for scientific reasons but because of length-based pricing schemes. A dean at a South African university offered the starkest account: publishing a single open-access paper in a journal like \textit{Nature} costs roughly 40,000 rand, while the government subsidy for a published paper returns only 18,000 rand---a guaranteed net loss of 22,000 rand per paper. The result, in his case, was to forego publication altogether.
 
Finally, for several interviewees the reliance on informal means of access was not merely a coping strategy but a component of their teaching. Three respondents reported that the repertoire of circumvention practices is a critical part of their pedagogy: from their perspective, learning to do science includes learning how to navigate the global inequities of institutional resources. The same dean noted that ``you have to teach [students] to navigate the real life situation of having no money and doing everything in the lowest cost way that we can.'' Taken together, the interviews paint a portrait of scientific life in the Global South defined not by the occasional obstacle but by a pervasive set of barriers driven by the scarcity of financial and personal resources and by geopolitical constraints. Though scholars creatively adapt---the work goes on---these constraints reproduce and even concentrate the capacity of better-resourced scholars in the Global North to both consume and produce science.

\subsection*{Study~2: Observational evidence for inequities in paper access}

Study 1 revealed that researchers in the Global South report substantially lower institutional access to published academic papers and encounter paywalls more frequently than their Global North counterparts, a pattern consistent with large-scale evidence that access barriers disproportionately affect scholars in lower-income countries~\cite{evans2009open, cuntz2025access}. Both our survey respondents and interview subjects noted that while informal workarounds can sometimes help, these channels are often unreliable. Hence, informed by the lived experience of researchers from the Global South countries, we hypothesize that the constraints experienced by individuals when accessing paywalled papers add up to shape broader patterns of knowledge production on the global scale. To begin to assess whether this is the case, we turn to citation behavior and measure citation disparities as a proxy for access disparities. Specifically, we analyze whether authors from the Global South are less likely to cite paywalled papers than their counterparts in the Global North, potentially reflecting unequal access. The descriptive analysis leads to a linear regression model that considers paper- and author-level characteristics. We also employ a neural embedding model to explore the potential consequences of this unequal access. This analysis builds on prior research documenting geographic imbalances in the flow and uptake of scientific knowledge~\cite{gomez2022leading}, and complements our subsequent experimental analysis by establishing the broader observational context.

\begin{figure}[htbp!]
    \centering
    \includegraphics[width=0.95\linewidth]{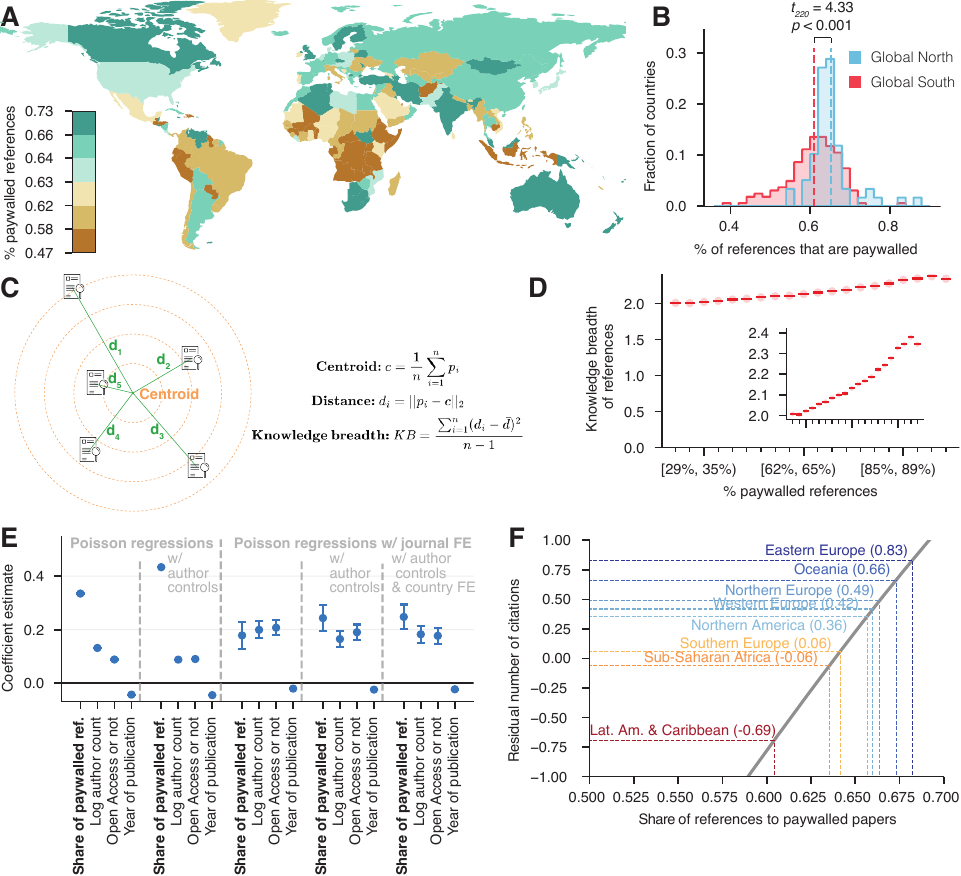}
    \caption{\small \textbf{Global disparity of citations to OA vs.\ paywalled articles.}
    \textbf{A}, Global distribution of the percentage of references that are made towards paywalled papers. Countries with a lower percentage are colored brown while countries with higher percentage are colored blue.
    \textbf{B}, The distribution of the percentage of references towards paywalled papers of each country. The dashed line denotes the mean value for global south (red) and global north countries (blue).
    \textbf{C}, We train a Doc2Vec model that embeds papers within a high dimensional vector space, and then calculate the knowledge breadth of reference lists as the variance of distances from each cited paper to the centroid of the reference list.
    \textbf{D}, Papers are divided into twenty percentile bins based on the percentage of references that are made to paywalled papers. For papers in each bin, we show the mean (line) and 95\% CI (shaded area) of knowledge breadth of their reference lists. {The inset shows the same relationship zoomed in to highlight variation across bins.}
    {\textbf{E}, Poisson regression models estimating the relationship between citation count and the share of references towards paywalled papers. Base model controls for the OA status, author count, the year of publication, and journal impact factor of focal papers. Alternative specifications additionally controls for author experience, journal fixed effects, and/or country fixed effects.
    \textbf{F}, The relationship between the share of references to paywalled papers and the number of citations that a paper receives (residual citations counts calculated using a Poisson regression which controls for year of publication, journal impact factor, the OA-status of papers, number of authors, author H-index, and author academic age).}
    }
    \label{fig:OA}
\end{figure}

We rely on the OpenAlex database, which records the OA status of over 250 million articles, to examine the disparities in geographic and institutional distribution of citations between OA and paywalled papers~\cite{priem2022openalex}. As expected, countries in the Global South are less likely to reference a paywalled article compared to those in the Global North (Figure~\ref{fig:OA}A). On average, 61\% of references from the Global South are made to paywalled articles, compared to 66\% for Global North countries (Figure~\ref{fig:OA}B; $t_220 = 4.33$, $p < 0.001$, Cohen's $d = 0.64$). This gap is the biggest in several regions that severely undercite paywalled articles, including Central Asia, Sub-Saharan Africa, Latin America and the Caribbean, and Southeast Asia (Figure~\ref{fig:OA}A and Supplementary Figure~1). The disparity narrows when restricting the analysis to references to papers published in Q1 journals only ($t_220 = 2.63$, $p = 0.009$, Cohen's $d = 0.40$), and further narrows when focusing on references to papers published in three premier publication venues \textit{Nature}, \textit{Science}, and \textit{Proceedings of the National Academy of Sciences} (Supplementary Figure~2; $t_219 = 1.37$, $p = 0.171$, Cohen's $d = 0.21$)~\cite{park2023papers}. {Note that OA operates under several categories, including diamond, gold, green, bronze, and hybrid~\cite{openalex_oa}. Further analysis shows that all OA types are associated with a higher likelihood of receiving citations from Global South publications (Supplementary Table~7).}

Next, we ask whether the disparities in access to paywalled articles have subsequent epistemic consequences. Intuitively, scientists with limited access to paywalled articles are restricted in the knowledge that they can access and build upon. Therefore, we expect papers that predominantly cite OA articles to draw upon a narrower range of knowledge compared to papers that cite more paywalled research. To this end, we measure the breadth of knowledge that a paper draws upon by projecting scientific papers into an embedding space of science using a machine learning model~\cite{le2014distributed,kang2025limited}. The model jointly learns vector representations of words and documents, placing words and papers with similar contextual information close to each other~\cite{le2014distributed}. For instance, words such as ``influenza'' and ``flu'' are close neighbors in the embedding space since they often appear in similar contexts. Likewise, papers covering similar topics are placed near each other even if they do not share the exact same vocabulary. This way, the embedding space captures the topical structure of scientific literature~\cite{kang2025limited}. We then measure, for each reference list, whether the cited papers occupy a small topical region (indicating narrower topical focus) or are more widely dispersed (indicating attention to a broader knowledge); see Figure~\ref{fig:OA}C for an illustrative example, and see Methods for more details. We show in Figure~\ref{fig:OA}D that papers with a higher percentage of references towards paywalled papers tend to draw upon a wider range of knowledge compared to those that cite more OA articles. Notably, papers that devote between 92\% to 96\% of references to paywalled articles exhibit the greatest knowledge breadth on average, while papers in the top ventile (more than 96\% paywalled references) draw on a somewhat narrower range of knowledge. Supplementary Figure~3 shows results obtained from an alternative specification of the model, showing similar trends. This non-monotonic pattern suggests that papers that predominantly cite paywalled research while also incorporating some OA literature achieve the widest coverage of knowledge.

{We also examine whether papers citing more paywalled articles tend to be more impactful, measured using citation impact~\cite{garfield1972citation}, which counts the total number of citations that a paper accumulates. We find significant positive correlations between the two across an array of regression specifications, while controlling for various paper-level covariates (Figure~\ref{fig:OA}E); see Supplementary Table~8 for negative binomial regression specifications and coefficient estimates and Supplementary Table~9 for additional Poisson regression models as robustness checks. The generalized variance inflation factors of these covariates are presented in Supplementary Table~10.
We next turn to the relationship between share of paywalled references and novelty~\cite{uzzi2013atypical}, which measures the extent to which a paper jointly references journals that are rarely cited together, as well as disruptiveness~\cite{funk2017dynamic}, which measures the extent to which a paper's references shift attention away from its predecessors. We find that the share of paywalled references is negatively, although not robustly, associated with greater novelty and is negatively associated with disruptiveness (Supplementary Table~11), therefore painting a nuanced picture between the share of paywalled references and paper characteristics.}

Using these regression estimates, we can assess how much regional variation in citation impact can be explained by the different level of access to paywall research articles across different regions. To obtain this back-of-the-envelope estimate, we first fit a {Poisson} regression model that calculates the residual citation counts after partialing out other confounders, including year of publication, {journal impact factor}, the OA-status of papers, number of authors, author H-index, author academic age, and author country. We then fit a third-order polynomial that captures the relationship between these residual citation counts and the regional percentage of references to paywalled papers. By inputting each region's average percentage of paywalled references into this polynomial, we estimate that papers submitted from Central Asia would receive an additional 0.56 lifetime citations (bootstrapped 95\% confidence interval is $[0.541, 0.566$]) on average if researchers from that region have the same level of access to paywall articles as those from Oceania, or an additional half a citation (bootstrapped 95\% confidence interval is $[0.481, 0.506]$) compared with papers from North America (a local view of the polynomial with is shown in Figure~\ref{fig:OA}H, only selected regions are marked to aid visualization; the full version can be found in Supplementary Figure~4).

\subsection*{Study~3: Investigating bias in requests for paywalled articles}

The findings above demonstrate that researchers in the Global South are less likely to reference paywalled papers, and that limited access to such articles is associated with narrower knowledge bases and reduced scholarly impact. While these results are consistent with the idea that access constraints hinder scientific participation, they cannot speak directly to the mechanisms underlying those constraints.
 
Motivated by the survey and interview findings in Study~1, we hypothesize that Global South researchers, already limited by funding constraints in their access to paywalled articles, face further hurdles when they turn to informal methods such as emailing authors to request copies of paywalled papers or data. That is, are published authors less likely to respond to paper-sharing inquiries from Global South researchers?

To test this, we conducted a large-scale randomized controlled trial in which fictional PhD students emailed authors of paywalled papers requesting access. We employed a $3 \times 3 \times 2$ factorial design, randomizing the sender's racial identity (German/White, Nigerian/Black, Pakistani/South Asian), institutional affiliation (top-50, 501--600, and 1000+ ranked universities), and the stated purpose of the request (journal article or class project). These three layers of randomization allow us to distinguish between taste-based racial discrimination, prestige-based discrimination, and instrumental motives (i.e., whether authors preferentially share with those more likely to cite their work). We preregistered hypotheses that non-White senders, those from less prestigious institutions, and those requesting papers for class projects would receive fewer responses (see Methods for the full experimental design, email templates, and data collection procedures).

Of the 18,000 emails sent, 9,850 (54.7\%) were opened, and after discarding 181 automated responses (e.g., out-of-office messages), 1,878 received a substantive reply (10.4\% of all emails sent; 19.0\% of opened emails).

Our primary analysis follows an intent-to-treat (ITT) framework, estimating reply rates as a proportion of all emails received rather than only those opened by the recipient. This is the standard approach for randomized experiments because conditioning on email opening would introduce post-treatment selection bias: the sender's name is visible in the recipient's inbox prior to opening, and the decision to open the email may itself be influenced by the treatment. As a robustness check, we repeat the analysis restricting to opened emails only in Supplementary Figure~6, where we find substantively similar results.

\begin{figure}[htbp!]
    \centering
    \includegraphics[width=\linewidth]{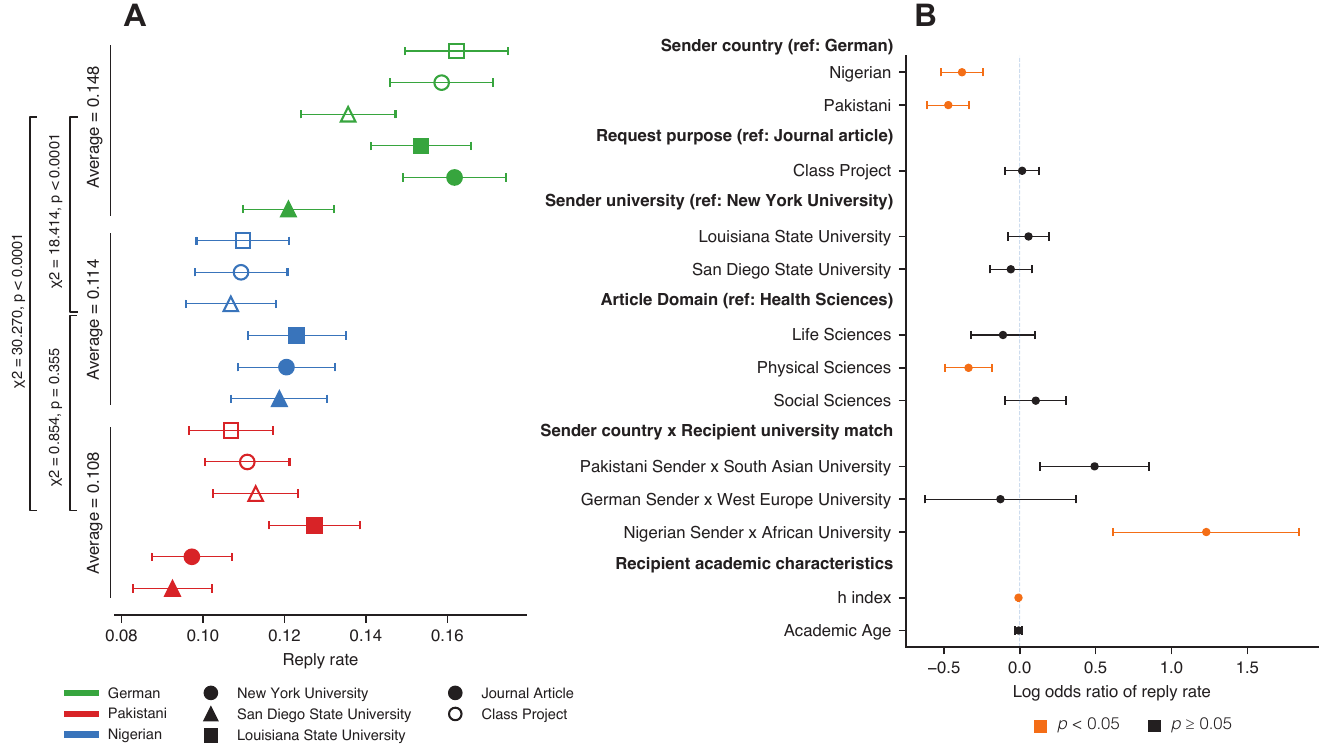}
    \caption{\small \textbf{Summary of paywalled paper audit experiment results.} (\textbf{A}) Reply rates for paywalled paper requests for each experimental condition. (\textbf{B}) Logistic regression estimated log-odds coefficients of independent variables predicting whether a given email was replied to.}
    \label{fig:OA_experiment}
\end{figure}

Figure~\ref{fig:OA_experiment}A illustrates the proportion of sent emails that were replied to for each experimental condition. Here, each color corresponds to a given racial group, each marker corresponds to a particular university, and filled markers correspond to requests for the purpose of writing a journal article, while empty ones correspond to those for the purpose of writing a class project. As expected, the German sender received the most replies on average, with 14.8\% of requests receiving responses. In contrast, racial minority groups received significantly fewer responses on average, with the Nigerian sender receiving replies to 11.4\% of emails ($\chi^2 = 18.414$, $p < 0.0001$), and the Pakistani sender receiving replies to only 10.8\% ($\chi^2 = 30.270$, $p < 0.0001$). However, we do not find significant differences between the two racial minority conditions ($\chi^2 = 0.854$, $p = 0.355$). Beyond the racial gap, these absolute rates are notable in their own right: for researchers who depend on informal requests as a primary means of accessing paywalled literature (Study~1), a channel that yields replies roughly one time in ten---as experienced by our Nigerian and Pakistani senders---represents a substantial practical barrier regardless of whether discrimination is present.

We model this response rate using logistic regression, where the sender's race, institution, and request purpose, as well as several other controls, such as the domain of the article in question, and the recipient's academic characteristics, namely, their h-index and academic age, appear on the right-hand side of the regression. We also control for a match between the sender's country of origin and the region of the recipient's university. We chose this as an alternative to inferring the racial group of the recipient due to the inherent uncertainty of name-based demographic classification techniques, which produce substantial misclassification rates, particularly for non-Western
names~\cite{imai2016improving, chintalapati2018predicting,
lockhart2023name}.

Figure~\ref{fig:OA_experiment}B shows the log-odds estimates of the above model. Relative to German senders, both Nigerian and Pakistani senders are 31.6\% ($p < 0.0001$) and 37.4\% ($p < 0.0001$) less likely to receive a reply, respectively. Yet, when the email recipient is from the same region as our fictitious sender, the racial minority senders receive more replies, suggesting that regional proximity or shared national origin may moderate racial disparities in access to paywalled papers (239\% and 63.7\% more likely for the Nigerian and Pakistani conditions, respectively). Interestingly, we do not observe a similar effect for the German sender, who received higher response rates regardless of regional match. Contrary to expectations, we also find no statistically significant effects of the sender's university or the stated purpose of their request on the likelihood of receiving a reply, ruling out prestige-based or instrumental motives for racial discrimination in this context.

Among article domains, physical sciences exhibited the lowest reply rates on average, being 28.6\% less likely to receive a reply relative to those in Health Sciences. Lastly, we observe a small but significant negative coefficient on recipient h-index (0.8\% less likely for each additional h-index value), implying that more highly cited researchers may be marginally less likely to respond to requests, potentially due to higher email volume or limited availability. See Supplementary Table~15 for several model specifications estimating reply likelihood (Figure~\ref{fig:OA_experiment}B corresponds to Model II in the table). 

Conditional on replying to the email request, 88.9\% of responses granted access to the paper in question, with the proportion of responses granting access to the paper being consistent across experimental conditions (see Supplementary Table~14). Nonetheless, we also include an additional model estimating the likelihood of receiving the article in question when requesting a paywalled article, and find largely similar results (see Supplementary Table~16). Our results suggest that bias occurs at the decision of whether to respond to the recipient or not, rather than at the decision of whether to respond positively or negatively. These experimental results provide a causal corroboration to the self-reported patterns in Study~1, where Global South researchers who had emailed authors for papers reported receiving a response more than half the time at markedly lower rates than their Global North peers (26\% vs.\ 44\%).

\subsection*{Study~4: Global citation disparity of papers with and without publicly available datasets}

{
While Studies 2 and 3 focused on disparities in access to published manuscripts, knowledge diffuses in forms other than just written manuscripts. In many empirical and computational fields, the data generated by a given experiment could be just as critical for replication and extension as the manuscript in which the experiment is described. As scientific research increasingly depends on large-scale, computationally demanding datasets~\cite{hey2009fourth,kitchin2014data,leonelli2020scientific}, the ability to access and reuse such data has become central to knowledge production.

Although publishers increasingly promote data sharing by requiring Data Availability Statements, authors are usually permitted to state that data are available ``upon request.'' Such policies create a formal appearance of openness while leaving actual access dependent on informal author discretion. In light of the disparities documented above, we hypothesize that when researchers in the Global South resort to informal channels to access knowledge, they may encounter similar barriers when requesting datasets as they do when requesting manuscripts. If this is indeed the case, such barriers may further exacerbate global disparities in research capacity and influence.

To examine whether such disparities exist, the current study analyzes Data Availability Statements as a case study. Specifically, we assess whether there are regional disparities in how often papers are cited for data-related purposes when the underlying dataset is available only upon request. We focus on Data Availability Statements for several reasons. First, journals, publishers, and funding bodies are increasingly encouraging or requiring Data Availability Statements in published articles~\cite{federer2018data,taichman2017data,hrynaszkiewicz2017standardising}. Second, unlike disciplinary data repositories, publication-linked Data Availability Statements are used across a wide range of fields, making them suitable for cross-disciplinary analysis~\cite{federer2018data,jiao2024data,tedersoo2021data}. Third, because data availability policies often allow different access arrangements, including public repository deposition, supplementary materials, and access upon request, papers published in the same journal and year may vary in their data availability arrangement, providing a useful setting for observational analysis~\cite{federer2018data,mcguinness2021descriptive,plos_data_policy}.

For this analysis, two pieces of information are crucial. First, for papers that report empirical analyses, we need to determine whether the underlying datasets are made available without restriction or only upon request. Second, for papers that cite those empirical studies, we need to determine whether the citing paper uses the dataset from the cited paper. Using the Semantic Scholar database, we retrieve Data Availability Statements for 200,000 papers and identify data citations to these papers, i.e., citations made for the purpose of using data, by classifying 4 million citation contexts with a custom deep learning classifier; see Methods for technical details. We hypothesize that papers whose datasets are available only upon request (hereafter referred to as ``upon-request papers'') receive fewer data citations than papers whose datasets are publicly available. We further hypothesize that researchers in the Global South are less likely to make data citations to upon-request papers compared to Global North researchers.

Our empirical analysis supports both hypotheses. More specifically, to test the first hypothesis, we use a logistic regression estimating the likelihood that a citation to a paper is a data citation. We find that citations to upon-request papers are significantly less likely to be data citations than citations to papers with publicly available data (Table~\ref{tab:logit_results}). This finding is robust across alternative model specifications (see Supplementary Table~18); generalized variance inflation factors for these models are reported in Supplementary Table~17. Holding characteristics of the cited paper constant, namely journal, publication year, Open Access status, and field of research, we estimate that citations made to upon-request papers have 46.3\% lower odds of being data citations compared to those made to papers whose data are publicly available. These findings confirm our first hypothesis.

To test the second hypothesis, we again estimate a logistic regression predicting whether a citation to a paper is a data citation, this time incorporating an indicator for whether the citing paper originates from the Global South and interacting this indicator with the cited paper's upon-request status. We find that the negative association between upon-request status and data citation is stronger when the citing paper is from the Global South. More specifically, although the cited paper's data availability status and the citing paper's region of origin are each independently associated with the likelihood that a citation is a data citation, the interaction between the two is associated with an additional 12\% reduction in the odds that a citation is a data citation, after controlling for publication year, journal, field of research, and Open Access status. These findings confirm our second hypothesis.}

\begin{table}[!htbp]
\centering
{\fontsize{7.2}{7.5}\selectfont
\setlength{\tabcolsep}{3pt}
\caption{{\textbf{Logit regression models estimating the likelihood that a citation to a paper is a data citation.} Each observation is a unique citation instance, defined as a citing paper--cited paper pair. Model~I estimates the association between the cited paper’s data availability status and the likelihood that a citation is a data citation. Model~II additionally considers regional heterogeneity. Rows report one coefficient each from logit regressions. $b$ denotes the log-odds coefficient. 
$p$ denotes two-sided $p$-values from Wald $z$-tests, and $p_{\mathrm{BH}}$ reports Benjamini-Hochberg adjusted $p$-values across the focal coefficients in both models. ``Upon request'' is a dummy variable indicating whether the cited paper mentions that its dataset is available upon request (1 means upon request). ``Open Access'' is a dummy variable indicating whether the cited paper is Open Access (1 means Open Access). ``Global South'' is a dummy variable indicating whether the majority of authors of the citing paper are affiliated with institutions situated in the Global South (1 means Global South). The rest are interaction terms. Each regression model additionally controls for year of publication fixed effects, journal of publication fixed effects, and field of study fixed effects.
}}
\label{tab:logit_results}
\begin{threeparttable}
\begin{tabular}{lcccccc cccccc}
\toprule
& \multicolumn{6}{c}{Model I. Likelihood of data citation} & \multicolumn{6}{c}{Model II. Regional heterogeneity} \\
\cmidrule(lr){2-7} \cmidrule(lr){8-13}
Term & $b$ & Odds ratio & $z$ & $p$ & $p_{\mathrm{BH}}$ & 95\% CI for $b$ & $b$ & Odds ratio & $z$ & $p$ & $p_{\mathrm{BH}}$ & 95\% CI for $b$ \\
\midrule
Upon request & -0.622 & 0.537 & -42.376 & $< 0.001$ & $<0.001$ & [-0.650, -0.593] & -0.583 & 0.558 & -35.648 & $<0.001$ & $<0.001$ & [-0.615, -0.551] \\
Open Access & 0.087 & 1.091 & 4.924 & $<0.001$ & $<0.001$ & [0.053, 0.122] & 0.041 & 1.042 & 2.267 & 0.023 & 0.033 & [0.006, 0.077] \\
Global South &  &  &  &  &  &  & -0.422 & 0.656 & -22.676 & $<0.001$ & $<0.001$ & [-0.458, -0.385] \\
Upon request $\times$ Global South &  &  &  &  &  &  & -0.128 & 0.880 & -3.605 & $<0.001$ & $<0.001$ & [-0.197, -0.058] \\
Open Access $\times$ Global South &  &  &  &  &  &  & 0.133 & 1.142 & 5.453 & $<0.001$ & $<0.001$ & [0.085, 0.181] \\
\midrule
McFadden pseudo-$R^2$ & \multicolumn{6}{c}{0.044} & \multicolumn{6}{c}{0.046} \\
Observations & \multicolumn{6}{c}{1534036} & \multicolumn{6}{c}{1534036} \\
\bottomrule
\end{tabular}
\end{threeparttable}
}
\end{table}

\subsection*{Study~5: Investigating racial and institutional biases in requests for data sharing}

Our global citation analysis of Data Availability Statements reveals that papers whose datasets are only available upon request tend to be cited less often by researchers in the Global South. This suggests that access to underlying data, like access to manuscripts, may be unequally distributed across geographic and institutional lines. Yet, citation patterns alone cannot confirm whether these disparities arise from bias in the data-sharing process itself.

To test this directly, we conducted a second large-scale randomized controlled trial in which fictional PhD students emailed corresponding authors of papers whose Data Availability Statements indicate that data is available upon request\footnote{This study is a conceptual replication of \cite{acciai2023estimating}.}. We employed a $2 \times 2 \times 2 \times 2$ factorial design, randomizing the sender's racial identity (English/White or South African/Black), the country of institutional affiliation (England or South Africa), the rank of the institution (globally ranked 150--155 or 1001--1200), and the stated purpose of the request (journal article or class project). In contrast to Study~3, this design allows us to independently vary both the sender's racial identity and their institutional affiliation, disentangling race from prestige and geographic signals. We preregistered four hypotheses: that the English/White sender would receive more responses than the South African/Black sender, that requests from English universities would outperform those from South African universities, that higher-ranked institutions would yield higher response rates, and that requests framed as supporting journal articles would be more successful than those for class projects (see Methods for the full experimental design, email templates, sampling procedures, and response coding).

Of the 16,000 emails sent, 13,503 were successfully delivered, 6,981 (51.7\%) were opened, and after discarding 171 automated responses (e.g., out-of-office messages), 1,813 (26.0\% of opened emails) received a substantive reply. Two independent annotators classified each reply as positive (data provided or forwarded), conditional (credentials or PI involvement requested), or negative (request declined, data lost, or access offered contingent on co-authorship or payment); inter-rater agreement was high (Cohen's $\kappa = 0.964$). See Methods for additional details. As in Study~3, our primary analysis follows an intent-to-treat framework, estimating reply rates as a proportion of all emails sent; see Supplementary Figure~7 for a robustness check restricting to opened emails only.

\begin{figure}[htbp!]
    \centering
    \includegraphics[width=\linewidth]{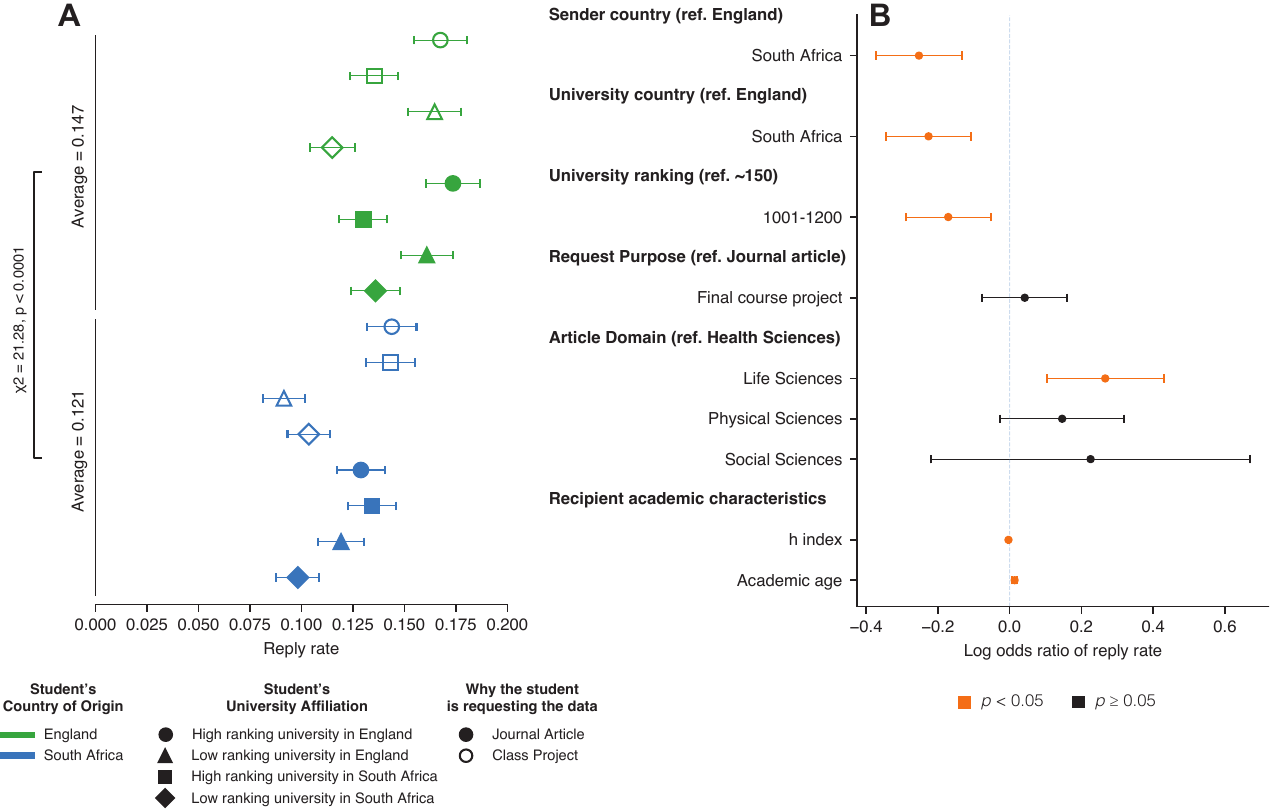}
     \caption{\small \textbf{Summary of dataset audit experiment results.} (\textbf{A}) Reply rates for datasets requests for each experimental condition. (\textbf{B}) Logistic regression estimated log-odds coefficients of independent variables predicting whether a given email was replied to.}
     \label{fig:data_availability_all_emails}
\end{figure}

Figure~\ref{fig:data_availability_all_emails}A reports response rates across experimental conditions. Consistent with our first three hypotheses, emails sent by the English/White sender received the highest average response rate (14.7\%), while those sent by the South African/Black sender elicited significantly fewer replies (12.1\%; $\chi^2 = 21.28, p < 0.001$). Institutional affiliation also substantially shaped response behavior. Requests originating from universities located in England were more likely to receive replies than those from South African universities (14.40\% vs. 12.46\%; $\chi^2 = 10.80, p = 0.001$), and emails sent from higher-ranked institutions generated higher response rates than those from lower-ranked institutions (14.46\% vs. 12.37\%; $\chi^2 = 12.49, p < 0.001$). By contrast, the stated purpose of the data request had no detectable effect on response likelihood (journal article: 13.5\%; course project: 13.3\%; $\chi^2 = 0.10, p = 0.75$). As in Study~3, the absolute response rates are themselves informative. Even the most favorably treated condition---the English/White sender from a high-ranked English university---received replies to only 14.7\% of requests, suggesting that informal data-sharing channels are unreliable for all requesters and particularly so for those from less privileged backgrounds. As a robustness check, we repeat the same analysis while focusing only on emails opened by the recipient (see Supplementary Figure~7).

We again model this response rate using logistic regression, where the sender conditions (race, institution, and request purpose) are our key predictors. We also include article-level controls, including the domain of the article in question, and recipient-level controls such as the recipient's h-index and academic age. Here, we do not control for a match between the sender's country of origin and the region of the recipient's university due to the limited number of recipients affiliated with a university in Africa (only 109 of the 16,000, or 0.68\%).

Figure~\ref{fig:data_availability_all_emails}B reports logit regression estimates from the full model. Relative to the English/White student, emails sent by the South African sender were associated with 22.3\% lower odds of receiving a reply ($p < 0.001$). Institutional affiliation also mattered independently of sender identity: requests originating from universities located in South Africa were associated with 20.2\% lower odds of receiving a response compared to those from universities in England ($p < 0.001$), and emails sent from lower-ranked institutions (global rank 1001--1200) exhibited 15.7\% lower odds of receiving a reply relative to those from higher-ranked universities (global rank 150--155; $p = 0.005$). The stated purpose of the request did not significantly affect response likelihood. Requests concerning life sciences articles were associated with 30.5\% higher odds of receiving a response relative to health sciences articles ($p = 0.001$), while effects for physical and social sciences were smaller and statistically insignificant. Finally, recipient characteristics show that higher h-index values are modestly associated with lower responsiveness, with each additional h-index point corresponding to a 0.3\% reduction in the odds of replying ($p = 0.004$), whereas greater academic age is associated with higher responsiveness, with each additional year since first publication corresponding to a 1.4\% increase in the odds of response ($p < 0.001$). As in Study~3, these experimental results mirror the self-reported patterns in Study~1, where Global South researchers who had requested data described as ``available upon request'' reported success more than half the time at less than half the rate of their Global North peers (15\% vs.\ 34\%).

We report both baseline model specifications as well as models reporting heterogeneous effects in Supplementary Tables 23 and 24. While the direction of the interaction coefficients in Supplementary Table 24 suggests that the relative importance of institutional rank versus institutional country may differ by sender race — with South African senders potentially more affected by lower institutional rank (Models III and VI) and English senders more affected by institutional country — none of the interaction terms remain statistically significant after Benjamini-Hochberg corrections. As such, we do not find robust evidence that these experimental factors operate differently across sender identities.

Similar to Study 3, conditional on receiving a reply, the vast majority of replies were positive in nature (94.5\%) with no statistically significant differences between any of the experimental conditions concerning the proportion of positive replies ($p > 0.05$ for all pairwise comparisons). These results again suggest that bias occurs at the decision of whether to respond to the recipient or not, rather than at the decision of whether to respond positively or negatively. While conditional and negative replies were sparse, they nonetheless mirror the conditional and transactional sharing practices that interviewees in Study~1 described. For instance, one paper's Data Availability statement reads as follows: ``The authors declare that all data supporting the findings of this study are available in the article, ...., or on request from the corresponding author.'' When contacted, that author responded with a request for co-authorship, stating ``The data can indeed be made available in the context of a scientific collaboration.'' Another paper's Data Availability statement reads ``The datasets ... are available from the corresponding author on reasonable request.'', but when contacted, the recipient replied ``Requests are limited to those with valid reasons for collaboration proposals.'' Even more egregiously, one recipient replied to the dataset request by stating ``I can provide you with the raw data from the paper, ..., for a fee of 1000 US dollars. Payment can be made using Ethereum.  Additionally, I am willing to assist you with your research as a co-author of your paper, free of charge.''

\section*{Discussion}

This study presents the first large-scale experimental evidence that informal paper- and data-sharing practices in science are shaped by both racial and institutional biases, though their influence varies by context. When requesting access to paywalled articles (Study~3), the racial identity of the requester, signaled through their name and country of origin, significantly affected reply rates, even when controlling for institutional affiliation. When requesting datasets (Study~5), both racial cues and institutional affiliation---particularly whether the sender was associated with a Global North or Global South university---strongly predicted the likelihood of receiving a response. These nuanced results suggest that while racial discrimination plays a prominent role in responses to both types of academic materials, institutional cues is an additional salient factor in decisions about data sharing.

A possible reason that the relative strength of institutional identity differs in the two experiments is that the perceived effort or risk differs between the two types of requests. Sharing a paywalled article may feel low-cost and routine, and as such, is more susceptible to unconscious or taste-based discrimination. In contrast, data requests may involve a higher perceived burden, such as evaluating the legitimacy of the requester or protecting sensitive information, prompting authors to rely relatively more heavily on institutional signals as a heuristic for trustworthiness or credibility. Sharing data may also entail higher \textit{risks} than sharing papers: risks of discovering errors, or of failed replications. For some authors, those risks may not justify the effort. Another possibility is that the norms surrounding data sharing are more ambiguous and contested than those surrounding article sharing, subsequently producing greater variability in authors' responses and stronger reliance on institutional cues. These findings point to the multifaceted nature of informal gatekeeping, and emphasize the need for reforms that address not only racial disparities but also institutional inequalities in access to scientific resources.
 
Beyond the relative disparities documented above, the absolute response rates deserve attention in their own right. The German/White sender in Study~3 serves as a reference category for detecting discrimination, but researchers affiliated with well-resourced Global North institutions rarely need to request paywalled papers from authors in the first place, since their libraries typically provide access. It is researchers in the Global South who disproportionately depend on this informal channel (Study~1), and for them the relevant figure is not the gap but the baseline rate of success. That rate is strikingly low: Nigerian and Pakistani senders received substantive replies to roughly 11\% of their requests. Even in the absence of any racial disparity, a channel that fails nearly nine times out of ten is a precarious foundation for research access. The racial gap compounds this problem but does not define it. In other words, eliminating discrimination in informal sharing would be necessary but insufficient: the channel itself is inadequate for the researchers who most rely on it. This reinforces the need for structural reforms---such as expanded open access mandates, institutional subscription support, and enforceable data-sharing policies---that reduce dependence on informal requesting altogether, rather than merely ensuring that such requests are answered equitably.
 
More broadly, our findings highlight a critical epistemic vulnerability in contemporary science: {publication does not necessarily make scientific knowledge equally accessible or usable. Even after research is published, knowledge created in the underlying work may remain under authors’ informal control. As demonstrated through five studies,} this is true in at least two different ways: for datasets available only ``upon request,'' the authors are the sole gatekeepers responsible for the dissemination of data; for papers published in paywalled journals, scientists who do not have subscriptions often turn to the authors as a last resort (Study~1). {
Moreover, our evidence suggests that this gatekeeping is far from neutral: Having an African or South Asian name reduces the likelihood of receiving a response compared to those with a British or German name (Studies 3 and 5). Similarly, being affiliated with an institution in South Africa reduces one's likelihood of receiving a response compared to those whose affiliation is in the UK (Study 5).
These disparities matter because individual access decisions can aggregate into broader inequalities in scientific participation. At the individual level, these barriers may cause scientists to shift their research focus entirely (Study~1). At the aggregate level, Studies~2 and 4 show that restricted access arrangements are associated with unequal patterns of engagement with published knowledge: Global South researchers cite paywalled papers less often and are less likely to cite a paper for the purposes of using its dataset when that dataset is only available upon request. Together, these findings show how biases operating through everyday interpersonal exchanges can translate into downstream inequalities in who is able to use and build upon science.
} 
Our results therefore call for the dismantling of such barriers to foster a more equitable and inclusive scientific ecosystem.
 
How can we facilitate open public sharing of papers and data? A close examination of the email responses we received in Study~5 sheds light on the central role of incentives and power in informal access channels. In some cases, authors offered access only conditionally, making requests contingent on co-authorship, collaboration, or monetary compensation. One author stated, ``We are always open to potential collaborations,'' while another demanded a \$1,000 payment via Ethereum. Although such responses are not the modal outcome, they illustrate how informal access mechanisms can enable exploitative demands that are shaped by underlying structural power dynamics between data holders and requesters. Researchers with greater institutional prestige, funding, or security may be better positioned to impose conditions, while those seeking access -- often from less resourced institutions -- have limited bargaining power and few alternatives.
 
These practices stand in tension with long-established scientific norms. The Mertonian ideals of communality and disinterestedness~\cite{merton1942note} prescribe that scientific knowledge should be shared openly and pursued for its collective value rather than private gain. Yet in practice, data frequently functions as a form of academic capital, controlled, rationed, and strategically leveraged for publications, collaborations, and prestige. When data sharing becomes transactional or contingent, informal gatekeeping mechanisms emerge that reproduce existing hierarchies and shift the scientific enterprise away from its foundational commitments to openness and collective advancement.
 
Addressing these challenges requires rethinking institutional incentives and enforcement mechanisms. First and foremost, each individual scientist should realize that making papers and datasets publicly available is indeed aligned with their own career objectives, since openly available datasets and papers are associated with greater visibility and impact~\cite{piwowar2013data,colavizza2020citation,langham2021open,mckiernan2016open}. There is more than one way to do so. Although not all scientists have the funding to publish papers in OA formats (which often entails exorbitant article processing charges), they can still ensure equitable access to knowledge by self-archiving, i.e., uploading the published version of the article to the Internet. Even when articles are published in paywalled journals, most publishers allow authors to self-archive after an embargo period, ranging between six months to a year~\cite{natureselfarchiving,oxfordselfarchiving,wileyselfarchiving}. Yet, many researchers are unaware of or under utilize this option~\cite{mbughuni2024self,tmava2023faculty}. While publishers may not have the financial incentive to promote self-archiving, funders and universities should be tasked with ensuring that all research that are associated with them is openly accessible. Some governmental funders, for example, have recently mandated public access to all funded research~\cite{lenharonih,brainard2022white}. 
 
Compared to papers, mandating all datasets to be publicly available is challenging due to the commercial (if data is proprietary) and legal or privacy (if the dataset contains individually identifying information) concerns~\cite{lazer2009computational}. But even in such cases, innovative solutions exist to ensure accessibility~\cite{ICPSR}. Despite this, many journals that require data sharing upon publication fail to enforce their own policies. Our results show that roughly 88\% of authors who promise data upon request fail to deliver it. 
Journals, as stewards of scientific dissemination, are uniquely positioned to lead reform, especially since current data-sharing policies remain inconsistently enforced. While stricter enforcement may seem appealing, social science research suggests that policies out of sync with prevailing norms risk backlash and noncompliance~\cite{acemoglu2017social}. A more sustainable strategy involves gradual norm change. For instance, the journal \textit{Science} has progressively tightened its data-sharing requirements through iterative policy reform~\cite{science2005policy,science2011policy,science2025policy}. This shift not only increases transparency but also removes the opportunity for biased gatekeeping. As these norms diffuse through the publication ecosystem, via manuscript resubmissions, cascading editorial policies, or top-down mandates from funders, their effects may be felt beyond elite journals. Recent studies suggest that higher-impact journals are more likely to require public data sharing~\cite{resnik2019effect,vasilevsky2017reproducible}, and our findings reinforce why such policies are essential. Meanwhile, journals such as \textit{Scientific Data} ensure that authors can be properly credited for their efforts in curating a dataset through proper publications and citations.
 
Our study demonstrates that informal, request-based data-sharing mechanisms are not only unreliable but also unequally applied. Ensuring equitable access to scientific data demands that journals, funders, and institutions move beyond voluntary compliance toward more robust, auditable, and enforceable data-sharing infrastructures. Only then can the promise of open and inclusive science be fully realized.
 
{
\subsection*{Limitations}
 
Several limitations should be noted. First, our audit experiments test a limited set of racial and national identities. Study~3 includes one White/European and two non-White/Global South senders, while Study~5 compares one White/English sender with one Black/South African sender. These identities do not capture the full spectrum of racial, ethnic, and national backgrounds that researchers hold, and the magnitude and direction of bias may differ for groups not represented here, such as East Asian, Latin American, or Indigenous researchers. Relatedly, our experimental design bundles racial cues with national origin---senders differ simultaneously in name, photo, and stated country---making it difficult to fully isolate the independent contribution of race versus nationality.
 
Second, our observational analyses use citation behavior as a proxy for access disparities. While this approach builds on prior work and is supported by the consistency between our observational and experimental findings, citation decisions are shaped by many factors beyond access, including disciplinary norms, language barriers, relevance judgments, and awareness of the cited work. We therefore interpret these patterns as suggestive rather than definitive evidence of access constraints.
 
Third, Study~4 relies on an automated classifier to infer citation intent from citation contexts and on keyword matching to identify data availability statements; both steps introduce some degree of measurement error, and citation contexts are not available for every paper indexed by Semantic Scholar. Similarly, the interview sample in Study~1, while purposively selected, is small and drawn from a limited set of disciplines and countries in the Global South. These qualitative findings are intended to motivate and contextualize our quantitative analyses rather than to support standalone generalizable claims.
 
Fourth, our binary classification of countries into Global North and Global South, while following UN Trade and Development conventions, necessarily obscures substantial within-group heterogeneity in institutional resources, research infrastructure, and access to subscriptions. Future work could employ more granular classifications, such as the World Bank income categories, to capture this variation.
 
Fifth, the audit experiments examine requests sent by fictional PhD students. Faculty, postdoctoral researchers, or industry scientists making similar requests may elicit different response patterns, and our findings should not be generalized to all academic career stages without further investigation. Additionally, Study~5 allowed a minimum of only one week for responses, which may underestimate true response rates, though we note that Study~3 allowed at least two weeks and produced substantively similar patterns.
 
Finally, while our email tracking system records whether emails were opened, we cannot distinguish between emails filtered to spam, overlooked in a crowded inbox, or deliberately ignored. The mechanisms underlying non-response thus remain partially opaque, and our estimates of bias should be understood as reflecting the combined effect of all stages in the decision process from email receipt to reply.}

\section*{Methods}

\subsection*{Study 1: Survey}

\subsubsection*{Sampling and recruitment}
 
We used Semantic Scholar to retrieve the emails of corresponding authors from published articles to construct a sampling frame of active researchers at institutions worldwide. Researchers were eligible for inclusion if they had at least one publication indexed in Semantic Scholar between 2020 and 2025, were listed as corresponding author on at least one publication, and had a retrievable institutional email address. We stratified our sampling by region (six categories: Sub-Saharan Africa, North Africa and Middle East, South Asia, Southeast Asia, Central Asia, and Latin America and the Caribbean for the Global South; all of Europe, North America, Oceania, and developed Asian economies for the Global North) and by discipline (Health Sciences, Life Sciences, Physical Sciences and Engineering, and Social Sciences) to ensure geographic and disciplinary diversity. Personalized email invitations were sent to 66,000 researchers, of which, we received completed survey responses from 376 participants.
 
\subsubsection*{Survey instrument}
 
The survey was administered via Qualtrics and consisted of 22 questions organized into four blocks: (1) demographics (country of institutional affiliation, nationality, sex, career stage, institution name, discipline, and publication count); (2) one open-ended question about the biggest challenges respondents face in accessing materials needed for their research; (3) structured questions about access to publisher databases (matrix format with ``I have access,'' ``I do not have access,'' and ``Not sure'' for each of twelve publishers), frequency of paywall encounters and dataset access needs, coping strategies used when unable to access papers, self-reported success rates when emailing authors for papers and datasets, reasons for not receiving requested data, consequences of access barriers on research, and the importance of data availability for research and teaching; and (4) an optional question about willingness to participate in a follow-up interview. The survey was pilot-tested with five researchers before deployment. The complete survey instrument is available in Supplementary Note~1.
 
\subsubsection*{Classification of countries}
 
Following UN Trade and Development (UNCTAD)~\cite{UNCTD}, we classify countries into Global South and Global North as described in the Methods of Study~2. Survey respondents who entered institution names rather than country names were manually classified based on the institution's location.
 
\subsubsection*{Statistical analysis}
 
We compare access rates, paywall encounter frequencies, and self-reported email success rates between Global North and Global South respondents using $\chi^2$ tests of independence. For publisher access, we test whether the proportion reporting ``I have access'' differs by region. For email success rates, we dichotomize respondents who have attempted email requests into those reporting above-50\% success versus below-50\% success, and compare proportions across regions using both $\chi^2$ and Fisher's exact tests. We report exact $p$-values throughout. Multi-select questions (coping strategies, consequences) are reported as descriptive percentages. All analyses are conducted using Python (scipy.stats).
 
\subsubsection*{Ethics}
 
The survey study was approved by the Institutional Review Board of New York University Abu Dhabi (HRPP-2024-45). All participants provided informed consent before beginning the survey. No identifiable information was collected as part of the survey itself; respondents who volunteered for follow-up interviews were directed to contact the research team via a separate email, ensuring that their survey responses remained anonymous.

\subsection*{Study 1: Semi-Structured Interviews}
 
To make sense of how researchers in and from the Global South navigate access barriers to paywalled publications and restricted datasets, we conducted 13 semi-structured interviews with faculty, doctoral and postdoctoral researchers, and research scientists. In-depth interviewing brings into focus the practical dilemmas faced by researchers in a way that survey and bibliometric data alone cannot: how researchers manage the problem of restricted publications and data, what creative workarounds they find, and how those barriers and solutions shape their capacity to do science.
 
Questions of generalizability, sample size, and the slippage between what people say and what people do have generated considerable debate among qualitative and quantitative researchers in recent years~\cite{jerolmack2014talk, lamont2014methodological, vaisey2014whence, tavory2020interviews}. Rather than demanding broad empirical generalizability or deeply patterned landscapes of meaning from our interview data, we conducted process-oriented interviews that bring into focus ``the unfolding of action''---how a task was accomplished, who and what means were involved, and how respondents adapted to what came next~\cite{tavory2020interviews}. Process-oriented questions, when coupled with additional interviews, help to empirically establish the phenomena of interest, and when coupled with other data, clarify the practical dilemmas people face as they navigate social life. Our interview strategy was attentive to intra-case variation, contingency, and paths not taken---that is, to the fact that interviewees found different solutions to similar problems.
 
We recruited interviewees from among the survey respondents who indicated willingness to participate in a follow-up interview. To preserve the anonymity of the survey, willing respondents were directed to contact the research team directly by email rather than through a link to their survey responses (see Survey Methods above); interviewees were therefore not linked to their survey data. From this pool, we invited respondents for in-depth, semi-structured interviews conducted over Zoom, continuing until new interviews replicated rather than extended the existing picture. The 13 interviewees spanned career stages from PhD students to full professors, including one dean, across fields including physics, biorobotics and mechanical engineering, architecture and the built environment, computer science and machine learning, molecular neuroscience, mathematics and statistics, and public health. They were affiliated with universities in Latin America, West and Southern Africa, South and West Asia (including the Middle East), and the Caucasus, with three interviewees of Global South origin based at institutions in Singapore, Italy, and the United Kingdom. Interviewed participants were compensated with 50 USD.
 
We approached the interviews as a ``case'' of interest rather than a statistically representative sample of researchers in the Global South, a criterion that would be unachievable at the scale of this study~\cite{small2009how}. Treating participant selection as a case meant that each new interview yielded new dimensions and variations of the access problem and its potential solutions until new cases replicated rather than extended the existing picture---what we describe as saturation.
 
Interviews were conducted remotely via Zoom and lasted between 30 and 60 minutes. After reading a short consent form and an explanation of our data-privacy procedures, all interviewees consented to be recorded. We followed a loosely structured interview guide that began with the participant's scientific and professional background and moved to the resources needed to do their work, the problems they faced, and the strategies they deployed to circumvent those problems. The full interview guide is provided in Supplementary Note~2.
 
Our analysis of the interviews first began by using Zoom’s auto-transcribe function to derive a simple text transcript of each interview. Since the corpus of interviews was relatively small, we opted not to use qualitative research software or AI tools to code and thematize the interviews. Instead, W.H. read and reread all of the transcripts to gain intimate fluency with the particularities of each interview. Based on this deep reading of all the interviews, W.H. then organized and thematized the interviews in Microsoft Word along a taxonomy of themes that reflected the unfolding processes of interest. We recognize that the decision to hand-code the interview transcripts in Microsoft Word rather than employ qualitative research software or AI assisted tools may appear archaic by the standards of contemporary mixed-methods social science. This was a deliberate methodological decision on our part. With a corpus of thirteen interviews, the analytical priority was not systematic pattern extraction across a large number of interviews but rather attaining deep interpretive understanding across a small, but focused body of interviews. Given the limited sample size of interviews, it was feasible to obtain granular familiarity with how individual respondents made sense of their circumstances that repeated close reading produces and that software-mediated coding can sometimes obscure. This approach is consistent with the interpretive traditions of historical and ethnographic research, in which intimate knowledge of one's sources is itself a methodological resource rather than a failure of rigor (see \cite{biernacki2014humanist}). We report our adherence to the Consolidated Criteria for Reporting Qualitative Research (COREQ)~\cite{tong2007consolidated} at the end of this file.

\subsection*{Methods of Study~2}

\subsubsection*{Data}
Study~2 relies on a local snapshot of the OpenAlex dataset~\cite{priem2022openalex} that was downloaded in 2023~\cite{openalexsnapshot}. This dataset indexes over 250 million papers published between 1800 and 2023. It has been shown that OpenAlex is one of the most comprehensive bibliometric databases~\cite{walters2025comparing}. Since our bibliometrics records do not extend beyond 2023, we restrict our analysis to papers published no later than 2020, while consider all citations to such papers as recorded in OpenAlex. This way, all papers have at least three full years to accumulate citations. The OpenAlex dataset records for each paper whether it is OA or not. We use this information when studying inequalities in paper access in Study 2.

In Study~2, we used the precomputed novelty and disruptiveness scores of papers provided by SciSciNet~\cite{lin2023sciscinet}.

\subsubsection*{Papers' country of origin}
Following UN Trade and Development (UNCTAD), we divide countries into two broad regions: the Global South countries (or, the ``Global South'') and the developed economies (or, the ``Global North'')~\cite{UNCTD}. The Global South countries consist of countries within Africa, Latin America, the Caribbean, and Asia, with the exceptions of Israel, Japan, the Republic of Korea, Australia, and New Zealand. The Global North economies consist of these five countries, in addition to all countries in Europe and North America.

\subsubsection*{Measuring the knowledge breadth of reference lists}
To measure the knowledge breadth of reference lists, we use a method similar to Hao et al.~\cite{hao2024ai}, by first applying the Doc2Vec algorithm on the abstracts of scientific papers in the OpenAlex database. Due to the limit in computation resources that the authors have access to, we train a Doc2Vec model using 42 million research papers that are randomly sampled from OpenAlex, which takes approximately 430GB of memory and 4 days to compute 40 epochs on a high-performance computing cluster. We then use the trained Word2Vec model to infer the embedding of all papers indexed in the OpenAlex database that have an abstract, including papers not used to train the model. We follow previous research and embed all papers in a 100-dimensional space~\cite{kang2025limited}, which is reported in the main manuscript. Additionally, we train a separate embedding model of 50 dimensions, and report the result in the Supplementary Figure~3.

Then, for each reference list, we compute the centroid of all referenced papers in the science embedding space and measure the Euclidean distance from each cited paper to the centroid. The knowledge extent is then calculated as the variance of the distances. In practice, the length of reference lists differs significantly, introducing potential biases to the measurement of knowledge breadth. As a result, reference lists consisting of fewer than 10 embedded references are excluded from the analysis. Reference lists consisting of more than 10 embedded references are downsampled by randomly selecting 10 so that all reference lists in this analysis have the same length.

{
\subsection*{Methods of Study~4}
\subsubsection*{Data}
Study~4 relies on a snapshot of the Semantic Scholar database dated to 2026-01-30. We switched to Semantic Scholar for this study since, unlike OpenAlex, it provides citation context that helps us to classifier citation intent.

\subsubsection*{Extraction of data availability information}
We filter the papers by parsing the fulltext and looking for any of the folloiwng section titles: ``data availability'', ``data and code availability'', ``data availability statement'', or ``data and code availability statement''. For papers with a data availability statement, we remove the ones that states ``data availability statement: not applicable'' or that ``no data was used'' or ``no data were used''. Among the rest, we classify them as ``upon request papers'' if their data availability statement contains any of the following phrases: ``available upon request'', ``available on request'', ``available upon reasonable request'', ``available on reasonable request'', ``available from the corresponding author upon request'', ``available from the corresponding author on request'', ``available from the corresponding author upon reasonable request'', or ``available from the corresponding author on reasonable request''.

\subsubsection*{Classification of citation intent}
Of all papers with a data availability statement, we gather all their citations and the corresponding citation context provided by Semantics Scholar, if available. To classify the citation intent of these contexts, we used a two-stage approach.
In the first stage, we use a large language model, namely, GPT-5.4-mini (2026-03-17 version), to classify whether a citation context explicitly mentions data usage using the following prompt ``Does the sentence itself indicate use of the dataset used in the cited paper? Answer only yes or no. yes = the sentence itself explicitly mentions or clearly describes using the cited paper's dataset. no = otherwise.'' We applied this prompt to 40,000 citation contexts to obtain binary yes/no labels.
In the second stage, we used the LLM-labeled citation contexts to fine-tune a ModernBERT model. Of the 40,000 labeled citation contexts, 4,000 were used as a validation set, 4,000 as a test set, and the remaining 32,000 as the training set, corresponding to 10\%, 10\%, and 80\% of the data, respectively. The fine-tuned ModernBERT model was then applied to all citation contexts to classify whether each context explicitly mentioned data usage.
To evaluate the final classifications, we manually sampled a balanced set of 200 classified citation contexts and assessed their accuracy. Two authors independently labeled the sampled citation contexts, achieving a Cohen’s kappa score of 0.89, indicating almost perfect agreement. After resolving annotation disagreements, we compared the ModernBERT labels with the resulting ground-truth labels and obtained an accuracy of 79.5\%.
Finally, since a cited paper $c_1$ could be cited multiple times by a citing paper $c_2$, we aggregate citation context-level classifications to citation pair-level classifications by denoting a citation ($c_1$, $c_2$) as a ``data citation'' if any sentences in $c_2$ that cites $c_1$ does so by explicitly mentioning data usage.
}

\subsection*{Study 3: Paywalled Article Access Study}

\subsubsection*{Overview}

The first randomized controlled trial (RCT) investigates whether access to paywalled scientific articles differs based on the racial identity and institutional affiliation of the requester. We emailed the corresponding authors of recently published paywalled papers, requesting access to the paper while experimentally manipulating the name, country of origin, university affiliation, and stated purpose of the request. Our hypotheses were preregistered prior to beginning the experiment (\url{https://osf.io/e5kfa/}).

In the preregistration, we proposed including inferred race and gender of the email recipients as control variables, following common practices in audit studies. However, during implementation we determined that automated inference of race and gender from names was highly unreliable in a global sample, producing substantial misclassification for non-Western names. We therefore chose not to include these noisy measures in the final models. Instead, we include a more reliable and substantively relevant control: institutional region match (i.e., whether the sender's country region matches that of the recipient's institution), which captures geographic proximity without relying on error-prone demographic inference. We note this deviation from the preregistration for transparency.

\subsubsection*{Sampling}

We used OpenAlex~\cite{priem2022openalex} to identify articles published between 2022 and 2024 in Q1-ranked journals that were not Open Access. For each article, we scraped the title, corresponding author name, institutional affiliation, and contact email from the publisher's website. Articles were filtered to include only those with a listed corresponding author and a valid contact email. This process yielded a total of 67,033 articles, which were reduced to 33,741 after removing duplicate corresponding authors (see Supplementary Table~12 for the number of articles collected from each publisher). From this pool, we randomly sampled 18,000 articles, with 1,000 randomly assigned to each of the 18 experimental conditions.

\subsubsection*{Experimental Design}

The experiment follows a $3 \times 3 \times 2$ factorial design with the following factors:
\begin{itemize}
    \item \textbf{Race/Nationality (3 levels):} German/White, Nigerian/Black, Pakistani/South Asian.
    \item \textbf{Institutional affiliation (3 levels):} New York University (Times Higher Education 2025 global rank: top 50), Louisiana State University (rank 501--600), and San Diego State University (rank 1000+)~\cite{timeshighereducationWorldUniversity}.
    \item \textbf{Request purpose (2 levels):} Literature review for a journal article or for a class project.
\end{itemize}

These three layers of randomization were chosen to distinguish between several distinct mechanisms of discrimination. First, we randomize the sender's racial identity to test for taste-based discrimination. We intentionally include two distinct Global South identities and one Global North identity. The German sender serves as a baseline reference category representing a White/Global North identity. Including both a Nigerian and a Pakistani sender allows us to assess whether responses toward Global South senders are homogeneous or vary across distinct racialized and national identities within the Global South. Second, because race and socioeconomic status are often correlated, we randomize institutional prestige independently of race. If recipients discriminate on the basis of perceived academic quality rather than race per se, we would expect institutional affiliation to predict response rates above and beyond racial identity. Third, we randomize whether the request is for a journal article or a class project to test for instrumental motives: if authors are primarily motivated by the prospect of receiving a citation, they should be more willing to share with someone writing a journal article.

We preregistered hypotheses that non-White senders, those from less prestigious institutions, and those requesting papers for class projects would receive fewer favorable responses.

\subsubsection*{Email Construction}

Three Google email (Gmail) accounts were created for the following fictional PhD student senders:
\begin{itemize}
    \item \textbf{Karl Wagner} (German/White)
    \item \textbf{Oluwasen Adeyemi} (Nigerian/Black)
    \item \textbf{Faisal Malik} (Pakistani/South Asian)
\end{itemize}

Each account included a professional profile picture generated using GPT-4o (see Supplementary Figure~5 for the photos used). Each email included the sender's full name, country of origin, and institutional affiliation in the body of the email, as well as a small photo in the email signature. Importantly, the country and institution associated with each sender were explicitly stated in the text, rather than being merely implied by the sender's name or email address. The body of the email was constructed as follows:

\begin{quote}
    Dear Professor [Recipient Name],

    My name is [Student Name] and I am a student from [Student Country] currently studying at [Student University]. I am currently working on [course project/journal article] and while writing the literature review, I came across your paper titled [Paper Title] which I believe would be useful to include in my literature review.

    Unfortunately, I do not seem to have access to the paper. Would you be willing to share the PDF of the paper with me?

    Best Regards and thank you in advance,

    [Student Name]
\end{quote}

\subsubsection*{Data Collection and Timeline}

The 1,000 emails per condition were sent out between April 26th and June 8th, 2025, in rotating batches across weeks to control for temporal effects. Responses were collected until June 22nd, 2025, giving respondents at least 14 days to reply. We used the Mailsuite Gmail extension to track whether each email was opened by the recipient. The number of emails opened in each experimental condition is reported in Supplementary Table~13.

Of the 18,000 total emails sent, 9,850 (54.7\%) were opened. Of those, 2,059 received a reply. We discarded 181 automated responses (e.g., out-of-office messages), leaving 1,878 substantive replies (10.4\% of all emails sent; 19.0\% of opened emails). Each reply was manually examined to determine whether it included the requested paper. The proportion of replies granting access across experimental conditions is reported in Supplementary Table~14.

\subsubsection*{Recipient Characteristics}

For each recipient, we used OpenAlex~\cite{priem2022openalex} to retrieve their h-index, academic age (measured as the number of years between 2025 and the year of their first publication), and the region of their primary institutional affiliation. Descriptive statistics for the matched sample are as follows: $N = 15{,}200$, h-index mean $= 25.7$, academic age mean $= 12.0$ years.

\subsubsection*{Statistical Analysis}

Our primary dependent variable is whether the recipient replied to the email. We estimate logistic regression models predicting the probability of reply for all sent emails. Independent variables include the sender's race, institutional affiliation, and request purpose. Controls include the domain of the article (retrieved from OpenAlex), the recipient's h-index and academic age, and whether the sender's country of origin matches the region of the recipient's institution. This region-match variable was chosen as an alternative to inferring the racial identity of the recipient, given the unreliability of name-based demographic classification in a global sample. We report multiple model configurations in Supplementary Table~15, including a baseline model with only the experimental variables, a model including the full set of controls, and a model estimating heterogeneous effects. We also report models using receipt of the requested paper (rather than any reply) as the outcome variable in Supplementary Table~16.

\subsection*{Study 5: Data Availability Request Study}

\subsubsection*{Overview}

The second randomized controlled trial investigates bias in access to datasets that authors claim are ``available upon reasonable request.'' We emailed the corresponding authors of papers containing such Data Availability Statements, varying the racial identity, country of institutional affiliation, institutional rank, and stated purpose of the requester. Our hypotheses were preregistered prior to beginning the experiment (\url{https://osf.io/4vhar/}). This study is a conceptual replication of \cite{acciai2023estimating}.

\subsubsection*{Sampling}

We compiled a set of articles published between 2022 and 2024 across four high-profile journals that mandate a Data and/or Code Availability section: \textit{Nature}, \textit{Nature Communications}, \textit{Nature Human Behaviour}, and \textit{Scientific Reports}. For each article, we scraped the Data Availability Statement and filtered for articles containing the phrase ``upon request.'' We also scraped the article title, journal name, corresponding author name, and contact email from each article's website. This process yielded a final set of 16,000 articles (see Supplementary Table~19 for the number of articles collected from each journal).

\subsubsection*{Experimental Design}

The experiment follows a $2 \times 2 \times 2 \times 2$ factorial design with the following factors:
\begin{itemize}
    \item \textbf{Race/Nationality (2 levels):} English/White or South African/Black.
    \item \textbf{Institution country (2 levels):} England or South Africa.
    \item \textbf{Institution rank (2 levels):} Globally ranked 150--155 (high rank) or 1001--1200 (low rank), according to the 2026 QS World University Rankings~\cite{topuniversitiesWorldUniversity}.
    \item \textbf{Request purpose (2 levels):} Journal article or final course project.
\end{itemize}

The crossing of institution country and institution rank yields four specific universities:
\begin{itemize}
    \item \textbf{England, high rank:} University of Exeter (global rank 155)
    \item \textbf{England, low rank:} University of Wolverhampton (global rank 1001--1200)
    \item \textbf{South Africa, high rank:} University of Cape Town (global rank 150)
    \item \textbf{South Africa, low rank:} University of the Free State (global rank 1001--1200)
\end{itemize}

In contrast to Study~3, this design independently varies both the sender's racial identity and their institutional affiliation, allowing us to disentangle race from prestige and geographic signals. Each of the 16 experimental conditions was assigned 1,000 randomly selected articles from the pool of 16,000.

We preregistered four main hypotheses. First, emails sent by the English/White sender would receive more responses than those sent by the South African/Black sender. Second, requests originating from universities located in England would receive more responses than those from South African universities. Third, senders affiliated with higher-ranked universities would receive more responses than those from lower-ranked institutions. Fourth, data requests framed as supporting a journal article would be more successful than those for a course project.

\subsubsection*{Email Construction}

Two Gmail accounts were created for the following fictional PhD student senders:
\begin{itemize}
    \item \textbf{James Whitfield} (English/White)
    \item \textbf{Kabelo Molefe} (South African/Black)
\end{itemize}

Each account included a professional profile picture (see Supplementary Figure~5). Each email stated the sender's full name, country of origin, and institutional affiliation in the body of the email, with the sender's name and university repeated in the email signature. The body of the email was constructed as follows:

\begin{quote}
    Dear Professor [Recipient Name],

    My name is [Sender Name] and I am a student from the [Sender Country of Origin] currently studying at [Sender University]. I recently came across your paper published in [Journal Name] titled [Paper Title]. I am currently working on a [Request Purpose] and was interested in pursuing a follow-up to your work. It is mentioned that some of the data could be available upon request in the manuscript.

    Would you be willing to share this data with me?

    Best regards and thank you in advance,

    [Sender Name]

    Ph.D. student - [Sender University]
\end{quote}

\subsubsection*{Data Collection and Timeline}

The 1,000 emails per condition were sent out between January 5th and February 4th, 2026. Responses were collected until February 11th, 2026, giving respondents at least one week to reply. We used the Mailsuite Gmail extension to track whether each email was opened by the recipient. The number of emails opened in each experimental condition is reported in Supplementary Table~20.

Of the 16,000 total emails sent, 13,503 were successfully delivered (the remainder bounced due to inactive email addresses). Of those delivered, 6,981 (51.7\%) were opened, and of those, 1,813 (26.0\%) received a reply. The reply rates for each experimental condition are reported in Supplementary Table~21.

\subsubsection*{Response Coding}

After discarding automated responses, two independent annotators manually examined each reply and classified it into one of three categories:
\begin{itemize}
    \item \textbf{Positive:} The recipient provided the data directly, provided information on how to access the data, or forwarded the email to the appropriate person with access.
    \item \textbf{Conditional:} The recipient requested additional information about the purpose of the request, asked the sender to provide credentials, or asked for the sender's advisor or principal investigator to submit the request instead.
    \item \textbf{Negative:} The recipient declined the request outright, citing reasons such as the data being used in an ongoing project, offering data only in exchange for co-authorship or financial compensation, or stating that the data had been lost.
\end{itemize}
Inter-rater agreement was high (Cohen's $\kappa = 0.964$). Disagreements were resolved through discussion. The number of positive, conditional, and negative replies per experimental condition is reported in Supplementary Table~22.

\subsubsection*{Recipient Characteristics}

For each recipient whose email was opened, we used OpenAlex~\cite{priem2022openalex} to retrieve their latest institutional affiliation, h-index, and academic age (measured as the number of years between 2025 and the year of their first publication). Descriptive statistics for the matched sample are as follows: $N = 6{,}685$, h-index mean $= 48.2$, academic age mean $= 28.9$ years. Response rates by article domain and journal are illustrated in Supplementary Figure~8.

\subsubsection*{Statistical Analysis}

Our primary dependent variable is whether the recipient replied to the email. We estimate logistic regression models predicting the probability of reply for all sent emails. Independent variables include the sender's racial identity, institutional country, institutional rank, and request purpose. Controls include the domain and journal of the article, and the recipient's h-index and academic age. We do not control for a match between the sender's country of origin and the region of the recipient's university due to the limited number of recipients affiliated with an African institution (109 of 16,000, or 0.68\%).

We report multiple model configurations in Supplementary Tables~23 and 24, including baseline models with only the experimental variables, models with the full set of controls, and models estimating heterogeneous effects. Supplementary Table~24 specifically examines interactions between sender identity and institutional characteristics to assess whether the relative importance of institutional rank versus institutional country differs by sender race.

\section*{Ethics statement}
The research was approved by the Institutional Review Board of New York University Abu Dhabi (HRPP-2024-45). All research was performed in accordance with relevant guidelines and regulations.

\section*{Data and Code Availability}
All the code and data required to replicate our observational findings, as well as anonymized data concerning the audit experiments, are deposited in the following GitHub repository: \url{https://github.com/comnetsAD/bias_academia}. The codebook used for classifying email responses in Study~5 is provided in Supplementary Table~25.

For Study 1, the full survey instrument, interview guide, and interview consent script are provided in Supplementary Notes 1--3, and aggregate survey statistics are reported in Supplementary Tables~1--5, 26. Interview transcripts are not available, consistent with the consent protocol for Study~1, under which interview data are accessed only by the research team and no material may be shared that could reasonably identify participants or their institutions. Transcripts remain securely stored on NYU-approved encrypted servers, subject to audit by the NYUAD Institutional Review Board; de-identified quotations are presented in the Results.

\bibliographystyle{naturemag}
\bibliography{sample}

\newpage

\section*{Author Contributions}

T.R. and Y.Z. conceived the study. T.R., Y.Z., A.K. and H.I. designed the audit experiments. K.M., F.L, and H.I. identified articles of interest and collected the emails of the respective corresponding authors. H.I. conducted the audit experiment. H.I. analyzed the data and produced the visualizations concerning the audit experiments. F.L. collected and analyzed the data, and produced the visualizations concerning the observational analyses. H.I and  A.K conducted the global survey. A.K. and W.A. conducted the interviews. H.I., F.L., W.A., A.K., Y.Z., and T.R. wrote the manuscript.

\section*{Competing Interests}
The authors declare no competing interests.

\clearpage
\setcounter{figure}{0}
\setcounter{table}{0}
\renewcommand{\figurename}{Supplementary Figure}
\renewcommand{\tablename}{Supplementary Table}

\begin{center}
{\fontsize{14}{14}\selectfont{Supplementary Information for}}
\end{center}

\begin{center}
{\fontsize{16}{16}\selectfont{Causal evidence of racial and institutional biases in accessing paywalled articles and scientific data}}
\end{center}
\smallskip\smallskip
\begin{center}
{\fontsize{12}{12}\selectfont{Hazem Ibrahim$^{1+}$, Fengyuan Liu$^{1+}$, Khalid Mengal$^1$, Wisam Alshaibi$^2$, \\ Aaron R. Kaufman$^2$, Yasir Zaki$^1$$^\ast$, Talal Rahwan$^1$$^\ast$}}
\end{center}

\begin{center}
{\fontsize{12}{12}\selectfont{
$^+$Joint first-author

\vspace{0.2em}
$^1$Computer Science, New York University Abu Dhabi

\vspace{0.2em}
$^2$Social Science Division, New York University Abu Dhabi

\vspace{0.2em}
$^\ast$Corresponding author. E-mail: \{yasir.zaki,talal.rahwan\}@nyu.edu}}
\end{center}

\ \\\\
\ \\\\

This document is structured as follows:
\smallskip\smallskip
\begin{itemize}\itemsep0.5em

\item \textbf{Supplementary Note 1: Survey Instrument} (\emph{page~\pageref{survey}})

\item \textbf{Supplementary Note 2: Interview Guide} (\emph{page~\pageref{interviewguide}})

\item \textbf{Supplementary Note 3: Interview Consent Script} (\emph{page~\pageref{interviewconsent}})

\item \textbf{Supplementary Tables} (\emph{page~\pageref{tables}})

\item \textbf{Supplementary Figures} (\emph{page~\pageref{figures}})

\end{itemize}
\clearpage

\section*{Supplementary Note 1: Survey Instrument}
\label{survey}

{

The following survey was administered via Qualtrics to researchers at institutions worldwide.

\subsection*{Consent}

\begin{quote}
\textbf{Principal Investigators:} Yasir Zaki / Talal Rahwan

\textbf{Study Title:} Experiences of Researchers in Accessing Academic Papers and Research Datasets

\medskip

You are being asked to provide consent to participate in a research study. Participation is voluntary. You can say yes or no. If you say yes now you can still change your mind later.

\medskip

\textbf{Purpose of research:} This research is being conducted to better understand the experiences of researchers at institutions across the world in accessing academic papers and research datasets, including barriers they may face and strategies they use to overcome them. You are being asked to participate because you are a researcher affiliated with an academic or research institution.

\textbf{Procedures to be followed:} You will be asked to complete an online survey about your institutional context, your experiences accessing paywalled academic papers and restricted research datasets, and the impact of any access barriers on your research. The survey questions are about your professional experiences with accessing scholarly resources. At the end of the survey, you will be asked whether you would be willing to participate in a follow-up interview; this is entirely optional and not required for survey participation.

\textbf{Duration of participation:} Participation will involve approximately 5--10 minutes of your time.

\textbf{Gratitude Award/Allowance:} You will not receive financial compensation for participation in the survey itself. Some participants who complete this study will be invited to participate in a 30-minute follow-up interview, which will be compensated at \$50. Any token gift, voucher, or other item of appreciation provided to you is not a fee, wage, or remuneration, and is given solely as a gesture of thanks.

\textbf{Risks or inconveniences:} We believe the risks for participation in this study are minimal; however, a possible inconvenience may be the time it takes to complete the study. Additionally, the most common risk of answering surveys is a possible breach of your privacy or confidentiality. To manage this risk, no identifiable information is collected as part of the survey itself, and all data is stored securely with password protection. Some questions ask about challenges you may face in your professional role, which may prompt reflection on frustrating experiences. You can skip any question that you do not wish to answer, or stop the survey at any time.

Your participation in this research is voluntary, and you will not be penalized or lose benefits if you refuse to participate or decide to stop.

\textbf{Benefits to the participant or to others:} You will not benefit directly from participating in this study. We hope to learn more about disparities in access to scientific knowledge, and your participation will help us to do that. The results may inform policies related to open access publishing and data-sharing practices.

\textbf{Privacy and Confidentiality:} Your privacy will be respected throughout this study and all information you provide will be treated as confidential to the extent permitted by applicable laws. Only information necessary for the study purpose will be collected and handled in accordance with NYU's privacy standards and applicable data protection laws. You have the right to request access to your data or request its withdrawal, modification or erasure from the research record. No identifiable information (e.g., your name and email address) will be collected as part of the survey itself. If you choose to volunteer for a follow-up interview, you will be asked to provide a contact email on a separate page that is not linked to your survey responses. Your responses will be collected and stored using NYU Qualtrics, a secure, cloud-based survey platform approved for academic research. Qualtrics uses distributed servers that may be located outside the country where data collection occurs. Data are encrypted through transmission and while stored on Qualtrics servers. Only the approved research team will have access to identifiable survey data. Access is restricted to NYU-issued devices protected by password login and multi-factor authentication. Identifiable data downloaded from Qualtrics will be stored on NYU-approved platforms and will be de-identified or deleted from Qualtrics once no longer needed for the study purpose. All study records will be stored securely on NYU-approved encrypted servers. Access to identifiable information will be limited to the members of the research team and controlled through password protection and multi-factor authentication. Any data shared with collaborators or used in publications will be de-identified or aggregated so that you cannot be reasonably identified. You will not be personally identified in any publications or presentations resulting from this research. The information collected about you through this study will be accessed, stored and analyzed by the researcher/research team while they are in the United Arab Emirates. You should also know that the NYUAD Institutional Review Board (IRB) may inspect study records as part of its auditing program, but these reviews will only focus on the researchers and not on your responses or involvement. The IRB is a group of people who review research studies to protect the rights and welfare of research participants.

\textbf{Contact Information:} If you have further questions about this study or if you have a research-related problem, you may contact the principal investigators, Yasir Zaki or Talal Rahwan (nyuad.ai-and-society@nyu.edu). If you have any questions concerning your rights as a research participant, you may contact the New York University Abu Dhabi Institutional Review Board (IRB) at irbnyuad@nyu.edu.

\medskip

\textbf{Agreement to Participate:} By clicking the button below, you acknowledge:
\begin{enumerate}
    \item Your participation in the survey is voluntary.
    \item You are 18 years of age or older.
    \item You are aware that you may withdraw from the study at any time for any reason.
\end{enumerate}

$\bigcirc$ I consent \quad $\bigcirc$ I do not consent
\end{quote}

\bigskip
\noindent\rule{\textwidth}{0.4pt}

\subsection*{Section 1: Demographics}

\begin{enumerate}[label=\textbf{Q\arabic*.}, leftmargin=2.5em, itemsep=1em]

\item \textbf{In which country is your primary institutional affiliation located?}\\
\textit{[Free text]}

\item \textbf{What is your nationality?}\\
\textit{[Free text]}

\item \textbf{What is your sex?}
\begin{itemize}[nosep, leftmargin=1.5em]
    \item[$\bigcirc$] Male
    \item[$\bigcirc$] Female
    \item[$\bigcirc$] Prefer not to say
    \item[$\bigcirc$] Prefer to self describe: \underline{\hspace{3cm}}
\end{itemize}

\item \textbf{Which of the following best describes your current academic position?}
\begin{itemize}[nosep, leftmargin=1.5em]
    \item[$\bigcirc$] PhD student
    \item[$\bigcirc$] Postdoctoral researcher
    \item[$\bigcirc$] Lecturer / Assistant Professor
    \item[$\bigcirc$] Senior Lecturer / Associate Professor
    \item[$\bigcirc$] Full Professor
    \item[$\bigcirc$] Research scientist (non-faculty)
    \item[$\bigcirc$] Other (please specify): \underline{\hspace{3cm}}
\end{itemize}

\item \textbf{What university are you primarily affiliated with?}\\
\textit{[Free text]}

\item \textbf{Which of the following best describes your primary research discipline?}
\begin{itemize}[nosep, leftmargin=1.5em]
    \item[$\bigcirc$] Health Sciences (e.g., medicine, public health, nursing)
    \item[$\bigcirc$] Life Sciences (e.g., biology, ecology, neuroscience)
    \item[$\bigcirc$] Physical Sciences and Engineering (e.g., physics, chemistry, computer science, engineering)
    \item[$\bigcirc$] Social Sciences (e.g., economics, political science, psychology, sociology)
    \item[$\bigcirc$] Formal Sciences (e.g., mathematics, statistics)
    \item[$\bigcirc$] Humanities (e.g., history, philosophy, literature)
    \item[$\bigcirc$] Other (please specify): \underline{\hspace{3cm}}
\end{itemize}

\item \textbf{Approximately how many peer-reviewed publications have you authored or co-authored?}
\begin{itemize}[nosep, leftmargin=1.5em]
    \item[$\bigcirc$] 0
    \item[$\bigcirc$] 1--5
    \item[$\bigcirc$] 6--15
    \item[$\bigcirc$] 16--30
    \item[$\bigcirc$] 31--50
    \item[$\bigcirc$] More than 50
\end{itemize}

\end{enumerate}

\noindent\rule{\textwidth}{0.4pt}

\subsection*{Section 2: Open-Ended}

\begin{enumerate}[label=\textbf{Q\arabic*.}, leftmargin=2.5em, itemsep=1em, resume]

\item \textbf{In your own words, what are the biggest challenges you face in accessing the materials you need to conduct your research (e.g., published papers, datasets, code, equipment)? If possible, please describe a specific instance where difficulty accessing a material affected your work.}\\
\textit{[Free text, large text box]}

\end{enumerate}

\noindent\rule{\textwidth}{0.4pt}

\subsection*{Section 3: Structured Questions --- Paper Access}

\noindent\textit{The following questions are about your experience accessing published academic papers.}

\begin{enumerate}[label=\textbf{Q\arabic*.}, leftmargin=2.5em, itemsep=1em, resume]

\item \textbf{Does your institution currently provide access to any of the following publisher databases?}\\
\textit{[Matrix: rows = publishers listed below; columns = ``I have access'' / ``I do not have access'' / ``Not sure'']}

\smallskip
\begin{tabular}{p{6cm}ccc}
\toprule
 & I have access & I do not have access & Not sure \\
\midrule
Elsevier (ScienceDirect) & $\bigcirc$ & $\bigcirc$ & $\bigcirc$ \\
Springer / Nature & $\bigcirc$ & $\bigcirc$ & $\bigcirc$ \\
Wiley & $\bigcirc$ & $\bigcirc$ & $\bigcirc$ \\
Taylor \& Francis & $\bigcirc$ & $\bigcirc$ & $\bigcirc$ \\
SAGE & $\bigcirc$ & $\bigcirc$ & $\bigcirc$ \\
IEEE & $\bigcirc$ & $\bigcirc$ & $\bigcirc$ \\
ACM Digital Library & $\bigcirc$ & $\bigcirc$ & $\bigcirc$ \\
JSTOR & $\bigcirc$ & $\bigcirc$ & $\bigcirc$ \\
Oxford University Press & $\bigcirc$ & $\bigcirc$ & $\bigcirc$ \\
Cambridge University Press & $\bigcirc$ & $\bigcirc$ & $\bigcirc$ \\
ACS (American Chemical Society) & $\bigcirc$ & $\bigcirc$ & $\bigcirc$ \\
AAAS (Science; Science Advances; etc.) & $\bigcirc$ & $\bigcirc$ & $\bigcirc$ \\
\bottomrule
\end{tabular}

\smallskip
Other (please specify): \underline{\hspace{5cm}}

\item \textbf{In the past 12 months, how often have you encountered a paper you needed for your research but could not access through your institution?}
\begin{itemize}[nosep, leftmargin=1.5em]
    \item[$\bigcirc$] Never
    \item[$\bigcirc$] Rarely (1--2 times)
    \item[$\bigcirc$] Sometimes (3--5 times)
    \item[$\bigcirc$] Often (6--10 times)
    \item[$\bigcirc$] Very often (more than 10 times)
\end{itemize}

\item \textbf{When you cannot access a paper through your institution, which of the following strategies do you use? (Select all that apply)}
\begin{itemize}[nosep, leftmargin=1.5em]
    \item[$\square$] Email the author(s) directly
    \item[$\square$] Request through ResearchGate or Academia.edu
    \item[$\square$] Use Sci-Hub or similar tools
    \item[$\square$] Search for a preprint version (e.g., on arXiv, SSRN, bioRxiv)
    \item[$\square$] Ask a colleague at another institution to download it for you
    \item[$\square$] Request through your library's interlibrary loan service
    \item[$\square$] Use a VPN or account from a previous/partner institution
    \item[$\square$] Give up and use a different paper instead
    \item[$\square$] Give up and do not replace the paper
    \item[$\square$] Other (please specify): \underline{\hspace{3cm}}
\end{itemize}

\item \textbf{If you have emailed authors directly to request a copy of a paywalled paper, approximately how often do you receive a response?}
\begin{itemize}[nosep, leftmargin=1.5em]
    \item[$\bigcirc$] I have never emailed an author for this purpose
    \item[$\bigcirc$] Almost never (less than 10\% of the time)
    \item[$\bigcirc$] Rarely (10--30\% of the time)
    \item[$\bigcirc$] Sometimes (30--50\% of the time)
    \item[$\bigcirc$] Often (50--70\% of the time)
    \item[$\bigcirc$] Usually (70--90\% of the time)
    \item[$\bigcirc$] Almost always (more than 90\% of the time)
\end{itemize}

\item \textbf{Has difficulty accessing a published paper ever caused you to do any of the following? (Select all that apply)}
\begin{itemize}[nosep, leftmargin=1.5em]
    \item[$\square$] Dropped a line of inquiry or research question
    \item[$\square$] Changed the topic of a research project
    \item[$\square$] Changed the topic of a student's thesis or dissertation
    \item[$\square$] Submitted a paper for publication without citing work you knew was relevant
    \item[$\square$] Narrowed the scope of a literature review
    \item[$\square$] Delayed the completion of a research project
    \item[$\square$] None of the above
    \item[$\square$] Other (please specify): \underline{\hspace{3cm}}
\end{itemize}

\end{enumerate}

\noindent\rule{\textwidth}{0.4pt}

\subsection*{Section 3 (continued): Structured Questions --- Data Access}

\noindent\textit{The following questions are about your experience accessing research datasets associated with published papers.}

\begin{enumerate}[label=\textbf{Q\arabic*.}, leftmargin=2.5em, itemsep=1em, resume]

\item \textbf{In the past 12 months, how often have you needed to access a dataset associated with a published paper (e.g., for replication, extension, or secondary analysis)?}
\begin{itemize}[nosep, leftmargin=1.5em]
    \item[$\bigcirc$] Never
    \item[$\bigcirc$] Rarely (1--2 times)
    \item[$\bigcirc$] Sometimes (3--5 times)
    \item[$\bigcirc$] Often (6--10 times)
    \item[$\bigcirc$] Very often (more than 10 times)
\end{itemize}

\item \textbf{When a paper states that data is ``available upon request,'' how often do you actually receive the data when you request it?}
\begin{itemize}[nosep, leftmargin=1.5em]
    \item[$\bigcirc$] I have never requested data from an author
    \item[$\bigcirc$] Almost never (less than 10\% of the time)
    \item[$\bigcirc$] Rarely (10--30\% of the time)
    \item[$\bigcirc$] Sometimes (30--50\% of the time)
    \item[$\bigcirc$] Often (50--70\% of the time)
    \item[$\bigcirc$] Usually (70--90\% of the time)
    \item[$\bigcirc$] Almost always (more than 90\% of the time)
\end{itemize}

\item \textbf{If you have requested data from authors and did NOT receive it, what reasons (if any) were given? (Select all that apply)}
\begin{itemize}[nosep, leftmargin=1.5em]
    \item[$\square$] I have never been refused data
    \item[$\square$] No response to my request
    \item[$\square$] The author said the data was lost or no longer available
    \item[$\square$] The author said the data could not be shared due to privacy or legal restrictions
    \item[$\square$] The author asked for co-authorship in exchange for the data
    \item[$\square$] The author asked for financial compensation
    \item[$\square$] The author asked for my credentials or supervisor's approval
    \item[$\square$] The author said the data was being used for another ongoing project
    \item[$\square$] The author did not give a reason
    \item[$\square$] Other (please specify): \underline{\hspace{3cm}}
\end{itemize}

\item \textbf{Has difficulty accessing a research dataset ever caused you to do any of the following? (Select all that apply)}
\begin{itemize}[nosep, leftmargin=1.5em]
    \item[$\square$] Dropped a line of inquiry or research question
    \item[$\square$] Changed the topic of a research project
    \item[$\square$] Changed the topic of a student's thesis or dissertation
    \item[$\square$] Was unable to replicate a published finding
    \item[$\square$] Delayed the completion of a research project
    \item[$\square$] None of the above
    \item[$\square$] Other (please specify): \underline{\hspace{3cm}}
\end{itemize}

\item \textbf{How important is the availability of underlying data when you decide which published papers to cite or build upon in your own research?}
\begin{itemize}[nosep, leftmargin=1.5em]
    \item[$\bigcirc$] Not at all important
    \item[$\bigcirc$] Slightly important
    \item[$\bigcirc$] Moderately important
    \item[$\bigcirc$] Very important
    \item[$\bigcirc$] Extremely important
\end{itemize}

\item \textbf{How important is the availability of underlying data when you decide which published papers to cite or build upon in your teaching?}
\begin{itemize}[nosep, leftmargin=1.5em]
    \item[$\bigcirc$] Not at all important
    \item[$\bigcirc$] Slightly important
    \item[$\bigcirc$] Moderately important
    \item[$\bigcirc$] Very important
    \item[$\bigcirc$] Extremely important
    \item[$\bigcirc$] Not applicable (I do not teach)
\end{itemize}

\end{enumerate}

\noindent\rule{\textwidth}{0.4pt}

\subsection*{Section 4: Interview Opt-In}

\noindent\textit{Thank you for completing the survey. As part of this research, we are also conducting follow-up interviews (approximately 45--60 minutes, via Zoom) with researchers who are willing to share their experiences in more detail. If interviewed, you would be compensated with 50 USD.}

\begin{enumerate}[label=\textbf{Q\arabic*.}, leftmargin=2.5em, itemsep=1em, resume]

\item \textbf{Would you be willing to participate in a follow-up interview? The interview would be conducted in English.}
\begin{itemize}[nosep, leftmargin=1.5em]
    \item[$\bigcirc$] Yes
    \item[$\bigcirc$] No
\end{itemize}

\end{enumerate}

\medskip

\noindent\textit{[If ``Yes'' is selected, the following message is displayed:]}

\begin{quote}
Thank you for your willingness to participate in the follow-up interview. Please enter your email address below, and if selected, we will follow up directly. We will not store your email address, and will only use it to contact you with your consent.

\medskip
Email: \underline{\hspace{6cm}}
\end{quote}

}

\clearpage

\section*{Supplementary Note 2: Interview Guide}
\label{interviewguide}

{

The following question bank guided the semi-structured interviews in Study~1 (see Methods in the main text). The questions served as a pool of topics and prompts rather than a fixed script: not every question was asked in every interview, questions could be asked in any order, and the interviewer adapted to the flow of conversation and to each participant's expertise and experiences. Interviews were conversational and participant-led where appropriate; questions could be skipped or rephrased depending on context; participants could decline to answer any question; and probes were used selectively to clarify or deepen responses.

\subsection*{1. Background and Research Context}
\begin{itemize}[nosep, leftmargin=1.5em]
    \item Can you tell me about your current role and research area?
    \item How would you describe the research environment in your institution or country?
    \item What kinds of projects do you typically work on?
    \item How did you come to work in this field?
\end{itemize}
\noindent\textit{Optional probes:} training and career trajectory; institutional affiliation(s); collaborations (local vs.\ international).

\subsection*{2. Research Infrastructure and Resources}
\begin{itemize}[nosep, leftmargin=1.5em]
    \item What resources are most important for your work (e.g., funding, data, lab space, computing)?
    \item How accessible are these resources in your context?
    \item Are there any infrastructure-related constraints that shape what research you can do?
\end{itemize}
\noindent\textit{Optional probes:} internet access, software, equipment; administrative or procurement processes; maintenance and reliability of resources.

\subsection*{3. Data Availability and Access}
\begin{itemize}[nosep, leftmargin=1.5em]
    \item What kinds of data do you typically rely on in your research?
    \item How accessible are these data in your context?
    \item Are there important datasets that are difficult or impossible for you to access?
\end{itemize}
\noindent\textit{Optional probes:} government or administrative data access; costs associated with acquiring data; data quality, completeness, or reliability; restrictions on data sharing or use; challenges with cross-border data access.

\subsection*{4. Funding and Financial Constraints}
\begin{itemize}[nosep, leftmargin=1.5em]
    \item What has your experience been with research funding?
    \item What kinds of funding opportunities are available to you?
    \item Are there barriers to accessing funding?
\end{itemize}
\noindent\textit{Optional probes:} domestic vs.\ international funding; grant application processes; currency, overhead, or transfer issues.

\subsection*{5. Access to Academic Literature and Paywalled Research}
\begin{itemize}[nosep, leftmargin=1.5em]
    \item How do you typically access academic papers and research materials?
    \item Do you encounter barriers when trying to access paywalled articles or journals?
    \item How do these access constraints affect your research?
\end{itemize}
\noindent\textit{Optional probes:} institutional subscriptions and library access; use of open access resources; informal access strategies (e.g., requesting papers from authors); impact on literature reviews or staying current in the field.

\subsection*{6. Collaboration and Networks}
\begin{itemize}[nosep, leftmargin=1.5em]
    \item Who do you typically collaborate with?
    \item How easy or difficult is it to build collaborations, both locally and internationally?
    \item What role do international partnerships play in your work?
\end{itemize}
\noindent\textit{Optional probes:} power dynamics in collaborations; authorship and credit; barriers to entering global research networks.

\subsection*{7. Publication and Dissemination}
\begin{itemize}[nosep, leftmargin=1.5em]
    \item What has your experience been with publishing research?
    \item Are there barriers to publishing in major journals?
    \item How do you share your work with others?
\end{itemize}
\noindent\textit{Optional probes:} publication fees (APCs); peer review experiences; language barriers; visibility and citation.

\subsection*{8. Institutional and Administrative Context}
\begin{itemize}[nosep, leftmargin=1.5em]
    \item How do institutional policies or structures affect your research?
    \item Are there administrative processes that facilitate or hinder your work?
\end{itemize}
\noindent\textit{Optional probes:} bureaucracy and approvals; teaching or service burdens; incentive structures.

\subsection*{9. Political, Social, or Regulatory Environment}
\begin{itemize}[nosep, leftmargin=1.5em]
    \item Are there broader political or regulatory factors that influence your research?
    \item Are there topics that are easier or harder to study in your context?
\end{itemize}
\noindent\textit{Optional probes (used cautiously):} government restrictions; data access limitations; ethical review processes.

\subsection*{10. Career Development and Training}
\begin{itemize}[nosep, leftmargin=1.5em]
    \item What opportunities exist for training or professional development?
    \item What challenges do early-career researchers face in your context?
\end{itemize}
\noindent\textit{Optional probes:} mentorship; mobility (travel, visas); brain drain and retention.

\subsection*{11. Inequality and the Global Research System}
\begin{itemize}[nosep, leftmargin=1.5em]
    \item How do you see your work fitting into the global research landscape?
    \item Do you perceive inequalities between researchers in different regions? If so, how do they manifest?
\end{itemize}
\noindent\textit{Optional probes:} agenda-setting power; resource asymmetries; recognition and prestige.

\subsection*{12. Adaptation and Strategies}
\begin{itemize}[nosep, leftmargin=1.5em]
    \item How have you adapted to constraints in your research environment?
    \item Are there strategies that have helped you overcome barriers?
\end{itemize}
\noindent\textit{Optional probes:} informal practices or workarounds; community or institutional support; personal trade-offs.

\subsection*{13. Reflections and Recommendations}
\begin{itemize}[nosep, leftmargin=1.5em]
    \item What changes would most improve your ability to do research?
    \item What do you wish people outside your context better understood?
\end{itemize}
\noindent\textit{Optional probes:} policy recommendations; advice to funders or institutions; advice to other researchers.

\subsection*{14. Closing}
\begin{itemize}[nosep, leftmargin=1.5em]
    \item Is there anything else you would like to share that we haven't discussed?
    \item Are there topics you think are important that I didn't ask about?
\end{itemize}

}

\clearpage

\section*{Supplementary Note 3: Interview Consent Script}
\label{interviewconsent}

{

\noindent\textbf{Verbal Consent Script for Interview Participation}

\medskip
\noindent\textbf{Principal Investigators:} Yasir Zaki / Talal Rahwan\\
\textbf{Study Title:} Barriers Scientists Face in Conducting Research in Local and Institutional Contexts

\medskip
\noindent\textit{[The following script will be read aloud to participants at the start of the interview. Verbal consent will be captured via audio recording, or noted by the interviewer if recording is declined.]}

\medskip
Hello, and thank you for taking the time to speak with me today. Before we begin, I'd like to briefly explain the study and your rights as a participant.

You are being asked to provide consent to participate in a research study. Participation is voluntary. You can say yes or no. If you say yes now, you can still change your mind later.

\medskip
\noindent\textbf{Purpose of research:} This research is being conducted to better understand structural, institutional, and resource-related challenges in global research environments, including the barriers scientists face in conducting research in their local or institutional contexts. You are being asked to participate because you are a researcher affiliated with an academic or research institution.

\medskip
\noindent\textbf{Procedures to be followed:} You will be asked to participate in a semi-structured interview lasting approximately 30 minutes. The interview questions are about your professional experiences, including challenges within your professional or institutional environment. With your permission, I would like to audio record this interview using Zoom. The recording will be used to create a transcript, including through Zoom's AI-based transcription tools. These tools may process audio data on external servers. You have the right to review recordings and request that they be deleted. If you decline to be recorded, I will take notes instead.

\medskip
\noindent\textbf{Duration of participation:} Participation will involve approximately 30 minutes of your time.

\medskip
\noindent\textbf{Gratitude award/allowance:} For your participation in this interview, you will be compensated the equivalent of \$50 US dollars. Any token gift, voucher, or other item of appreciation provided to you is not a fee, wage, or remuneration, and is given solely as a gesture of thanks.

\medskip
\noindent\textbf{Risks or inconveniences:} We believe the risks for participation in this study are minimal; however, a possible inconvenience may be the time it takes to complete the interview. Additionally, the most common risk of being interviewed is a possible breach of your privacy or confidentiality. To manage this risk, your name and any identifying information will be removed from transcripts and replaced with a code or pseudonym. Some questions ask about challenges within your professional or institutional environment, which may feel sensitive. You can skip any question that you do not wish to answer, or stop the interview at any time.

Your participation in this research is voluntary, and you will not be penalized or lose benefits if you refuse to participate or decide to stop.

\medskip
\noindent\textbf{Benefits to the participant or to others:} You will not benefit directly from participating in this study. We hope to learn more about structural and resource-related challenges in global research environments, and your participation will help us to do that.

\medskip
\noindent\textbf{Privacy and confidentiality:} Your privacy will be respected throughout this study and all information you provide will be treated as confidential to the extent permitted by applicable laws. Only information necessary for the study purpose will be collected and handled in accordance with NYU's privacy standards and applicable data protection laws. You have the right to request access to your data or request its withdrawal, modification or erasure from the research record.

At the time of recruitment, your contact information may have been linked to prior survey responses. This identifying information has already been deleted and replaced with a study ID. After this interview, your responses will be fully de-identified.

All study records will be stored securely on NYU-approved encrypted servers. Access to identifiable information will be limited to the research team and controlled through password protection and multi-factor authentication. Any data shared with collaborators or used in publications will be de-identified or aggregated so that you cannot be reasonably identified. Audio recordings will be transcribed by the research team, and your name and identifying information will be replaced with a code or pseudonym. You will not be personally identified in any publications or presentations resulting from this research. In any publications or presentations, we will not include information that could reasonably identify you or your institution without your explicit permission.

The information collected about you through this study will be accessed, stored and analyzed by the researcher/research team while they are in the United Arab Emirates.

You should also know that the NYUAD Institutional Review Board (IRB) may inspect study records as part of its auditing program, but these reviews will only focus on the researchers and not on your responses or involvement. The IRB is a group of people who review research studies to protect the rights and welfare of research participants.

\medskip
\noindent\textbf{Contact information:} If you have further questions about this study or if you have a research-related problem, you may contact the principal investigators, Yasir Zaki or Talal Rahwan (\texttt{nyuad.ai-and-society@nyu.edu}). If you have any questions concerning your rights as a research participant, you may contact the New York University Abu Dhabi Institutional Review Board (IRB) at \texttt{irbnyuad@nyu.edu}.

\medskip
\noindent\textbf{Agreement to participate:}

\noindent\textit{[Verbal consent will be obtained by reading the following questions aloud and recording the participant's responses. If audio recording is declined, the interviewer will note the participant's verbal responses.]}

\begin{itemize}[nosep, leftmargin=1.5em]
    \item Do you have any questions before we begin?
    \item Do you consent to participate in this interview?
    \item Do you consent to being audio recorded and having the recording transcribed using Zoom's transcription tools?
\end{itemize}

}

\clearpage
\section*{Supplementary Tables}
\label{tables}

\begin{table}[h!]
\centering
\small
\caption{\textbf{Demographic characteristics of survey respondents.}}
\label{tab:demographics}
\begin{tabular}{lrrr}
\toprule
 & Global North & Global South & Total \\
\midrule
\textbf{N} & 216 & 160 & 376 \\
\textbf{Countries represented} & 36 & 40 & 76 \\
\midrule
\textbf{Career stage} & & & \\
\quad Full Professor & 81 (37.5\%) & 50 (31.2\%) & 131 (34.8\%) \\
\quad Senior Lecturer / Assoc.\ Prof. & 44 (20.4\%) & 47 (29.4\%) & 91 (24.2\%) \\
\quad Lecturer / Asst.\ Prof. & 22 (10.2\%) & 35 (21.9\%) & 57 (15.2\%) \\
\quad Postdoctoral researcher & 24 (11.1\%) & 4 (2.5\%) & 28 (7.4\%) \\
\quad PhD student & 18 (8.3\%) & 14 (8.8\%) & 32 (8.5\%) \\
\quad Research scientist (non-faculty) & 13 (6.0\%) & 6 (3.8\%) & 19 (5.1\%) \\
\quad Other & 14 (6.5\%) & 4 (2.5\%) & 18 (4.8\%) \\
\midrule
\textbf{Discipline} & & & \\
\quad Physical Sci.\ \& Engineering & 104 (48.1\%) & 70 (43.8\%) & 174 (46.3\%) \\
\quad Health Sciences & 37 (17.1\%) & 23 (14.4\%) & 60 (16.0\%) \\
\quad Social Sciences & 30 (13.9\%) & 22 (13.8\%) & 52 (13.8\%) \\
\quad Life Sciences & 20 (9.3\%) & 18 (11.2\%) & 38 (10.1\%) \\
\quad Formal Sciences & 10 (4.6\%) & 11 (6.9\%) & 21 (5.6\%) \\
\quad Humanities & 6 (2.8\%) & 4 (2.5\%) & 10 (2.7\%) \\
\quad Other & 9 (4.2\%) & 12 (7.5\%) & 21 (5.6\%) \\
\midrule
\textbf{Sex} & & & \\
\quad Male & 155 (71.8\%) & 123 (76.9\%) & 278 (73.9\%) \\
\quad Female & 52 (24.1\%) & 37 (23.1\%) & 89 (23.7\%) \\
\quad Other / Prefer not to say & 9 (4.2\%) & 0 (0.0\%) & 9 (2.4\%) \\
\midrule
\textbf{Publications} & & & \\
\quad More than 50 & 123 (56.9\%) & 72 (45.0\%) & 195 (51.9\%) \\
\quad 31--50 & 34 (15.7\%) & 27 (16.9\%) & 61 (16.2\%) \\
\quad 16--30 & 16 (7.4\%) & 23 (14.4\%) & 39 (10.4\%) \\
\quad 6--15 & 29 (13.4\%) & 26 (16.2\%) & 55 (14.6\%) \\
\quad 1--5 & 14 (6.5\%) & 12 (7.5\%) & 26 (6.9\%) \\
\bottomrule
\end{tabular}
\end{table}

\begin{table}[h!]
\centering
\small
\caption{\textbf{Institutional access to publisher databases by region.} Percentages reflect the proportion of respondents reporting ``I have access'' to each publisher. $\chi^2$ tests compare access rates between Global North and Global South.}
\label{tab:publisher_access}
\begin{tabular}{lrrrrl}
\toprule
Publisher & GN (\%) & GS (\%) & Gap (pp) & $\chi^2$ & $p$ \\
\midrule
Elsevier (ScienceDirect) & 82.9 & 62.0 & 20.9 & 19.57 & $<$0.001 \\
Springer / Nature & 81.0 & 54.4 & 26.6 & 29.35 & $<$0.001 \\
Wiley & 66.7 & 41.8 & 24.9 & 21.97 & $<$0.001 \\
Taylor \& Francis & 48.1 & 31.0 & 17.1 & 10.39 & 0.001 \\
OUP & 42.6 & 20.3 & 22.3 & 19.55 & $<$0.001 \\
JSTOR & 39.4 & 21.5 & 17.9 & 12.57 & $<$0.001 \\
IEEE & 39.8 & 27.8 & 12.0 & 5.25 & 0.022 \\
SAGE & 38.0 & 29.1 & 8.9 & 2.79 & 0.095 \\
CUP & 38.9 & 19.0 & 19.9 & 16.13 & $<$0.001 \\
AAAS (Science) & 25.9 & 15.2 & 10.7 & 5.63 & 0.018 \\
ACM Digital Library & 22.2 & 15.2 & 7.0 & 2.47 & 0.116 \\
ACS & 16.7 & 13.9 & 2.8 & 0.34 & 0.563 \\
\bottomrule
\end{tabular}
\end{table}

\begin{table}[h!]
\centering
\small
\caption{\textbf{Strategies used when unable to access a paper through institutional channels.} Respondents could select all that apply.}
\label{tab:strategies}
\begin{tabular}{lrr}
\toprule
Strategy & GN (\%) & GS (\%) \\
\midrule
Search for a preprint version & 52.1 & 43.7 \\
Request through ResearchGate or Academia.edu & 47.9 & 47.7 \\
Give up and use a different paper & 45.7 & 35.8 \\
Use Sci-Hub or similar tools & 41.0 & 47.7 \\
Email the author(s) directly & 41.5 & 39.1 \\
Ask a colleague at another institution & 34.6 & 47.0 \\
Interlibrary loan & 25.5 & 13.9 \\
Give up and do not replace the paper & 22.3 & 15.9 \\
Use a VPN or account from a partner institution & 7.4 & 6.6 \\
\bottomrule
\end{tabular}
\end{table}

\begin{table}[h!]
\centering
\small
\caption{\textbf{Reported consequences of access barriers.} Respondents could select all that apply.}
\label{tab:consequences}
\begin{tabular}{lrrrr}
\toprule
 & \multicolumn{2}{c}{Paywall barriers} & \multicolumn{2}{c}{Data barriers} \\
\cmidrule(lr){2-3} \cmidrule(lr){4-5}
Consequence & GN (\%) & GS (\%) & GN (\%) & GS (\%) \\
\midrule
Any consequence reported & 64.4 & 70.9 & 58.3 & 62.6 \\
\midrule
Narrowed scope of literature review & 36.7 & 37.7 & --- & --- \\
Submitted without citing relevant work & 32.4 & 33.8 & --- & --- \\
Delayed project completion & 25.0 & 28.5 & 29.5 & 34.8 \\
Dropped a line of inquiry & 9.0 & 11.9 & 20.1 & 18.3 \\
Changed topic of research project & 5.3 & 7.3 & 17.3 & 18.3 \\
Changed topic of student thesis & 2.1 & 9.3 & 7.9 & 17.4 \\
\bottomrule
\end{tabular}
\end{table}

\begin{table}[h!]
\centering
\small
\caption{\textbf{Robustness check: primary findings stratified by interview opt-in status.} All comparisons are Global North vs.\ Global South within each subgroup. Effect directions are consistent across both subgroups; reduced significance in the opt-out subgroup is attributable to smaller Global South sample sizes ($n = 52$ vs.\ $n = 108$). $^{***}p<0.001$, $^{**}p<0.01$, $^{*}p<0.05$, $^{\dagger}p<0.1$.}
\label{tab:optin_robustness}
\begin{tabular}{lrrrrrr}
\toprule
 & \multicolumn{3}{c}{Opted in ($n = 211$)} & \multicolumn{3}{c}{Opted out ($n = 165$)} \\
\cmidrule(lr){2-4} \cmidrule(lr){5-7}
Variable & GN & GS & $p$ & GN & GS & $p$ \\
\midrule
\textit{Publisher access (\%)} & & & & & & \\
\quad Elsevier & 82.5 & 55.1 & $<$0.001$^{***}$ & 83.2 & 76.5 & 0.423 \\
\quad Springer / Nature & 82.5 & 50.5 & $<$0.001$^{***}$ & 79.6 & 62.7 & 0.036$^{*}$ \\
\quad Wiley & 65.0 & 36.4 & $<$0.001$^{***}$ & 68.1 & 52.9 & 0.090$^{\dagger}$ \\
\midrule
\textit{Paywall encounters (\%)} & & & & & & \\
\quad Often or very often & 31.1 & 53.7 & 0.002$^{**}$ & 23.0 & 36.5 & 0.104 \\
\midrule
\textit{Email success $>$50\% (\%)} & & & & & & \\
\quad Papers & 43.6 & 25.7 & 0.051$^{\dagger}$ & 44.3 & 25.9 & 0.164 \\
\quad Datasets & 33.3 & 11.7 & 0.016$^{*}$ & 35.4 & 25.0 & 0.581 \\
\midrule
\textit{Consequences (\%)} & & & & & & \\
\quad Any paywall consequence & 69.4 & 77.7 & 0.264 & 60.2 & 56.2 & 0.779 \\
\bottomrule
\end{tabular}
\end{table}

\begin{table}[h!]
\centering
\small
\caption{\textbf{Demographic characteristics of interview participants.} Each row corresponds to one of the 13 semi-structured interviews. Participants were drawn from survey respondents who opted in and were not linked to their survey responses; to preserve anonymity, only coarsened, country-level attributes are reported and no institution names, identifiers, or free-text responses are included. ``Institution country'' denotes the country of the participant's primary institutional affiliation; ``Nationality'' is self-reported. Participant identifiers (P1--P13) are arbitrary and do not encode interview order.}
\label{tab:interview_demographics}
\setlength{\tabcolsep}{6pt}
\begin{tabular}{llll}
\toprule
ID & Institution country & Nationality & Career stage  \\
\midrule
P1  & Colombia       & Colombian     & Full Professor                 \\
P2  & Brazil         & Brazilian     & Senior Lect.\ / Assoc.\ Prof.  \\
P3  & Nigeria        & Nigerian      & Lecturer / Asst.\ Prof.        \\
P4  & Nigeria        & Nigerian      & PhD student                    \\
P5  & South Africa   & South African & Full Professor                 \\
P6  & India          & Indian        & PhD student                    \\
P7  & Nepal          & Nepali        & PhD student                    \\
P8  & Singapore      & Chinese       & PhD student                    \\
P9  & Armenia        & Chinese       & Postdoctoral researcher        \\
P10 & T\"urkiye      & Turkish       & Lecturer / Asst.\ Prof.        \\
P11 & Qatar          & Jordanian     & PhD student                    \\
P12 & Italy          & Indian        & PhD student                    \\
P13 & United Kingdom & Iranian       & Postdoctoral researcher        \\
\bottomrule
\end{tabular}
\end{table}

\begin{table}[!htbp] \centering
\caption{Logistic regression estimates of the likelihood that a paper is cited by a paper originated from the Global South, as a function of the cited paper's open access (OA) status. Closed access serves as the reference category. Diamond OA means the cited paper is published in an OA journal with no article processing charges (APC). Gold OA means the cited paper is published in a fully OA journal but with an APC. Green OA means that the cited paper is not published in an OA journal but there is a free version of the paper in an OA repository (for example, a preprint server). Hybrid OA means that the cited paper is not published in an OA journal but is free under an open license. Bronze OA means that the cited paper is free to read but there are not identifiable OA license associted with the journal. Model (2) additionally controls for publication year. Standard errors in parentheses. We calculate the Benjamini-Hochberg adjusted $p$-value of all OA status dummy coefficients across two modeling choices, denoted by $p_{\text{BH}}$. Three stars (\(^{***}\)) indicate \(p_{\text{BH}}<0.001\).}
\label{tab:regression}
\begin{tabular}{lcc}
\hline\hline
 & (1) & (2) \\
\hline
Bronze & 0.184*** & 0.140*** \\
 & (0.001) & (0.001) \\
Diamond & 0.411*** & 0.046*** \\
 & (0.002) & (0.002) \\
Gold & 1.002*** & 0.599*** \\
 & (0.001) & (0.001) \\
Green & 0.366*** & 0.231*** \\
 & (0.001) & (0.001) \\
Hybrid & 0.533*** & 0.260*** \\
 & (0.001) & (0.001) \\
Year &  & 0.024*** \\
 &  & (0.000) \\
\hline
Observations & 86,042,813 & 86,042,813 \\
Pseudo $R^2$ & 0.013 & 0.046 \\
\hline\hline
\end{tabular}
\end{table}

\begin{table}[!htbp]
\centering
\scriptsize
\setlength{\tabcolsep}{3pt}
\caption{{
Negative binomial regressions estimating factors associated with the citation impact of papers (Study 2). 
The number of citations paper \(i\) receives, denoted by \(Y_i\), is modeled using a negative binomial regression model with log link: 
\(Y_i \sim \operatorname{NB2}(\mu_i, 1)\), where \(\mu_i\) is the expected citation count and the overdispersion parameter is fixed at 1. 
Model (1) estimates the baseline negative binomial specification,
\(\log(\mu_i)=\beta_0+\beta_1\mathrm{Percent}_i+\beta_2\mathrm{Year}_{i}+\beta_3\mathrm{OA}_i+\beta_4\log(\mathrm{AuthorCount}_i)+\beta_5\log(\mathrm{JIF}_i)\), 
where \(\mathrm{Percent}_i\) is the main variable of interest, denoting the percentage of references that are paywalled; \(\mathrm{Year}_{i}\) is centered publication year; \(\mathrm{OA}_i\) indicates whether the paper is open access; \(\log(\mathrm{AuthorCount}_i)\) is the log-transformed number of authors; and \(\log(\mathrm{JIF}_i)\) is the log-transformed journal impact factor. 
Model (2) additionally controls for author experience, including the maximum H-index among all authors, the first author's H-index, the maximum academic age among all authors, and the first author's academic age. 
The table reports the coefficient \(b\), the effect size as an incidence rate ratio (IRR), the $p$-value, the Benjamini-Hochberg adjusted \(p\)-value, and the 95\% confidence interval for the incidence rate ratio. Values of \(\mathrm{IRR}>1\) indicate higher expected citation counts, while values of \(\mathrm{IRR}<1\) indicate lower expected citation counts. 
}}
\label{tab:nb_citation_impact}
\begin{tabular}{lcccccccccc}
\hline
  & \multicolumn{5}{c}{(1)} & \multicolumn{5}{c}{(2)} \\
\hline
Variable & \(b\) & IRR & \(p\) & \(p_{\mathrm{BH}}\) & 95\% CI for IRR & \(b\) & IRR & \(p\) & \(p_{\mathrm{BH}}\) & 95\% CI for IRR \\
\hline
Share of paywalled references & 0.326 & 1.386 & $<0.001$ & $<0.001$ & [1.384, 1.388] & 0.467 & 1.595 & $<0.001$ & $<0.001$ & [1.592, 1.598] \\
Publication year & -0.060 & 0.942 & $<0.001$ &  & [0.942, 0.942] & -0.061 & 0.941 & $<0.001$ &  & [0.941, 0.941] \\
Open access & 0.154 & 1.167 & $<0.001$ &  & [1.166, 1.168] & 0.146 & 1.157 & $<0.001$ &  & [1.156, 1.157] \\
Author count (log) & 0.162 & 1.176 & $<0.001$ &  & [1.175, 1.177] & 0.092 & 1.096 & $<0.001$ &  & [1.095, 1.097] \\
Journal impact factor (log) & 1.279 & 3.592 & $<0.001$ &  & [3.590, 3.595] & 1.231 & 3.425 & $<0.001$ &  & [3.423, 3.427] \\
Maximum author H-index &  &  &  &  &  & 0.005 & 1.005 & $<0.001$ &  & [1.005, 1.005] \\
First-author H-index &  &  &  &  &  & 0.010 & 1.010 & $<0.001$ &  & [1.010, 1.011] \\
Maximum author academic age &  &  &  &  &  & 0.001 & 1.001 & $<0.001$ &  & [1.001, 1.001] \\
First-author academic age &  &  &  &  &  & -0.001 & 0.999 & $<0.001$ &  & [0.999, 0.999] \\
\hline
\end{tabular}
\end{table}

\begin{table}
\centering
\scriptsize
\caption{{
Robustness checks of the relationship between citation impact of papers and the percentage of paywalled references using Poisson regressions. The number of citations paper \(i\) receives, denoted by \(Y_i\), is modeled as \(Y_i \sim \operatorname{Poisson}(\mu_i)\). 
Model (1) estimates the baseline Poisson specification with log link, 
\(\log(\mu_i)=\beta_0+\beta_1\mathrm{Percent}_i+\beta_2\mathrm{Year}_{i}+\beta_3\mathrm{OA}_i+\beta_4\log(\mathrm{AuthorCount}_i)+\beta_5\log(\mathrm{JIF}_i)\). 
Model (2) additionally controls for author experience. Model (3) controls for journal fixed effects instead of journal impact factor, thereby comparing papers published within the same journal; the corresponding specification is \(Y_i \sim \operatorname{Poisson}(\mu_i)\), with \(\log(\mu_i)=\alpha_{j(i)}+\beta_1\mathrm{Percent}_i+\beta_2\mathrm{Year}_{i}+\beta_3\mathrm{OA}_i+\beta_4\log(\mathrm{AuthorCount}_i)\), where \(\alpha_{j(i)}\) denotes a fixed effect for the journal \(j\) in which paper \(i\) was published. 
Models (4) and (5) extend the journal fixed-effects specification by adding the author-experience controls, and Model (5) further includes country fixed effects. 
Standard errors are shown in parentheses. Models (3), (4), and (5), where journal fixed-effects are included, use journal-clustered standard errors. We calculate the Benjamini-Hochberg adjusted $p$-value of focal coefficients across all modeling choices, denoted by $p_{\text{BH}}$. Three stars (\(^{***}\)) indicate \(p_{\text{BH}}<0.001\).
}}
\label{tab:poisson_impact}
\begin{tabular}{lccccc}
\toprule
Variable & (1) & (2) & (3) & (4) & (5) \\
\midrule
Percentage of paywalled references & 0.335*** & 0.434*** & 0.179*** & 0.244*** & 0.248*** \\
 & (0.000) & (0.000) & (0.025) & (0.024) & (0.024) \\
Publication year & -0.042*** & -0.044*** & -0.020*** & -0.024*** & -0.022*** \\
 & (0.000) & (0.000) & (0.001) & (0.001) & (0.001) \\
Open Access status & 0.089*** & 0.091*** & 0.208*** & 0.191*** & 0.177*** \\
 & (0.000) & (0.000) & (0.014) & (0.015) & (0.015) \\
Author count (log) & 0.132*** & 0.088*** & 0.200*** & 0.165*** & 0.183*** \\
 & (0.000) & (0.000) & (0.016) & (0.016) & (0.016) \\
Journal impact factor (log) & 1.119*** & 1.069*** &  &  &  \\
 & (0.000) & (0.000) &  &  &  \\
Author controls &  & $\checkmark$ &  & $\checkmark$ & $\checkmark$ \\
Journal fixed effects &  &  & $\checkmark$ & $\checkmark$ & $\checkmark$ \\
Country controls &  &  &  &  & $\checkmark$ \\
\bottomrule
\end{tabular}
\end{table}

\begin{table}[htbp]
\centering
\caption{{Generalized variance inflation factors (GVIF) for the two model specifications in Study 2. The last two columns report the adjusted GVIFs, where $df$ denotes the degrees of freedom of each regressor.}}
\label{tab:gvif}
\begin{tabular}{@{}lccc@{}}
\toprule
Factor & GVIF & $\mathrm{GVIF}^{1/(2\,\mathrm{Df})}$ 
       & $\left(\mathrm{GVIF}^{1/(2\,\mathrm{Df})}\right)^2$ \\
\midrule
\multicolumn{4}{@{}l}{\textit{Panel A: Model controlling for journal impact factor}} \\
\addlinespace[2pt]
Percentage of paywalled references & 1.345 & 1.160 & 1.345 \\
Year of publication & 1.157 & 1.076 & 1.157 \\
Open Access & 1.226 & 1.107 & 1.226 \\
Number of authors & 1.467 & 1.211 & 1.467 \\
Journal impact factor & 1.285 & 1.134 & 1.285 \\
H-index (maximum of all authors) & 1.890 & 1.375 & 1.890 \\
H-index (first author) & 1.589 & 1.260 & 1.589 \\
Academic age (maximum of all authors) & 1.626 & 1.275 & 1.626 \\
Academic age (first author) & 1.723 & 1.313 & 1.723 \\

\midrule
\multicolumn{4}{@{}l}{\textit{Panel B: Model with journal fixed effects}} \\
\addlinespace[2pt]
Percentage of paywalled references & 1.346 & 1.160 & 1.346 \\
Year of publication & 1.327 & 1.152 & 1.327 \\
Open Access & 1.242 & 1.115 & 1.242 \\
Number of authors    & 1.567 & 1.252 & 1.567 \\
Journal fixed effects & 1.092 & 1.045 & 1.092 \\
H-index (maximum of all authors) & 1.883 & 1.372 & 1.883 \\
H-index (first author)  & 1.595 & 1.263 & 1.595 \\
Academic age (maximum of all authors) & 1.671 & 1.293 & 1.671 \\
Academic age (first author) & 1.728 & 1.314 & 1.728 \\
Country fixed effects & 1.385 & 1.005 & 1.009 \\
\bottomrule
\end{tabular}
\end{table}

\begin{table}[!htbp]
\centering
\scriptsize
\setlength{\tabcolsep}{3pt}
\caption{{OLS regressions estimating the relationship between percentage of paywalled references and paper novelty and disruptiveness. Models (1) and (5) are baseline models estimating the relationship of interest while controlling for year of publication, the Open Access status of academic papers, number of authors (log-transformed), the journal impact factor (log-transformed). Model (2) and (6) additionally include author-experience controls, including the maximum H-index among all authors, the first author's H-index, the maximum academic age among all authors, and the first author's academic age.
Model (3) and (7) replace the journal impact factor control in the baseline models with journal fixed effects. Finally, model (4) and (8) are fixed-effect models with author-experience controls. Models (3), (4), (7), and (8), where journal fixed-effects are included, use journal-clustered standard errors. We calculate the Benjamini-Hochberg adjusted $p$-value of focal coefficients across all modeling choices, denoted by $p_{\text{BH}}$. One star (\(^{*}\)) indicates $p_{\text{BH}}<0.05$ but  $p_{\text{BH}}>=0.01$; two stars (\(^{**}\)) indicate $p_{\text{BH}}<0.01$ but  $p_{\text{BH}}>=0.001$; three stars (\(^{***}\)) indicate \(p_{\text{BH}}<0.001\).
}}
\label{tab:novelty_disruption}
\begin{tabular}{lcccccccc}
\hline
 & \multicolumn{4}{c}{Novelty} & \multicolumn{4}{c}{Disruptiveness} \\
\hline
 & (1) & (2) & (3) & (4) & (5) & (6) & (7) & (8) \\
\hline
Share of paywalled references & -1.459*** & -0.937*** & -5.003 & -4.679 & -0.001*** & -0.001*** & -0.001*** & -0.001*** \\
  & (0.155) & (0.157) & (2.615) & (2.620) & (0.000) & (0.000) & (0.000) & (0.000) \\
Publication year & -0.091*** & -0.094*** & -0.164*** & -0.189*** & 0.000*** & 0.000*** & 0.000* & 0.000* \\
  & (0.003) & (0.003) & (0.039) & (0.039) & (0.000) & (0.000) & (0.000) & (0.000) \\
Open access & 2.170*** & 2.264*** & -0.689* & -0.770** & 0.001*** & 0.001*** & 0.000* & 0.000* \\
  & (0.068) & (0.068) & (0.294) & (0.293) & (0.000) & (0.000) & (0.000) & (0.000) \\
Author count (log) & -9.628*** & -8.425*** & -6.680*** & -6.068*** & -0.001*** & -0.001*** & 0.000*** & 0.000*** \\
  & (0.061) & (0.068) & (0.357) & (0.352) & (0.000) & (0.000) & (0.000) & (0.000) \\
Journal impact factor (log) & -1.439*** & -1.461*** &  &  & -0.001*** & -0.001*** &  &  \\
  & (0.058) & (0.060) &  &  & (0.000) & (0.000) &  &  \\
\hline
Author-experience controls &  & Yes &  & Yes &  & Yes &  & Yes \\
Journal fixed effects &  &  & Yes & Yes &  &  & Yes & Yes \\
\hline
\end{tabular}
\end{table}

\begin{table}[!t]
    \centering
    \begin{tabular}{lc}
    \hline
        \textbf{Publisher} & \textbf{Article Count} \\ 
        \hline
        Elsevier & 27677 \\ 
        American Chemical Society & 2482 \\ 
        Wiley & 1104 \\ 
        Springer & 1089 \\ 
        Taylor and Francis & 354 \\ 
        Royal Society of Chemistry & 337 \\ 
        SAGE & 266 \\ 
        Oxford University Press & 230 \\ 
        Routledge & 194 \\ \
        IEEE & 8 \\ \hline
    \end{tabular}
    \caption{Article counts by publisher for the paywalled article audit experiment.}
\end{table}

\begin{table}[!t]
    \centering
    \begin{tabular}{lllc}
    \hline
        \textbf{Sender} & \textbf{Purpose} & \textbf{University} & \textbf{Opened Email Count} \\ \hline
        German & Class project & Louisiana State University & 563 \\ 
        German & Class project & New York University & 521 \\ 
        German & Class project & San Diego State University & 594 \\ 
        German & Journal Article & Louisiana State University & 598 \\ 
        German & Journal Article & New York University & 576 \\ 
        German & Journal Article & San Diego State University & 543 \\ 
        Nigerian & Class project & Louisiana State University & 470 \\ 
        Nigerian & Class project & New York University & 473 \\ 
        Nigerian & Class project & San Diego State University & 494 \\ 
        Nigerian & Journal Article & Louisiana State University & 461 \\ 
        Nigerian & Journal Article & New York University & 456 \\ 
        Nigerian & Journal Article & San Diego State University & 491 \\ 
        Pakistani & Class project & Louisiana State University & 612 \\ 
        Pakistani & Class project & New York University & 618 \\ 
        Pakistani & Class project & San Diego State University & 577 \\ 
        Pakistani & Journal Article & Louisiana State University & 601 \\ 
        Pakistani & Journal Article & New York University & 590 \\ 
        Pakistani & Journal Article & San Diego State University & 580 \\  \hline
    \end{tabular}
    \caption{The number of opened emails per condition for the paywalled article audit experiment.}
\end{table}

\begin{table}[!ht]
    \centering
    \small
    \begin{tabular}{lllccc}
    \hline
        \textbf{Sender} & \textbf{Purpose} & \textbf{University} & \textbf{Positive Reply} & \textbf{Total Replied} & \textbf{Positive \%} \\ \hline
        German & Class Project & Louisiana State University & 118 & 138 & 85.50 \\ 
        German & Class Project & New York University & 120 & 134 & 89.55 \\ 
        German & Class Project & San Diego State University & 106 & 118 & 89.83 \\ 
        German & Journal Article & Louisiana State University & 116 & 132 & 87.87 \\ 
        German & Journal Article & New York University & 124 & 137 & 90.51 \\ 
        German & Journal Article & San Diego State University & 97 & 105 & 92.38 \\ 
        Nigerian & Class Project & Louisiana State University & 74 & 83 & 89.15 \\ 
        Nigerian & Class Project & New York University & 71 & 82 & 86.58 \\ 
        Nigerian & Class Project & San Diego State University & 71 & 83 & 85.54 \\ 
        Nigerian & Journal Article & Louisiana State University & 82 & 92 & 89.13 \\ 
        Nigerian & Journal Article & New York University & 83 & 90 & 92.22 \\ 
        Nigerian & Journal Article & San Diego State University & 85 & 90 & 94.44 \\ 
        Pakistani & Class Project & Louisiana State University & 82 & 96 & 85.41 \\ 
        Pakistani & Class Project & New York University & 91 & 102 & 89.21 \\ 
        Pakistani & Class Project & San Diego State University & 94 & 103 & 91.26 \\ 
        Pakistani & Journal Article & Louisiana State University & 99 & 115 & 86.08 \\ 
        Pakistani & Journal Article & New York University & 79 & 89 & 88.76 \\ 
        Pakistani & Journal Article & San Diego State University & 72 & 84 & 85.71 \\ \hline
    \end{tabular}
        \caption{The number and proportion of email replies which granted access to the paywalled paper for each experimental condition.}
\end{table}

\begin{table}[htbp!]
\centering
\scriptsize
\begin{tabular}{lccc}
\toprule
 & Model I & Model II & Model III \\
 \midrule
\textbf{Sender country (ref. German)} \\
Nigerian & -0.299*** & -0.380*** & -0.428** \\
 & (0.061) & (0.072) & (0.122) \\
Pakistani & -0.367*** & -0.471*** & -0.556*** \\
 & (0.059) & (0.071) & (0.121) \\
 \midrule
\textbf{Sender university (ref. New York University)} \\
Louisiana State University & 0.039 & 0.058 & 0.062 \\
 & (0.060) & (0.070) & (0.115) \\
San Diego State University & -0.113 & -0.059 & -0.184 \\
 & (0.061) & (0.071) & (0.115) \\
\midrule
\textbf{Request type (ref. Journal article)} \\
Class project & 0.003 & 0.016 & 0.015 \\
 & (0.049) & (0.057) & (0.057) \\
\midrule
\textbf{Matching continent (ref. No match)} \\
Pakistani Sender + South Asian University &  & 0.493 & 0.496 \\
 &  & (0.183) & (0.183) \\
German Sender + West Europe University &  & -0.127 & -0.134 \\
 &  & (0.255) & (0.255) \\
Nigerian Sender + African University &  & 1.229** & 1.232** \\
 &  & (0.312) & (0.312) \\
\midrule
\textbf{Article domain (ref. Health sciences)} \\
Life sciences &  & -0.110 & -0.106 \\
 &  & (0.107) & (0.107) \\
Physical sciences &  & -0.337*** & -0.328*** \\
 &  & (0.078) & (0.078) \\
Social sciences &  & 0.106 & 0.103 \\
 &  & (0.102) & (0.102) \\
\midrule
\textbf{Recipient academic characteristics} \\
Academic age &  & -0.007 & -0.006 \\
 &  & (0.012) & (0.012) \\
$h$-index &  & -0.008*** & -0.008*** \\
 &  & (0.002) & (0.002) \\
\midrule
\textbf{Sender country × Sender university} \\
Nigerian × LSU &  &  & -0.052 \\
 &  &  & (0.173) \\
Nigerian × SDSU &  &  & 0.192 \\
 &  &  & (0.173) \\
Pakistani × LSU &  &  & 0.042 \\
 &  &  & (0.168) \\
Pakistani × SDSU &  &  & 0.210 \\
 &  &  & (0.169) \\
\midrule
Intercept & -1.724*** & -1.293*** & -1.259*** \\
 & (0.058) & (0.157) & (0.164) \\
 \midrule
Observations & 15133 & 11915 & 11915 \\
 \midrule
Pseudo $R^2$ & 0.005 & 0.02 & 0.02 \\
\bottomrule
\end{tabular}
\caption{Logistic regression results on the likelihood of a paywalled article request receiving a reply with different model specifications after Benjamini-Hochberg corrections.}
\end{table}

\begin{table}[htbp!]
\centering
\scriptsize
\begin{tabular}{lccc}
\toprule
 & Model I & Model II & Model III \\
 \midrule
\textbf{Sender country (ref. German)} \\
Nigerian & -0.288*** & -0.362*** & -0.407* \\
 & (0.064) & (0.075) & (0.129) \\
Pakistani & -0.378*** & -0.470*** & -0.530*** \\
 & (0.062) & (0.075) & (0.127) \\
 \midrule
\textbf{Sender university (ref. New York University)} \\
Louisiana State University & 0.006 & 0.057 & 0.057 \\
 & (0.063) & (0.074) & (0.121) \\
San Diego State University & -0.106 & -0.025 & -0.117 \\
 & (0.064) & (0.074) & (0.120) \\
\midrule
\textbf{Request type (ref. Journal article)} \\
Class project & -0.016 & -0.007 & -0.008 \\
 & (0.052) & (0.060) & (0.060) \\
\midrule
\textbf{Matching continent (ref. No match)} \\
Pakistani Sender + South Asian University &  & 0.491 & 0.493 \\
 &  & (0.191) & (0.191) \\
German Sender + West Europe University &  & -0.349 & -0.354 \\
 &  & (0.289) & (0.289) \\
Nigerian Sender + African University &  & 1.142** & 1.142** \\
 &  & (0.327) & (0.327) \\
\midrule
\textbf{Article domain (ref. Health sciences)} \\
Life sciences &  & -0.023 & -0.020 \\
 &  & (0.114) & (0.114) \\
Physical sciences &  & -0.249* & -0.241 \\
 &  & (0.084) & (0.084) \\
Social sciences &  & 0.210 & 0.208 \\
 &  & (0.108) & (0.108) \\
\midrule
\textbf{Recipient academic characteristics} \\
Academic age &  & -0.008 & -0.008 \\
 &  & (0.012) & (0.012) \\
$h$-index &  & -0.010*** & -0.010*** \\
 &  & (0.002) & (0.002) \\
\midrule
\textbf{Sender country × Sender university} \\
Nigerian × LSU &  &  & -0.013 \\
 &  &  & (0.182) \\
Nigerian × SDSU &  &  & 0.142 \\
 &  &  & (0.181) \\
Pakistani × LSU &  &  & 0.018 \\
 &  &  & (0.178) \\
Pakistani × SDSU &  &  & 0.155 \\
 &  &  & (0.178) \\
\midrule
Intercept & -1.838*** & -1.443*** & -1.416*** \\
 & (0.061) & (0.164) & (0.171) \\
 \midrule
Observations & 15133 & 11915 & 11915 \\
 \midrule
Pseudo $R^2$ & 0.005 & 0.02 & 0.02 \\
\bottomrule
\end{tabular}
\caption{Logistic regression results on the likelihood of a paywalled article request receiving a positive reply with different model specifications after Benjamini-Hochberg corrections.}
\end{table}

\begin{table}[t]
\begin{center}
\begin{tabular}{llll}
\toprule
 & GVIF & $\text{GVIF}^{(1/2Df)}$ & $\text{GVIF}^{(1/2Df)^2}$ \\
\midrule
Upon request & 1.086 & 1.042 & 1.086 \\
Global South & 1.039 & 1.019 & 1.039 \\
Open Access & 1.926 & 1.388 & 1.926 \\
Year & 2.063 & 1.436 & 2.063 \\
Paper journal fixed effects & 2.713 & 1.010 & 1.021 \\
Agricultural and food sciences & 1.142 & 1.069 & 1.142 \\
Biology & 1.367 & 1.169 & 1.367 \\
Chemistry & 1.073 & 1.036 & 1.073 \\
Computer science & 1.227 & 1.108 & 1.227 \\
Economics or business & 1.102 & 1.050 & 1.102 \\
Engineering & 1.152 & 1.073 & 1.152 \\
Environmental science & 1.356 & 1.164 & 1.356 \\
Geology or geography & 1.104 & 1.051 & 1.104 \\
Humanities & 1.029 & 1.014 & 1.029 \\
Materials science & 1.160 & 1.077 & 1.160 \\
Mathematics & 1.054 & 1.027 & 1.054 \\
Medicine & 1.774 & 1.332 & 1.774 \\
Physics & 1.344 & 1.159 & 1.344 \\
Psychology & 1.456 & 1.207 & 1.456 \\
Sociology, education, or political science & 1.194 & 1.093 & 1.194 \\
\bottomrule
\end{tabular}
\end{center}
\caption{{\textbf{Generalized variance inflation factors (GVIF) for regression models in Study 4.} The last two columns report the adjusted GVIFs, where $df$ denotes the degrees of freedom of each regressor. Since a paper may belong to more than one field of study, each field of study is coded using a distinct dummy variable, which is set to 1 for a paper if the paper belongs to that field of study.}}
\end{table}

\begin{table}[!htbp] \centering
\scriptsize
\caption{Logit regression models estimating the likelihood that a citation to a paper is a data citation. We consider an array of models that controls for an increasing number of fixed-effect controls. The results of Model (4) and Model (8) are reported in Table 1. Standard errors are in parentheses. * $p<0.05$, ** $p<0.01$, *** $p<0.001$. }
\label{tab:regression}
\begin{tabular}{lcccccccc}
\hline\hline
 & (1) & (2) & (3) & (4) & (5) & (6) & (7) & (8) \\
\hline
Upon request & -0.778*** & -0.794*** & -0.766*** & -0.622*** & -0.736*** & -0.749*** & -0.721*** & -0.583*** \\
 & (0.014) & (0.014) & (0.014) & (0.015) & (0.016) & (0.016) & (0.016) & (0.016) \\
Open Access & 0.376*** & 0.046** & 0.148*** & 0.087*** & 0.326*** & -0.008 & 0.101*** & 0.041* \\
 & (0.009) & (0.017) & (0.018) & (0.018) & (0.010) & (0.018) & (0.018) & (0.018) \\
Global South &  &  &  &  & -0.437*** & -0.426*** & -0.366*** & -0.422*** \\
 &  &  &  &  & (0.017) & (0.017) & (0.018) & (0.019) \\
Upon request $\times$ Global South &  &  &  &  & -0.146*** & -0.159*** & -0.174*** & -0.128*** \\
 &  &  &  &  & (0.035) & (0.035) & (0.035) & (0.035) \\
Open Access $\times$ Global South &  &  &  &  & 0.104*** & 0.123*** & 0.089*** & 0.133*** \\
 &  &  &  &  & (0.023) & (0.024) & (0.024) & (0.024) \\
Year fixed effects & no & yes & yes & yes & no & yes & yes & yes \\
Journal fixed effects & no & no & yes & yes & no & no & yes & yes \\
Field of study fixed effects & no & no & no & yes & no & no & no & yes \\
\hline
Observations & 1,632,370 & 1,597,069 & 1,597,069 & 1,534,036 & 1,632,370 & 1,597,069 & 1,597,069 & 1,534,036 \\
Pseudo $R^2$ & 0.012 & 0.016 & 0.028 & 0.044 & 0.015 & 0.019 & 0.030 & 0.046 \\
\hline\hline
\end{tabular}
\end{table}

\begin{table}[!ht]
    \centering
    \begin{tabular}{l|c}
    \hline
        \textbf{Venue} & \textbf{Number of Articles} \\ \hline
        Nature Communications & 9595 \\ 
        Scientific Reports & 5767 \\ 
        Nature & 579 \\ 
        Nature Human Behaviour & 59 \\ \hline
    \end{tabular}
    \caption{Article venue counts for data access audit experiment.}
\end{table}

\begin{table}[!ht]
\centering
\scriptsize
\begin{tabular}{llll r r r}
\toprule
Sender country & Request purpose & University ranking & University country & Opened & Received & Open rate (\%) \\
\midrule
England & Final course project & 155 & England & 413 & 810 & 50.9 \\
England & Final course project & 150 & South Africa & 382 & 815 & 46.8 \\
England & Final course project & 1001--1200 & England & 373 & 787 & 47.3 \\
England & Final course project & 1001--1200 & South Africa & 380 & 807 & 47.0 \\
England & Journal article & 155 & England & 429 & 816 & 52.5 \\
England & Journal article & 150 & South Africa & 397 & 803 & 49.4 \\
England & Journal article & 1001--1200 & England & 393 & 797 & 49.3 \\
England & Journal article & 1001--1200 & South Africa & 410 & 806 & 50.8 \\
South Africa & Final course project & 155 & England & 431 & 816 & 52.8 \\
South Africa & Final course project & 150 & South Africa & 425 & 803 & 52.9 \\
South Africa & Final course project & 1001--1200 & England & 391 & 767 & 50.9 \\
South Africa & Final course project & 1001--1200 & South Africa & 450 & 824 & 54.6 \\
South Africa & Journal article & 155 & England & 431 & 801 & 53.8 \\
South Africa & Journal article & 150 & South Africa & 450 & 804 & 55.9 \\
South Africa & Journal article & 1001--1200 & England & 424 & 799 & 53.0 \\
South Africa & Journal article & 1001--1200 & South Africa & 408 & 784 & 52.0 \\
\bottomrule
\end{tabular}
\caption{The number and percentage of opened emails out of all emails sent for each experimental condition in the data request audit experiment.}
\end{table}

\begin{table}[!ht]
\centering
\scriptsize
\begin{tabular}{llll r r r}
\toprule
Sender country & Request purpose & University ranking & University country & Replied & Received & Reply rate (\%) \\
\midrule
England & Final course project & High & England & 132 & 810 & 16.3 \\
England & Final course project & High & South Africa & 108 & 815 & 13.3 \\
England & Final course project & Low & England & 128 & 787 & 16.3 \\
England & Final course project & Low & South Africa & 88 & 807 & 10.9 \\
England & Journal article & High & England & 135 & 816 & 16.5 \\
England & Journal article & High & South Africa & 101 & 803 & 12.6 \\
England & Journal article & Low & England & 119 & 797 & 14.9 \\
England & Journal article & Low & South Africa & 105 & 806 & 13.0 \\
South Africa & Final course project & High & England & 113 & 816 & 13.8 \\
South Africa & Final course project & High & South Africa & 95 & 803 & 11.8 \\
South Africa & Final course project & Low & England & 66 & 767 & 8.6 \\
South Africa & Final course project & Low & South Africa & 80 & 824 & 9.7 \\
South Africa & Journal article & High & England & 100 & 801 & 12.5 \\
South Africa & Journal article & High & South Africa & 99 & 804 & 12.3 \\
South Africa & Journal article & Low & England & 87 & 799 & 10.9 \\
South Africa & Journal article & Low & South Africa & 76 & 784 & 9.7 \\
\bottomrule
\end{tabular}
\caption{The number and percentage of emails that were replied to out of all emails sent for each experimental condition in the data request audit experiment.}
\end{table}

\begin{table}[!ht]
\centering
\scriptsize
\begin{tabular}{llcc r r r r r}
\toprule
Sender & Purpose & Rank & Univ. & Replied & Pos. replies & Cond. replies & Neg. replies &  Pos. (\%) \\
\midrule
England & Course & High & UK & 132 & 128 & 2 & 2 & 97.0 \\
England & Course & High & SA & 108 & 104 & 2 & 2&96.3 \\
England & Course & Low & UK & 128 & 121 & 4 & 3&94.5 \\
England & Course & Low & SA & 88 & 84 & 2 & 2&95.5 \\
England & Article & High & UK & 135 & 129 & 3 & 3&95.6 \\
England & Article & Low & UK & 119 & 113 & 3 & 3&95.0 \\
England & Article & Low & SA & 105 & 100 & 2 & 3&95.2 \\
South Afr. & Course & High & UK & 113 & 109 & 1 & 1&96.5 \\
South Afr. & Course & High & SA & 95 & 88 & 2 & 5&92.6 \\
South Afr. & Course & Low & UK & 66 & 62 & 2 & 2&93.9 \\
South Afr. & Course & Low & SA & 80 & 73 & 2 & 5&91.3 \\
South Afr. & Article & High & UK & 100 & 92 & 1 & 7&92.0 \\
South Afr. & Article & High & SA & 99 & 91 & 5 & 3&91.9 \\
South Afr. & Article & Low & SA & 76 & 72 & 3 & 1&94.7 \\
\bottomrule
\end{tabular}
\caption{The number received emails that were positive, conditional and negative in nature for each experimental condition.}
\end{table}

\clearpage

\begin{table}[!htbp]\centering
\centering
\label{tab:reply_rates_datasets}
{
\def\sym#1{\ifmmode^{#1}\else\(^{#1}\)\fi}
\footnotesize
\begin{tabular}{l*{2}{c}}
\toprule
                    &\multicolumn{1}{c}{(1)}&\multicolumn{1}{c}{(2)}\\
                    &\multicolumn{1}{c}{Model  I}&\multicolumn{1}{c}{Model II}\\
\midrule
\textbf{Sender Country}             &                     &                     \\
(Ref = England)            &                     &                     \\
South Africa       &      -0.235\sym{***}&      -0.250\sym{***}\\
                    &     (0.051)         &     (0.061)         \\
\textbf{Request Purpose}             &                     &                     \\
(Ref = Course Project)            &                     &                     \\
Journal Article     &       0.017         &      -0.0448         \\
                    &     (0.051)         &     (0.060)         \\
                    
\textbf{University Ranking}             &                     &                     \\
(Ref = 150-155)            &                     &                     \\
1001-1200                 &      -0.180\sym{**} &      -0.173\sym{*} \\
                    &     (0.051)         &     (0.060)         \\
                    
\textbf{University Country}             &                     &                     \\
(Ref = England)            &                     &                     \\
South Africa        &      -0.167\sym{**} &      -0.225\sym{**} \\
                    &     (0.051)         &     (0.061)         \\
                    
\textbf{Recipient Characteristics}             &                     &                     \\
\addlinespace
H-Index             &                     &      -0.003\sym{*}  \\
                    &                     &     (0.001)         \\
\addlinespace
Academic Age        &                     &       0.014\sym{***}\\
                    &                     &     (0.003)         \\
                    
\textbf{Article Domain}             &                     &                     \\
(Ref = Health Sciences)            &                     &                     \\
Life Sciences       &                     &       0.265\sym{*}  \\
                    &                     &     (0.083)         \\
\addlinespace
Physical Sciences   &                     &       0.146  \\
                    &                     &     (0.088)         \\
\addlinespace
Social Sciences     &                     &       0.299         \\
                    &                     &     (0.227)         \\
\midrule
\addlinespace
Intercept            &      -1.592\sym{***}&      -2.029\sym{***}\\
                    &     (0.054)         &     (0.125)         \\
\midrule
Observations        &       13458         &        9971         \\
\midrule
Pseudo R2           &       0.004         &        0.009             \\
\bottomrule
\end{tabular}
}
\caption{Baseline logistic models for reply rate based on sender and recipient characteristics after Benjamini-Hochberg corrections.}
\end{table}

\begin{table}[htbp!]
\centering
\scriptsize
\begin{tabular}{lcccc}
\toprule
 & Model III & Model IV & Model V & Model VI \\
\midrule
\textbf{Sender Country (ref. England)} \\
South Africa & -0.149 & -0.379*** & -0.251*** & -0.279 \\
 & (0.084) & (0.084) & (0.061) & (0.102) \\
\midrule
\textbf{University Country (ref. England)} \\
University country: South Africa & -0.227** & -0.342*** & -0.158 & -0.277 \\
 & (0.061) & (0.080) & (0.083) & (0.099) \\
\midrule
\textbf{University Ranking (ref. 150-155)} \\
1001-1200 & -0.084 & -0.173 & -0.107 & -0.024 \\
 & (0.079) & (0.061) & (0.082) & (0.095) \\
 \midrule
\textbf{Request Purpose (ref. Journal article)} \\
Final course project & 0.046 & 0.045 & 0.046 & 0.046 \\
 & (0.060) & (0.060) & (0.060) & (0.060) \\
\midrule
\textbf{Article Journal (ref. Nature)} \\
Nature Communications & -0.099 & -0.093 & -0.103 & -0.097 \\
 & (0.166) & (0.166) & (0.166) & (0.166) \\
Scientific Reports & -0.028 & -0.021 & -0.033 & -0.022 \\
 & (0.171) & (0.171) & (0.171) & (0.171) \\
 \midrule
\textbf{Article Domain (ref. Health Sciences)} \\
Life Sciences & 0.284* & 0.287* & 0.281* & 0.289* \\
 & (0.085) & (0.085) & (0.085) & (0.085) \\
Physical Sciences & 0.167 & 0.167 & 0.166 & 0.170 \\
 & (0.090) & (0.090) & (0.090) & (0.090) \\
Social Sciences & 0.294 & 0.292 & 0.294 & 0.305 \\
 & (0.228) & (0.228) & (0.228) & (0.229) \\
\midrule
\textbf{Recipient academic characteristics} \\
Academic age & 0.014*** & 0.014*** & 0.014*** & 0.014*** \\
 & (0.003) & (0.003) & (0.003) & (0.003) \\
$h$-index & -0.003* & -0.003* & -0.003* & -0.003 \\
 & (0.001) & (0.001) & (0.001) & (0.001) \\
 \midrule
\textbf{Interaction effects} \\
Sender country: South Africa × University ranking: Low & -0.217 &  &  & -0.210 \\
 & (0.123) &  &  & (0.123) \\
Sender country: South Africa × University country: South Africa &  & 0.280 &  & 0.274 \\
 &  & (0.123) &  & (0.123) \\
 University country: South Africa × University ranking: 1001-1200 &  &  & -0.145 & -0.136 \\
 &  &  & (0.122) & (0.122) \\
 \midrule
Intercept & -2.074*** & -1.988*** & -2.059*** & -2.061*** \\
 & (0.208) & (0.207) & (0.208) & (0.210) \\
\midrule
Observations & 9941 & 9941 & 9941 & 9941 \\
\midrule
Pseudo $R^2$ & 0.01 & 0.01 & 0.01 & 0.01 \\
\bottomrule
\end{tabular}

\caption{Logistic models estimating heterogeneous effects for reply rate based on sender and recipient characteristics.}
\end{table}

\begin{table}[htbp!]
\centering
\caption{\textbf{Codebook for classification of email responses.} In Study~3, response coding did not require subjective judgment: replies were classified as positive if the requested paper was included as an attachment or download link, and negative otherwise. In Study~5, two independent annotators classified each non-automated reply into one of three categories using the decision rules below. Disagreements were resolved through discussion. Inter-rater reliability was Cohen's $\kappa = 0.964$.}
\label{tab:codebook}
\scriptsize
\begin{tabular}{p{2cm} p{3.5cm} p{4cm} p{4cm}}
\hline
\textbf{Category} & \textbf{Definition} & \textbf{Decision rules} & \textbf{Examples} \\
\hline
Positive & The recipient provides the requested dataset, provides information on how to access the data, or forwards the request to a colleague with access. & 
Classify as positive if: (1)~data files are attached or a working download link is provided; (2)~the recipient directs the sender to a public repository where the data can be accessed; or (3)~the recipient forwards or copies the email to another researcher who holds the data, indicating willingness to facilitate access. &
``Please find the data attached''; ``The data is available at [link]''; ``I have forwarded your request to the first author who collected the data.'' \\
\hline
Conditional & The recipient does not refuse the request but imposes conditions or requests additional information before sharing. &
Classify as conditional if the reply does not include the data and instead: (1)~asks for the sender's institutional credentials or proof of affiliation; (2)~asks for the sender's advisor or PI to submit the request; (3)~requests further detail about the intended use of the data; or (4)~expresses willingness to share but defers to a later date without providing a concrete timeline. If a reply both provides data \textit{and} asks a clarifying question, classify as positive. &
``Could you have your supervisor contact me?''; ``Can you tell me more about what you plan to do with the data?''; ``Please send me your institutional credentials.'' \\
\hline
Negative & The recipient declines the request outright or imposes conditions that effectively preclude access. &
Classify as negative if the reply: (1)~explicitly refuses to share the data; (2)~states that the data is lost, corrupted, or no longer available; (3)~conditions access on co-authorship, financial payment, or a formal collaboration agreement; or (4)~states that the data is being used in ongoing work and cannot be shared. If the recipient expresses regret but does not offer any path to access, classify as negative. &
``Unfortunately the data is no longer available''; ``We can share the data in the context of a collaboration''; ``I can provide the raw data for a fee of 1000 US dollars.'' \\
\hline
\end{tabular}
\end{table}

\begin{table}[h!]
\centering
\footnotesize
\caption{\textbf{Source data for Figure~1B--D.} Counts and within-region
percentages for each response option, for Global North (GN; $n=216$) and Global South
(GS; $n=160$) respondents. For the author-email items (Figure~1D), percentages are
computed among respondents who reported having attempted such requests; respondents
selecting ``I have never emailed an author for this purpose'' or ``I have never requested
data from an author'' are shown for completeness but excluded from the Figure~1D
denominators, as are respondents who did not answer the item (papers item: 28 GN, 9 GS;
datasets item: 77 GN, 45 GS). Statistical comparisons for these distributions are
reported in the main text.}
\begin{tabular}{lcc}
\toprule
 & Global North, $n$ (\%) & Global South, $n$ (\%) \\
\midrule
\multicolumn{3}{l}{\textit{Figure~1B: Encountered a paper you could not access (past 12 months)}}\\
Never                              & 28 (13.0) & 9 (5.6)   \\
Rarely (1--2 times)                & 69 (31.9) & 32 (20.0) \\
Sometimes (3--5 times)             & 61 (28.2) & 42 (26.2) \\
Often (6--10 times)                & 26 (12.0) & 35 (21.9) \\
Very often (more than 10 times)    & 32 (14.8) & 42 (26.2) \\
\midrule
\multicolumn{3}{l}{\textit{Figure~1C: Needed to access a dataset associated with a published paper (past 12 months)}}\\
Never                              & 77 (35.6) & 45 (28.1) \\
Rarely (1--2 times)                & 64 (29.6) & 30 (18.8) \\
Sometimes (3--5 times)             & 44 (20.4) & 37 (23.1) \\
Often (6--10 times)                & 15 (6.9)  & 27 (16.9) \\
Very often (more than 10 times)    & 16 (7.4)  & 21 (13.1) \\
\midrule
\multicolumn{3}{l}{\textit{Figure~1D (papers): Response rate when emailing authors for paywalled papers (triers: GN $n=116$, GS $n=101$)}}\\
Almost never (less than 10\% of the time) & 17 (14.7) & 28 (27.7) \\
Rarely (10--30\% of the time)      & 22 (19.0) & 23 (22.8) \\
Sometimes (30--50\% of the time)   & 26 (22.4) & 24 (23.8) \\
Often (50--70\% of the time)       & 16 (13.8) & 12 (11.9) \\
Usually (70--90\% of the time)     & 20 (17.2) & 8 (7.9)   \\
Almost always (more than 90\% of the time) & 15 (12.9) & 6 (5.9) \\
\quad Above 50\% (combined)        & 51 (44.0) & 26 (25.7) \\
\quad Never emailed (excluded)     & 72        & 50        \\
\midrule
\multicolumn{3}{l}{\textit{Figure~1D (datasets): Receipt rate when requesting ``available upon request'' data (triers: GN $n=90$, GS $n=80$)}}\\
Almost never (less than 10\% of the time) & 22 (24.4) & 25 (31.2) \\
Rarely (10--30\% of the time)      & 17 (18.9) & 24 (30.0) \\
Sometimes (30--50\% of the time)   & 20 (22.2) & 19 (23.8) \\
Often (50--70\% of the time)       & 8 (8.9)   & 3 (3.8)   \\
Usually (70--90\% of the time)     & 16 (17.8) & 3 (3.8)   \\
Almost always (more than 90\% of the time) & 7 (7.8) & 6 (7.5)  \\
\quad Above 50\% (combined)        & 31 (34.4) & 12 (15.0) \\
\quad Never requested (excluded)   & 49        & 35        \\
\bottomrule
\end{tabular}
\end{table}

\clearpage
\section*{Supplementary Figures}
\label{figures}

\begin{figure}[!ht]
    \centering
    \includegraphics{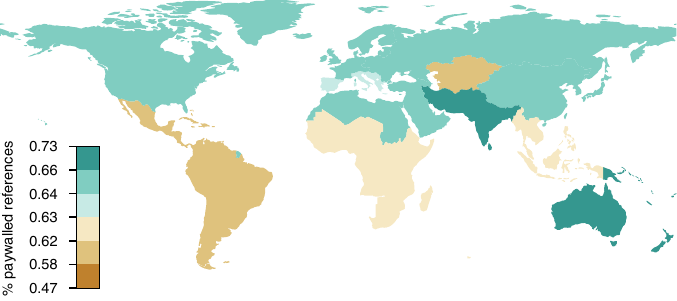}
    \caption{\textbf{Global disparity of citations to Open Access papers.} Same as Figure 2A but the percentage of references towards paywalled papers is calculated for each region (instead of country).
    }
    \label{fig:OA_regression_field}
\end{figure}

\begin{figure}[!ht]
    \centering
    \includegraphics{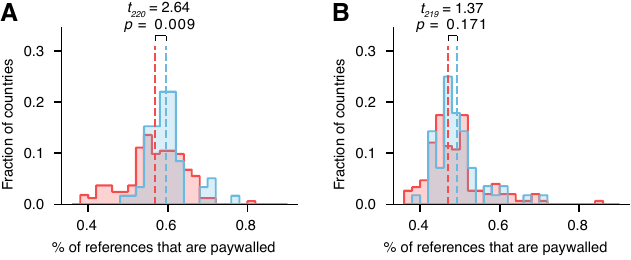}
    \caption{{\textbf{The distribution of the percentage of references towards paywalled papers.} Same as Figure 2B but the analysis is restricted to references to papers in top journals, i.e., Q1 journals (\textbf{A}) or \textit{Nature}, \textit{Science}, and \textit{Proceedings of the National Academy of Sciences} (\textbf{B}). Journal quartile ranking data is obtained from Scimago Journal \& Country Rank (\url{https://www.scimagojr.com/}). The dashed line denotes the mean value for global south (red) and global north countries (blue).}}
    \label{fig:OA_regression_field}
\end{figure}

\begin{figure}[!ht]
    \centering
    \includegraphics{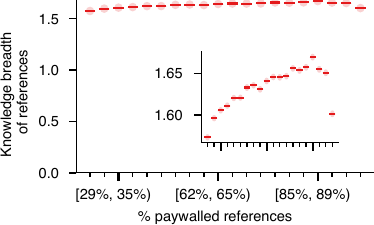}
    \caption{{\textbf{Relationship between the percentage of references that are paywalled and the knowledge breadth of the references.} Same as Figure~2D but with an alternative model specification using a 50-dimensional embedding space, instead of 100.
    }}
    \label{fig:embedding_50dim}
\end{figure}

\begin{figure}[!ht]
    \centering
    \includegraphics{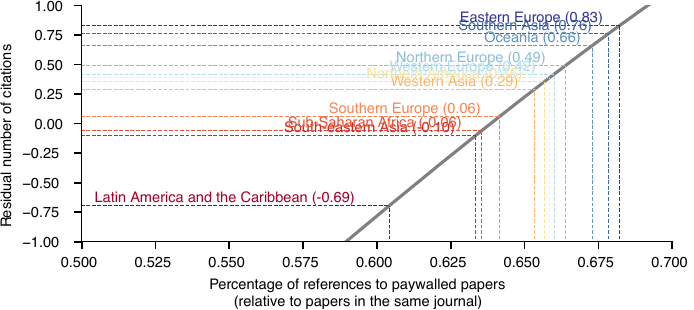}
    \caption{{\textbf{The relationship between the percentage of references to paywalled papers and the number of citations that a paper receives.} Same as Figure~2D but with all regions marked along the regression line.}}
\end{figure}

\begin{figure}[ht]
    \centering
    \includegraphics[width=\linewidth]{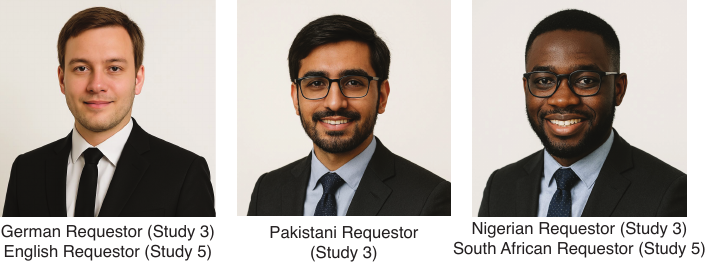}
    \caption{Photos used for fictional students in audit experiments.}
    \label{fig:sender_pictures}
\end{figure}

\begin{figure}
    \centering
    \includegraphics[width=\linewidth]{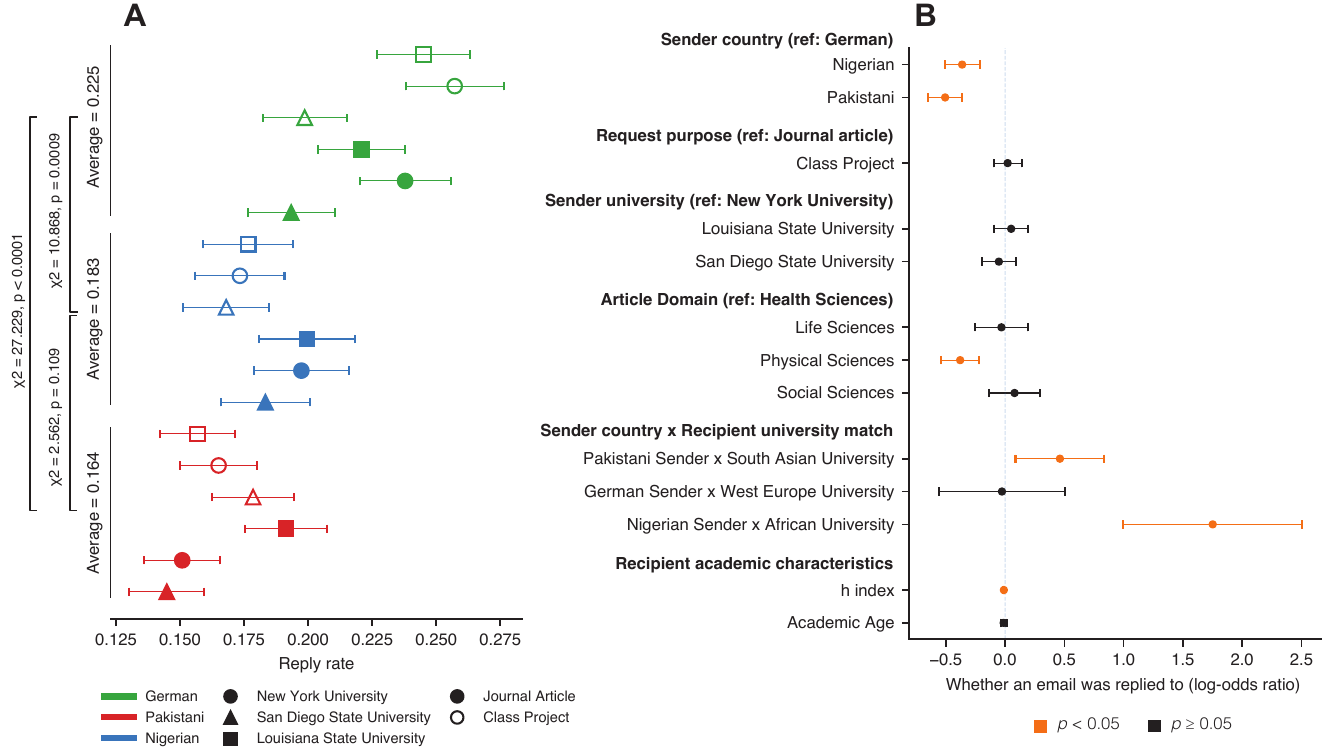}
    \caption{\small \textbf{Summary of experimental results for opened emails in paywalled paper audit.} (\textbf{A}) Reply rates for paywalled paper requests for each experimental condition. (\textbf{B}) Logistic regression estimated log-odds coefficients of independent variables predicting whether a given opened email was replied to.}
    \label{fig:OA_experiment_opened_emails}
\end{figure}

\begin{figure}[!ht]
    \centering
    \includegraphics[width=\linewidth]{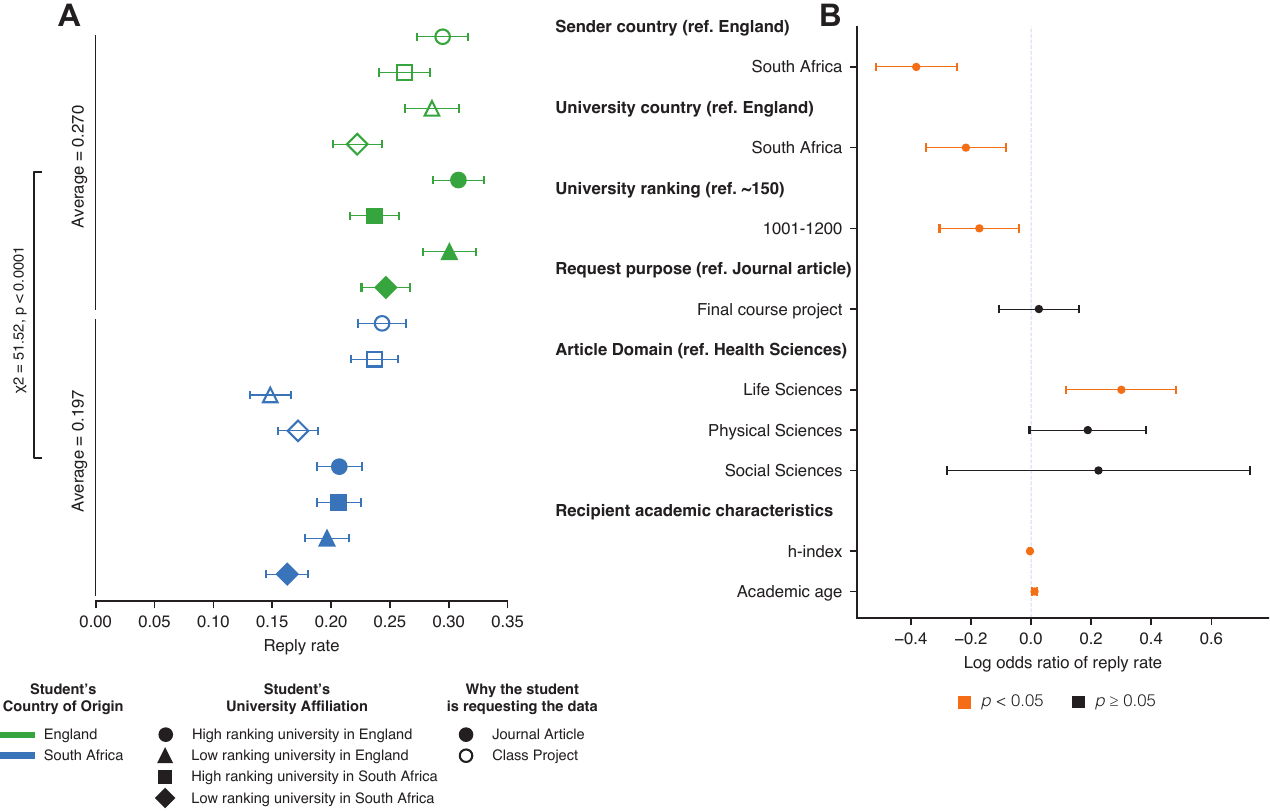}
    \caption{\small \textbf{Summary of experimental results for opened emails in dataset request audit. (\textbf{A}) Reply rates for dataset requests for each experimental condition. (\textbf{B}) Logistic regression estimated log-odds coefficients of independent variables predicting whether a given opened email was replied to.}}
    \label{fig:dataset_opened}
\end{figure}

\begin{figure}[!ht]
    \centering
    \includegraphics[width=\linewidth]{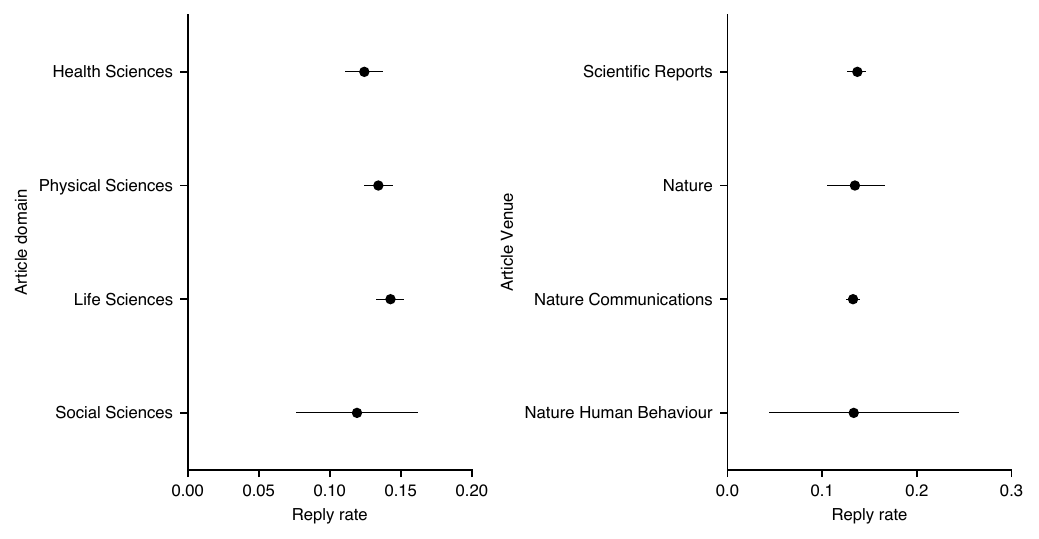}
    \caption{\textbf{Response rates to data request emails. The percentage of emails which received a reply, split by the article's domain (A) and journal (B).}}
    \label{fig:domain_field_journal}
\end{figure}

\label{bibliography}

\clearpage
\section*{COREQ (Consolidated Criteria for Reporting Qualitative Research)
  Checklist}

\noindent\textbf{Study component:} Semi-structured interview component of
Study~1 (``Global disparities in access to paywalled papers and data'').\\

\bigskip

\noindent\rule{\linewidth}{0.6pt}

\noindent\textbf{\large Domain 1: Research Team and Reflexivity}

\noindent\rule{\linewidth}{0.4pt}

\medskip

\noindent\textit{\textbf{Personal characteristics}}

\medskip

\noindent\textbf{Item 1 — Interviewer / facilitator} \hfill
\textit{Manuscript location: Author Contributions; Methods, Study~1}

\smallskip
\noindent All three pre-interviews, and four of semi-structured interviews were conducted by Aaron
R.\ Kaufman (A.K.), the fourth listed author. Seven interviews were conducted by W.A., the fifth listed author.  A.K.\ and W.A. jointly conducted one interview. No additional facilitators or co-interviewers were present during any session.

\bigskip

\noindent\textbf{Item 2 — Credentials} \hfill
\textit{Manuscript location: Title page; Methods, Study~1}

\smallskip
\noindent A.K.\ holds a PhD in Political Science (Harvard University, 2016).
He is trained in both quantitative and qualitative social science research
methods. His doctoral training included coursework and mentored practice in
qualitative interviewing, inductive thematic analysis, and interpretive
social inquiry. He also received additional training in qualitative coding
methodology in preparation for this study, as noted in the manuscript.

\medskip
\noindent W.A.\ holds a PhD in Sociology (University of California, Los
Angeles, UCLA). He is trained in qualitative sociological research methods,
including in-depth interviewing, archival analysis, and interpretive
approaches to political and historical inquiry. His published research in
peer-reviewed sociology venues (including \textit{American Journal of
Sociology}) demonstrates sustained engagement with qualitative data
collection and analysis.

\bigskip

\noindent\textbf{Item 3 — Occupation} \hfill
\textit{Manuscript location: Title page}

\smallskip
\noindent At the time of data collection (2024--2025), A.K.\ was Associate
Professor of Political Science in the Social Science Division of New York
University Abu Dhabi (NYUAD), UAE. W.A.\ was Assistant Professor of
Sociology, also in the Social Science Division of NYUAD.

\bigskip

\noindent\textbf{Item 4 — Gender} \hfill
\textit{Manuscript location: COREQ statement (below); interviewing team named in Author Contributions}

\smallskip
\noindent Both interviewers (A.K.\ and W.A.) are male.

\bigskip

\noindent\textbf{Item 5 — Experience and training} \hfill
\textit{Manuscript location: Methods, Study~1}

\smallskip
\noindent A.K.\ has over a decade of experience in social science research
combining quantitative and qualitative methods, including prior experience
conducting semi-structured interviews in studies of political behavior and
representation. In preparation for this study, he completed additional
training in qualitative coding methodology and reviewed the theoretical
rationale for process-oriented qualitative interviewing with the wider
research team.

\medskip
\noindent W.A.\ is a qualitative sociologist whose research on political
mobilization, human rights discourse, and Middle Eastern politics draws
extensively on in-depth interview and archival methods. He brings
disciplinary expertise in political, historical, and cultural sociology to
the interview design and execution. W.A.\ led the majority of the
semi-structured interviews (seven of thirteen) and jointly conducted one
further interview with A.K., providing the study's primary qualitative
methodological grounding.

\medskip
\noindent Prior to beginning data collection, both interviewers participated
in two internal pilot run-throughs of the interview protocol with members of
the research team to align on question sequencing, probe strategies, and
inter-interviewer consistency. These pilots were not included in the
analytical sample.

\bigskip

\noindent\textit{\textbf{Relationship with participants}}

\medskip

\noindent\textbf{Item 6 — Relationship established prior to study} \hfill
\textit{Manuscript location: Methods, Study~1: Survey; Methods, Study~1:
  Interviews}

\smallskip
\noindent No personal or professional relationship existed between either
interviewer (A.K.\ or W.A.) and any interview participant prior to study
commencement. All participants were recruited from the pool of survey
respondents who had independently elected to provide a contact email for a
potential follow-up interview (see Item~10). First contact was made by the
research team via a standardized recruitment email; neither A.K.\ nor W.A.\
had any prior acquaintance with any participant.

\bigskip

\noindent\textbf{Item 7 — Participant knowledge of the interviewer} \hfill
\textit{Manuscript location: Methods, Study~1; Consent Document}

\smallskip
\noindent Via the recruitment email and the verbal consent process
administered at the start of each interview, participants were informed of
the name, title, and institutional affiliation of the interviewer conducting
their session (A.K.\ or W.A., as applicable) and that the interviewer was a
member of the research team at NYUAD. They were told the study concerned
structural and resource-related barriers researchers face in accessing
scientific publications and data in their institutional and national
contexts, and that the purpose of the interview was to gain deeper
understanding of how researchers in the Global South navigate those barriers
in practice. Participants were not informed of specific quantitative findings
from the broader study (Studies 2--5) prior to being interviewed, in order
to minimize anchoring effects. They were also informed that their
participation was voluntary, that their responses would be de-identified and
used solely for academic research, and that the study had received
institutional ethics approval (NYUAD IRB, protocol HRPP-2024-45).

\bigskip

\noindent\textbf{Item 8 — Interviewer characteristics} \hfill
\textit{Manuscript location: Methods, Study~1}

\smallskip
\noindent A.K.\ is a quantitative political scientist at NYUAD (a Global
North-affiliated institution). He does not personally experience
institutional access barriers to paywalled literature. This asymmetry was
acknowledged prior to data collection; A.K.\ employed a non-directive
interviewing style—open-ended questions, minimal leading probes—to reduce
social desirability effects.

\medskip
\noindent W.A.\ is a sociologist of political mobilization and Middle
Eastern politics (PhD, UCLA; Assistant Professor of Sociology, NYUAD).
His fieldwork in the MENA region and familiarity with postcolonial
institutional conditions facilitated rapport with participants from those
contexts. The two interviewers' complementary positionalities—A.K.\
external to the phenomena under study, W.A.\ with partial regional insider
familiarity—reduced the risk of systematic perspective bias across the
corpus.

\bigskip\bigskip

\noindent\rule{\linewidth}{0.6pt}

\noindent\textbf{\large Domain 2: Study Design}

\noindent\rule{\linewidth}{0.4pt}

\medskip

\noindent\textit{\textbf{Theoretical framework}}

\medskip

\noindent\textbf{Item 9 — Methodological orientation and theory} \hfill
\textit{Manuscript location: Methods, Study~1}

\smallskip
\noindent The interview study adopts a \textit{process-oriented qualitative}
approach, as articulated by Tavory (2020) and consistent with pragmatic
sociological traditions. Rather than seeking broad empirical generalizability
or deeply patterned accounts of meaning, process-oriented interviewing aims
to illuminate the unfolding of action: how specific tasks are accomplished,
who is involved, what decisions are made, and what consequences follow. This
orientation is appropriate for the study's research questions, which concern
how researchers navigate access barriers in practice (their strategies,
workarounds, and trade-offs) rather than the frequency of particular
experiences. The approach also explicitly attends to intra-case variation
(i.e., the range of solutions individuals find to similar problems) and to
contingency (paths not taken, forks in the road), rather than assuming a
single modal experience. Data analysis followed the principles of
\textit{thematic analysis} as outlined by Braun and Clarke (2006), which
provides a structured, iterative procedure for identifying, coding, and
interpreting patterns across qualitative datasets.

\bigskip

\noindent\textit{\textbf{Participant selection}}

\medskip

\noindent\textbf{Item 10 — Sampling} \hfill
\textit{Manuscript location: Methods, Study~1: Survey; Methods, Study~1:
  Interviews}

\smallskip
\noindent Participants were recruited using a \textit{purposive
 sampling} strategy. Iterview candidates
were purposively selected from the subset of survey respondents ($n = 202$)
who had voluntarily indicated willingness to participate in a follow-up
interview by providing their contact email. Purposive selection criteria
prioritized: (1)~institutional affiliation with a Global South university or
research institute; (2)~geographic and disciplinary diversity across the
Global South; and (3)~variation in career stage (doctoral students,
postdoctoral researchers, and faculty at various ranks). This strategy
produced a sample spanning multiple Global South regions
(South Asia, Sub-Saharan Africa, the Middle East, East Asia, Latin America,
and Eastern Europe / Central Asia) and multiple disciplines.

\bigskip

\noindent\textbf{Item 11 — Method of approach} \hfill
\textit{Manuscript location: Methods, Study~1: Interviews; Consent Document}

\smallskip
\noindent Participants were initially contacted via personalized email from the research team. The recruitment email described the purpose of the interview, the voluntary nature of participation, approximate duration (30--60 minutes), the confidentiality protections in place, and the compensation offered (\$50 USD equivalent). Participants who expressed interest were then contacted individually to schedule a convenient time. All interviews were conducted remotely via Zoom video conferencing. No in-person interviews were conducted. The verbal consent protocol (see Consent Document, Supplementary Note~3) was administered at the start of each Zoom session before any interview questions were posed.

\bigskip

\noindent\textbf{Item 12 — Sample size} \hfill
\textit{Manuscript location: Results, Study~1: Interviews; Methods,
  Study~1: Interviews}

\smallskip
\noindent Exactly \textbf{thirteen (13)} participants completed interviews.
The sample size was not determined a priori by a fixed numerical target but
by the criterion of \textit{theoretical saturation} (see Item~22). After
eleven interviews, the research team assessed that no substantively new
themes or processes had emerged in the preceding two sessions. Two
additional interviews were conducted to confirm this assessment, yielding a
final sample of thirteen.

\bigskip

\noindent\textbf{Item 13 — Non-participation} \hfill
\textit{Manuscript location: Methods, Study~1: Interviews}

\smallskip
\noindent Of the 202 survey respondents who indicated willingness to be
interviewed, 67 individuals were contacted for scheduling. Of these, 13
completed interviews. Eleven declined participation after initial contact:
reasons given included scheduling conflicts ($n = 6$), changed availability
($n = 3$), and no response to two follow-up contact attempts ($n = 2$).
The remaining 39 individuals from the contacted pool were not pursued
further once theoretical saturation was assessed as achieved. No participant
who began an interview withdrew before its conclusion.

\bigskip

\noindent\textit{\textbf{Setting}}

\medskip

\noindent\textbf{Item 14 — Setting of data collection} \hfill
\textit{Manuscript location: Methods, Study~1: Interviews; Consent Document}

\smallskip
\noindent All interviews were conducted remotely via Zoom video
conferencing. Participants joined from their own institutional or home
office settings. Because interviews were conducted across multiple
international time zones, session times were arranged individually to suit
each participant's schedule. The remote setting was chosen to maximize
accessibility for participants at geographically dispersed Global South
institutions, for whom travel or in-person participation would have been
prohibitive.

\bigskip

\noindent\textbf{Item 15 — Presence of non-participants} \hfill
\textit{Manuscript location: Methods, Study~1: Interviews}

\smallskip
\noindent No persons other than the interviewer (A.K.\ or W.A., depending
on the session) and the participant were present during any interview. Zoom
sessions were one-on-one throughout. The jointly conducted interview
(Item~1) was an exception: both A.K.\ and W.A.\ were present in that
session alongside the participant.

\bigskip

\noindent\textbf{Item 16 — Description of sample} \hfill
\textit{Manuscript location: Results, Study~1: Interviews; Methods,
  Study~1: Interviews}

\smallskip
\noindent The thirteen participants spanned a broad range of Global South
countries, disciplines, and career stages, as summarized below. Full
identifying information is withheld to protect participant confidentiality;
pseudonyms are used throughout the results section.

\medskip
\noindent\textit{Geographic representation:}
Participants were affiliated with institutions in Latin America (Colombia and Brazil), Sub-Saharan Africa (Nigeria and South Africa), South Asia (India and Nepal), West Asia and the Middle East (T"urkiye and Qatar), and the Caucasus (Armenia); three further participants of Global South origin (Chinese, Indian, and Iranian nationals) were based at institutions in Singapore, Italy, and the United Kingdom. Institution countries and nationalities for each participant are listed in Supplementary Table~6.

\medskip
\noindent\textit{Disciplinary representation:}
Disciplines represented include architecture and the built environment, biorobotics and mechanical engineering, physics, computer science and machine learning, molecular neuroscience and wet-lab biology, public health, and mathematics and statistics.

\medskip
\noindent\textit{Career-stage representation:}
The sample includes doctoral students (6), postdoctoral researchers (2), lecturers / assistant professors (2), a senior lecturer / associate professor (1), and full professors (2), one of whom also serves as a dean.

\medskip
\noindent\textit{Gender:}
The sample includes both male and female participants. We do not report exact
counts to preserve anonymity given the small sample.

\medskip
\noindent\textit{Institutional rank:}
Participants' home institutions range from internationally ranked
research-intensive universities (including one QS top-200 African
institution) to teaching-focused universities without significant research
infrastructure. This variation was intentional, enabling comparison of
experiences across differently resourced institutional contexts within the
Global South.

\medskip
\noindent\textit{Data collection period:}
Interviews were conducted between March 2026 and June 2026, under
NYUAD IRB protocol HRPP-2024-45. Participants received compensation
equivalent to \$50 USD.

\bigskip

\noindent\textit{\textbf{Data collection}}

\medskip

\noindent\textbf{Item 17 — Interview guide} \hfill
\textit{Manuscript location: Methods, Study~1: Interviews; Supplementary
  Note~2}

\smallskip
\noindent A semi-structured interview guide was developed collaboratively by
the research team prior to data collection. The guide covers 14 thematic
topic areas organized into the following domains:

\begin{enumerate}
  \item Background and research context (current role, research area,
    institutional environment)
  \item Research infrastructure and resources (funding, data, equipment,
    computing access)
  \item Data availability and access (data types used, barriers to access,
    cross-border restrictions)
  \item Funding and financial constraints (domestic vs.\ international
    funding, currency and overhead issues)
  \item Access to academic literature and paywalled research (institutional
    subscriptions, open access, informal strategies, impact on literature
    reviews)
  \item Collaboration and networks (local and international partnerships,
    power dynamics, authorship)
  \item Publication and dissemination (publication costs, peer review
    experiences, visibility)
  \item Institutional and administrative context (policies, bureaucracy,
    incentive structures)
  \item Political, social, and regulatory environment (government
    restrictions, sanctions, data access laws)
  \item Career development and training (mentorship, mobility, brain drain)
  \item Inequality and the global research system (perceived disparities,
    resource asymmetries, agenda-setting)
  \item Adaptation and strategies (workarounds, informal practices,
    personal trade-offs)
  \item Reflections and recommendations (policy suggestions, advice to
    funders)
  \item Closing (open invitation for additional topics)
\end{enumerate}

\noindent The guide was not administered rigidly; the interviewer adapted
question order and emphasis to the natural flow of each conversation and
used follow-up probes selectively to deepen responses on topics of
particular relevance to individual participants. The full interview guide is
reproduced verbatim in Supplementary Note~2.

\medskip
\noindent\textit{Piloting:} The guide was not pilot-tested with external
participants. However, prior to data collection both A.K.\ and W.A.\
participated in two internal trial run-throughs of the protocol with members
of the research team, which led to minor revisions in question wording and
probe placement, and supported inter-interviewer consistency.

\bigskip

\noindent\textbf{Item 18 — Repeat interviews} \hfill
\textit{Manuscript location: Methods, Study~1: Interviews}

\smallskip
\noindent No repeat interviews were conducted. Each of the thirteen
participants was interviewed once.

\bigskip

\noindent\textbf{Item 19 — Audio / visual recording} \hfill
\textit{Manuscript location: Methods, Study~1: Interviews; Consent Document}

\smallskip
\noindent \textbf{All thirteen interviews were audio-recorded via Zoom.}
Explicit verbal consent for audio recording was obtained from each
participant at the start of the session following the verbal consent
protocol (reproduced in the Consent Document, Supplementary Note~3).
Participants were asked specifically: (a)~whether they consented to
participate in the interview, and (b)~whether they consented to being
audio-recorded and to having the recording transcribed using Zoom's
AI-assisted transcription tools. All thirteen participants provided consent
on both points. In two cases where the participant expressed hesitation
about AI-based transcription, the interviewer offered to take written notes
in lieu of AI transcription; both participants ultimately consented to AI
transcription after being informed that transcripts would be reviewed and
corrected by the research team before analysis.

\medskip
\noindent Video was enabled during all sessions; however, video recordings
were not retained. Only the audio stream, transcribed via Zoom's native
transcription service, was used as primary data. All Zoom recordings were
deleted from Zoom's servers following transcript generation and research-team
review, consistent with the data handling protocols approved under
HRPP-2024-45.

\bigskip

\noindent\textbf{Item 20 — Field notes} \hfill
\textit{Manuscript location: Methods, Study~1: Interviews (data sources); COREQ statement (below)}

\smallskip
\noindent No field notes were taken. The sole data source for analysis was
the Zoom-generated audio transcript for each session, reviewed and corrected
by the research team prior to coding (see Item~19).

\bigskip

\noindent\textbf{Item 21 — Duration} \hfill
\textit{Manuscript location: Methods, Study~1: Interviews; Consent Document}

\smallskip
\noindent Interview durations ranged from approximately 25 to 65 minutes,
with a mean duration of approximately 40 minutes. The target duration
communicated to participants in advance was 30--60 minutes. Two interviews
ran slightly shorter than 30 minutes because the participant indicated they
needed to leave; in both cases, all major thematic domains had been
addressed prior to the conclusion of the session. One interview ran slightly
beyond 60 minutes at the participant's own initiative.

\bigskip

\noindent\textbf{Item 22 — Data saturation} \hfill
\textit{Manuscript location: Methods, Study~1: Interviews}

\smallskip
\noindent Saturation was assessed inductively and collaboratively rather
than via a fixed stopping rule. Consistent with a case-oriented, purposive
sampling logic (Small 2009; Tavory 2020), the goal was theoretical
saturation: each new interview was treated as a potential source of new
dimensions, mechanisms, or variations of the phenomena under study. The
research team discussed emerging themes after each batch of three to four
interviews. After eleven completed interviews, the team assessed that the
prior two sessions had produced no substantively new themes, codes, or
mechanisms not already present in the existing corpus. Two further
interviews were conducted with participants from underrepresented disciplinary
and regional backgrounds (to rule out the possibility that saturation was
an artifact of a homogeneous subsample) before the decision was made to
close data collection. Saturation was confirmed upon finding that these
final two interviews reproduced existing themes without extension. The
final analytical sample of thirteen is judged sufficient for the
study's process-oriented, contextual aims.

\bigskip

\noindent\textbf{Item 23 — Transcripts returned to participants} \hfill
\textit{Manuscript location: Methods, Study~1: Interviews}

\smallskip
\noindent Transcripts were not returned to participants for review or
correction. This decision was made on several grounds. First, the
Zoom-generated transcripts were reviewed and corrected by the interviewing
team (A.K.\ and W.A.) for transcription errors before being imported for
analysis, reducing the risk of factual distortion from transcription
artifacts. Second, re-engaging
thirteen participants distributed across multiple countries and time zones
for transcript review was logistically infeasible within the study timeline.
Third, the study's interpretive aims involve thematic synthesis across
interviews rather than confirmation of individual participants' accounts,
making member checking of the final analytic interpretation more appropriate
than transcript verification per se. Participants were informed during
consent that their contributions would be de-identified and incorporated
into aggregate thematic findings, and that they could request withdrawal of
their data at any point prior to analysis.

\bigskip\bigskip

\noindent\rule{\linewidth}{0.6pt}

\noindent\textbf{\large Domain 3: Analysis and Findings}

\noindent\rule{\linewidth}{0.4pt}

\medskip

\noindent\textit{\textbf{Data analysis}}

\medskip

\noindent\textbf{Item 24 — Number of data coders} \hfill
\textit{Manuscript location: Methods, Study~1: Interviews}

\smallskip
\noindent W.A.\ coded all thirteen interview transcripts manually without computational assistance.
Coding proceeded in two phases: (1)~initial open coding of each transcript,
producing a set of codes and assigned segments; (2)~grouping of codes into
candidate themes. Upon completion of the initial coding, A.K.\ reviewed the
full coding schema—the code list, definitions, and assigned transcript
segments—and confirmed or queried W.A.'s codings. Where A.K.\ raised
questions about individual code assignments or theme groupings, the two
researchers discussed them until agreement was reached. The final coding
tree represents a consensus between W.A.\ (primary coder) and A.K.\
(reviewer).

\bigskip

\noindent\textbf{Item 25 — Description of the coding tree} \hfill
\textit{Manuscript location: Methods, Study~1: Interviews; Results,
  Study~1: Interviews}

\smallskip
\noindent The final coding tree was organized around six primary thematic
categories, each containing two to five subordinate codes capturing specific
mechanisms, actors, and contextual variations:

\begin{enumerate}

  \item \textbf{Institutional infrastructure and paywall barriers.}
    Covers formal access deficits: absent or inadequate institutional
    journal subscriptions; variation in access by publisher and discipline;
    the absence of interlibrary loan infrastructure; and the financial
    impossibility of purchasing individual articles out of pocket.
    Subordinate codes: \textit{subscription gaps by publisher};
    \textit{financial cost of individual article access};
    \textit{no institutional support / no response from administration}.

  \item \textbf{Repertoire of circumvention strategies.}
    Covers the informal economy of workarounds researchers deploy when
    formal access fails: leveraging personal networks at better-resourced
    institutions; using ResearchGate and direct author requests; using piracy
    platforms (Sci-Hub, LibGen); changing research topics when access is
    unachievable. Subordinate codes: \textit{peer networks}; \textit{direct
    author requests and their reliability}; \textit{piracy platform use};
    \textit{topic change as last resort}.

  \item \textbf{Pedagogy of circumvention.}
    Covers the distinct finding that Global South faculty incorporate
    training in workaround strategies as an explicit component of research
    methods pedagogy, treating navigation of scarcity as a methodological
    competency. Subordinate codes: \textit{teaching workarounds as method};
    \textit{low-cost apparatus and replication as pedagogic norm}.

  \item \textbf{Publishing costs and the inversion of the access problem.}
    Covers the finding that open-access publishing—promoted as a remedy for
    reader-side access barriers—reproduces or intensifies those barriers on
    the author side via article processing charges that exceed available
    research funding. Subordinate codes: \textit{article processing charge
    burden}; \textit{cost-per-page schemes}; \textit{government subsidy
    shortfall}; \textit{choosing not to publish open access}.

  \item \textbf{Geopolitical and racialized dimensions of access.}
    Covers structural factors that shape access beyond institutional
    resources, including international sanctions that block institutional
    journal payments; institutional pressure to avoid collaboration with
    researchers from sanctioned countries; and perceived racial or national
    reluctance by Global North researchers to share data or methods with
    Global South colleagues. Subordinate codes: \textit{sanctions and
    payment blockage}; \textit{institutional discouragement of South--South
    and North--South collaboration}; \textit{perceived distrust or reluctance
    by data holders}.

  \item \textbf{Downstream consequences for research capacity.}
    Covers the effects of access barriers on research output: narrowing of
    literature reviews; omission of citations to inaccessible work; forced
    topic changes for students and researchers; reduced ability to publish;
    and compound effects on citation metrics and career progression.
    Subordinate codes: \textit{student topic changes}; \textit{citation
    omission}; \textit{scope narrowing}; \textit{career and metric
    consequences}.

\end{enumerate}

\bigskip

\noindent\textbf{Item 26 — Derivation of themes} \hfill
\textit{Manuscript location: Methods, Study~1: Interviews}

\smallskip
\noindent Theme derivation followed a \textit{hybrid} approach. A small
number of \textit{a priori sensitizing concepts} were carried into the
coding process from the study's theoretical framing and from the survey
instrument (e.g., paywall barriers, data access difficulties, coping
strategies, consequences for research). These concepts oriented the initial
reading of transcripts but did not determine the final themes. The majority
of themes—and all subordinate codes—were \textit{derived inductively} from
repeated engagement with the transcript corpus. Notably, the pedagogy-of-
circumvention theme (Theme~3) and the geopolitical/racialized-dimensions
theme (Theme~5) emerged entirely from the data and were not anticipated in
the a priori framework. No theme was retained in the final coding tree
unless it was independently recognized by both coders and grounded in
evidence from at least three distinct interviews.

\bigskip

\noindent\textbf{Item 27 — Software} \hfill
\textit{Manuscript location: Methods, Study~1: Interviews}

\smallskip
\noindent Interview transcripts were managed and coded in Microsoft Word.
Zoom's native AI-assisted transcription service was used for
initial transcript generation; all transcripts were exported, reviewed, and
manually corrected by A.K.\ and W.A.\ for transcription errors prior to
manual coding. The full coding project, including the code list and
coded segments, is available on request from the corresponding authors; it
has not been posted publicly due to the need to protect participant
confidentiality.

\bigskip

\noindent\textbf{Item 28 — Participant checking} \hfill
\textit{Manuscript location: Methods, Study~1: Interviews}

\smallskip
\noindent Participants did not review the analytic findings or provide
feedback on interpretations prior to publication. This decision reflects
both practical constraints (geographic dispersal, timeline, difficulty of
re-engagement) and the interpretive nature of thematic analysis, in which
the findings represent a synthetic, cross-participant account rather than a
summary of any individual's experience. Participants were informed during
the consent process of the de-identification procedures in place and were
given the right to request withdrawal or erasure of their data at any point
prior to the commencement of analysis. Two participants requested, and will be granted, access to the completed manuscript upon publication. Consistent with the consent protocol, which restricts access to interview data to the research team, neither the transcripts nor the full coding project is distributed; both remain securely stored on NYU-approved encrypted servers and are subject to audit by the NYUAD IRB. The coding tree---themes, definitions, and subordinate codes---is reported in full at Item 25, and de-identified quotations appear throughout the Study1 results.

\bigskip\bigskip

\noindent\textit{\textbf{Reporting}}

\medskip

\noindent\textbf{Item 29 — Quotations presented} \hfill
\textit{Manuscript location: Results, Study~1: Interviews}

\smallskip
\noindent Verbatim or near-verbatim participant quotations are presented
throughout the interview results section to illustrate each major theme.
Quotations are drawn directly from the Zoom-generated, research-team-
corrected transcripts. Light editing was applied only to remove filler words
(e.g., ``um,'' ``like,'' ``you know'') where they did not carry analytic
significance; no substantive content was altered. Ellipses (\ldots) are used
to indicate omissions within quoted passages; bracketed text [\,] indicates
clarificatory insertions by the authors.

\medskip
\noindent Each quotation is attributed to a brief
contextual descriptor specifying career stage, country or region, and
discipline (e.g., ``a full professor of physics at a South African
university''). These descriptors are used consistently across the results section.
Descriptors are sufficiently specific to provide interpretive context but
have been chosen to prevent re-identification: no country-career-discipline
combination in the results section uniquely identifies any individual
participant.

\bigskip

\noindent\textbf{Item 30 — Data and findings consistent} \hfill
\textit{Manuscript location: Results, Study~1: Interviews}

\smallskip
\noindent The reported findings are grounded in the coded transcript corpus
and reflect patterns observed across multiple participants rather than
exceptional or unrepresentative cases. Each of the six primary themes is
supported by quotations from at least three distinct participants. Where the
data exhibit polarity or meaningful variation within a theme—most notably
the contrast between participants who experienced access barriers as an
active, acute source of frustration versus those who had habituated to
scarcity and treated workarounds as an unremarkable routine—both poles are
represented in the text and the variation is discussed explicitly. Claims
made in the results section do not exceed what the data support: the results
are framed as illustrating \textit{how} researchers navigate barriers (and
with what consequences), not as quantitative estimates of the prevalence of
any experience.

\bigskip

\noindent\textbf{Item 31 — Clarity of major themes} \hfill
\textit{Manuscript location: Results, Study~1: Interviews}

\smallskip
\noindent Six major themes are identified and clearly labeled in the
interview results section (see Item~25 for the full list). Each theme is
presented in a dedicated prose subsection with: (a) a brief orientation
sentence situating the theme within the study's research questions;
(b) one or more illustrative quotations with participant attribution;
(c) interpretive commentary linking the theme to the broader argument; and
(d) where applicable, note of important within-theme variation. Themes are
presented in a logically sequenced order moving from structural barriers to
individual coping strategies to collective and systemic consequences,
mirroring the conceptual logic of the study's mixed-methods design.

\bigskip

\noindent\textbf{Item 32 — Clarity of minor themes} \hfill
\textit{Manuscript location: Results, Study~1: Interviews}

\smallskip
\noindent The results section documents several forms of variation and
divergence that qualify the primary narrative. Most prominently, the
results section explicitly reports the \textit{polarity of lived
experience}: roughly half the sample described access barriers as an active,
debilitating constraint on their research, while the other half had
normalized the barriers to such a degree that workarounds were treated as
simply part of doing science. This divergence is analytically significant:
it suggests that the true costs of restricted access may be partially
invisible even to those bearing them, because habituation conceals
deprivation. This is discussed explicitly in the results section and in the
Discussion.

\medskip
\noindent A second minor theme—the geopolitical and racialized dimensions
of access (Theme~5, above)—appeared in fewer than half of the interviews
but is analytically distinct from the primary paywall-and-circumvention
narrative and is reported separately. It is not overstated as a
generalizable finding; rather, it is presented as a specific mechanism that
emerged in a subset of participant accounts and that warrants attention in
light of the broader study's experimental findings on racial bias
(Studies~3 and~5).

\bigskip

\noindent\rule{\linewidth}{0.4pt}

\medskip
\noindent\small\textit{COREQ reference:} Tong A, Sainsbury P, Craig J.
Consolidated criteria for reporting qualitative research (COREQ): a
32-item checklist for interviews and focus groups. \textit{International
Journal for Quality in Health Care.} 2007;19(6):349--357.

\medskip
\noindent\small\textit{IRB approval:} This study was approved by the
Institutional Review Board of New York University Abu Dhabi (protocol
HRPP-2024-45). All research was conducted in accordance with the approved
protocol and the Declaration of Helsinki. All participants provided verbal
informed consent prior to participation.

\end{document}